\documentclass[12pt]{article}
\usepackage{ae,aecompl}
\usepackage{cite}
\usepackage{amsmath,amsfonts,amssymb}
\usepackage{graphicx,color}
\usepackage{epsfig,epsf}
\usepackage{bm}

\usepackage{relsize}
\usepackage{pifont}

\usepackage[bookmarks, breaklinks, colorlinks,urlcolor=Darkgreen, citecolor=red, 
linkcolor=Darkgreen]{hyperref}

\usepackage{url}
\usepackage{multirow}
\usepackage{slashed}
\usepackage{graphbox}
\usepackage{subfigure}

\usepackage[T1]{fontenc}
\usepackage{array,booktabs, makecell}
\usepackage[title]{appendix}
\usepackage{tikz}
\usepackage{tikz-feynman}
\tikzfeynmanset{compat=1.0.0}

\usepackage{ulem}
\usepackage{soul}

\def\dsp{\displaystyle}
\def\bea{\begin{eqnarray}}
\def\eea{\end{eqnarray}} 
\def\be{\begin{equation}}
\def\ee{\end{equation}} 
\def\nn {\nonumber}
\def \Re{\text{Re}}

\def\gev{\ensuremath{\mathrm{Ge\kern -0.1em V}}}
\def\mev{\ensuremath{\mathrm{Me\kern -0.1em V}}}

\def\abs#1{\left| #1 \right|}

\def\ptmiss{\not\!\!{p_T}}

\definecolor{Darkgreen}{RGB}{0,120,120}

\newcommand{\fbi}{fb$^{-1}$ }

\newcommand{\xmark}{\ding{55}}%
\newcommand{\cmark}{\ding{51}}%

\definecolor{amethyst}{rgb}{0.6, 0.4, 0.8}

\definecolor{antiquefuchsia}{rgb}{0.57, 0.36, 0.51}
\definecolor{armygreen}{rgb}{0.29, 0.43, 0.13}
\newcommand{\pp}[1]{{\color{armygreen} #1}}

\begin{document}

\begin{flushright}
IITH-PH-0002/21\\
SI-HEP-2021-22\\[0.2cm]
%\today
\end{flushright}

	%\renewcommand*{\thefootnote}{\fnsymbol{footnote}}

%%%%%%%%%%%%%%%%%%%%%%%%%%%%%%%%%%%%%%%%%%%%%%%%%%%%%%%
%\mbox{}\hfill{IFIC/19-32}
%\vskip 2cm	

\begin{center}
	
	{\Large\bf 
		Distinguishing signatures of scalar leptoquarks \\ at hadron and muon colliders}
	\\[10mm]
	{
		Priyotosh Bandyopadhyay$^a$\,\footnote{Email: bpriyo@phy.iith.ac.in}, Anirban Karan$^{a\, b}$\,\footnote{Email: kanirban@ific.uv.es}, Rusa Mandal$^{c}$\,\footnote{Email: rusa.mandal@iitgn.ac.in} and 	Snehashis Parashar$^a$\,\footnote{Email: ph20resch11006@iith.ac.in}
	}
	\\[10pt]
	$^a$\,{\small\it  Indian Institute of Technology Hyderabad, Kandi, Sangareddy-502284, Telangana, India} \\[2mm]
	$^b$\,{\small\it  Instituto de F\'isica Corpuscular (CSIC - Universitat de Val\`encia)\\	
		Apt. Correus 22085, E-46071 Val\`encia, Spain} \\[2mm]
%	$^c$\,{\small\it Center for Particle Physics Siegen (CPPS),
%		Theoretische Physik 1,   \\ Universit\"at Siegen, 57068 Siegen, Germany}\\
	$^c$\,{\small\it Indian Institute of Technology, Gandhinagar, Gujarat-382355, India}

\end{center}
%%%%%%%%%%%%%%%%%%%%%%%%%%%%%%%%%%%%%%%%%%%%%%%%%%%%%%%

\vspace{5mm}

\begin{abstract}
	
While the hunt for new states beyond the standard model (SM) goes on for various well motivated theories, the leptoquarks are among the most appealing scenarios at recent times due to a series of tensions observed in $B$-meson decays. We consider  $SU(2)$ singlet and  triplet scalar leptoquarks separately, which contribute to charged and neutral current $B$-meson decays. Focusing on the single production of these two scalar leptoquarks, we perform a PYTHIA-based simulation considering all the dominant SM backgrounds at the current and future setups of the Large Hadron Collider (LHC).  The mono-$b$-jet + $\ptmiss$ finalstate gives the strongest signal for the singlet leptoquark at the 30 TeV LHC or Future Circular Collider (FCC), with a possibility of $5\sigma$ signal significance with $\gtrsim 1000$ \fbi of integrated luminosity, for the chosen benchmark scenarios. The finalstate consisting of a $c$-jet and two $\tau$-jets provides highest reach for the singlet leptoquark, probing   an $\mathcal{O}(10^{-1})$ value of the  Yukawa-type couplings for up to $3.0$ TeV leptoquark mass. For the triplet leptoquark, $1-{\rm jet}+2\mu + \ptmiss$ topology is the most optimistic signature at the LHC, probing leptoquark couplings to fermions at $\mathcal{O}(10^{-1})$ value for the leptoquark mass range up to $ 4.0$ TeV. The invariant mass edge distribution is found to be instrumental in determination of the leptoquark mass scale at the LHC. We also perform  the  analysis at the proposed multi-TeV muon collider, where an $\mathcal{O}(10^{-1})$   leptoquark Yukawa coupling  can be probed for a $5.0$ TeV leptoquark mass. 
	
\end{abstract}
{\hypersetup{linkcolor=black}
\newpage
\tableofcontents}

%%%%%%%%%%%%%%%%%%%%%%%%%%%%%%%%%%%%%%%%%%%%%%%%%%%%%%%%%
\section{Introduction}
\label{sec:Intro}

Leptoquarks among the most promising beyond the Standard Model (SM) candidates have been extensively searched at the experiments in past few years and the hunt is very much on at recent colliders. These colour charged bosons couple to quarks and leptons at the tree level and carry electromagnetic charge as well. Although the idea of quark-lepton unification was put forward in the 70's~\cite{Georgi:1974sy,Pati:1974yy}, leptoquarks have drawn a significant attention recent days in order to explain the tensions observed in $B$-decays by several experimental collaborations~\cite{Aaij:2019wad,Aaij:2021vac,Aaij:2017vbb,LHCb:2021zwz,Aaij:2020nrf,HFLAV:2019otj}.

Leptoquark can be pair or singly produced at the colliders and because of its strong interaction nature, pair production generally dominates at the LHC~\cite{Blumlein:1996qp,Belyaev,Kramer:1997hh,Plehn:1997az,Eboli:1997fb,Kramer:2004df}. However the single production, which is governed by the model dependent Yukawa-type couplings, can also be significant in higher mass region~\cite{Hammett:2015sea,Mandal:2015vfa}. Such Yukawa-type couplings are directly related to the low energy processes like meson decays.   Hence, at present when no signatures of these new particles have been found at the LHC, which in turn is regularly pushing the
lower limit of the allowed masses upwards, it is important to study the single production in connection with the possible hints of new physics (NP) seen in $B$-decays.

In this paper, we focus on two scalar leptoquarks namely $S_3$ and $S_1$ having $(\bar{\bf 3}, {\bf 3}, 1/3)$ and $(\bar{\bf 3}, {\bf 1},1/3)$ quantum numbers, respectively, under the SM gauge group $(SU(3)_c,\,SU(2)_L,\,U(1)_Y)$. This choice is motivated with the possibility to address the discrepancies observed in either or both the $b\to s \mu \mu$ (neutral current (NC)) and $b\to c \tau \bar{\nu}$ (charged current (CC)) transitions~\cite{Babu:2020hun,Marzocca:2018wcf,Saad:2020ihm,Gherardi:2020qhc,Becirevic:2018afm,Bigaran:2019bqv,Crivellin:2019dwb,Crivellin:2017zlb,Aydemir:2019ynb,Mandal:2018kau,Iguro:2020keo,Lee:2021jdr,Bordone:2020lnb}. Phenomenology of scalar leptoquarks~\cite{Dorsner:2021chv,Bandyopadhyay:2018syt,Bhaskar:2020gkk,Bhaskar:2021pml,Bhaskar:2021gsy,DaRold:2021pgn,Hiller:2021pul,Haisch:2020xjd,Chandak:2019iwj,Bhaskar:2020kdr,Alves:2018krf,Dorsner:2019vgp,Mandal:2018qpg,Padhan:2019dcp,Baker:2019sli,Nadeau:1993zv,Atag:1994hk,Atag:1994np,Buchmuller:1986zs,Hewett:1987bh,Hewett:1987yg,Cuypers:1995ax,Bandyopadhyay:2016oif,Saad:2020ucl} and R-parity violating scalars (which resemble the leptoquark scenarios)~\cite{BhupalDev:2021ipu,Altmannshofer:2020axr,Altmannshofer:2017poe} at the colliders has been studied in literature with main emphasis on the pair production. The distinct features of scalar and vector leptoquarks carrying all possible combinations of the SM gauge quantum numbers are explored at the lepton-photon collider \cite{Bandyopadhyay:2020klr}, electron-proton collider \cite{Bandyopadhyay:2020jez} and at the LHC \cite{Bandyopadhyay:2020wfv,Dutta:2021wid} as well. The couplings to first generation of quarks are leptons are stringently constrained from Kaon and lepton physics~\cite{Mandal:2019gff,Davidson:1993qk,Dorsner:2016wpm,Crivellin:2021egp} and the recent ATLAS searches performed with a centre-of-mass energy of 13\,TeV and an integrated luminosity of 139\,fb$^{-1}$~\cite{Aad:2020iuy} exclude mass up to $1.8\,$TeV decaying into an electron and a quark. The limits are relatively weaker while looking for finalstates into third generation of fermions~\cite{CMS:2020gru,Aad:2021rrh}. 

The finalstate topologies studied in this work is directly related to the channels where certain tensions have been observed in $B$-decays and thus the phenomenology of $S_1$ leptoquark aims at modes with $\tau$ lepton and neutrinos in the finalstate whereas for $S_3$ leptoquark mostly muons and neutrinos are present. This provides very interesting and distinguishable signatures for the direct searches which can probe the most favored parameter space. Apart from the current setup of the LHC, this work also presents outcomes for the potential of the high luminosity LHC and the high energy LHC projects~\cite{Cepeda:2019klc} to measure the properties of the considered leptoquarks. In view of the European Strategy Update for Particle Physics released its recommendation to investigate the technical and financial feasibility of a future hadron collider (FCC) at CERN with a center-of-mass energy of at least 100 TeV~\cite{FCC:2018byv}, we provide the analysis for such a setup as well.

Recently, there is a growing interest in the community for a multi-TeV muon collider which can succeed the LHC~\cite{Ankenbrandt:1999cta,Neuffer:2018yof,Delahaye:2019omf,Han:2021kes}. Due to less synchrotron radiation of muon compared to electron, no initial state QCD radiation, centre-of-mass frame and significantly reduced background environment in contrast to hadron colliders, a muon collider has potential to look for new states beyond the SM~\cite{Bandyopadhyay:2020otm,Huang:2021nkl,Sen:2021fha,Costantini:2020stv}. The advantage is eminent for the NP mediators having direct connection to the $b\to s \mu \mu$ anomalies~\cite{Huang:2021biu,Asadi:2021gah}. Hence,  in this article, we explore the phenomenology of $S_3$ leptoquark at muon collider as well. The pair production of leptoquark provides interesting signatures such as di-muon plus jets at muon collider which has spectacular sensitivity for the leptoquark coupling and mass parameters.  

The rest of the paper is organized as follows. In \autoref{sec:framework}, we discuss the theoretical framework behind the choice of the benchmark scenarios (in \autoref{sec:theory}) and then specify the setup used for the phenomenological study at colliders (in \autoref{sec:setup_colliders}). We perform the LHC simulation for the single production of the scalar leptoquark $S_1$ in \autoref{sec:S1}; starting with the kinematic distributions (in \ref{sec:kine_S1}) and then with two separate subsections (\ref{sec:S1_btau} and \ref{sec:S1_ctau}) differing due to the flavour of the jets in the finalstates. The invariant mass edge distribution for $S_1$ is discussed in \autoref{sec:edge}. Similar analysis for the phenomenology at the current and future LHC for the scalar leptoquark $S_3$ is described in \autoref{sec:S3}. Several subsections are devoted to study the signatures arising from the different components of this electroweak triplet leptoquark and we separately analyze lepton flavour violating signatures in the decay for $S_3$ in \autoref{sec:lfv}. We perform the simulation at a multi-TeV muon collider for the scalar leptoquarks in \autoref{sec:muon}. Finally, \autoref{sec:reachplots} presents comparison of all the results for both of these leptoquarks highlighting the prospects at current and future colliders and our concluding remarks are mentioned in \autoref{sec:summary}.

%%%%%%%%%%%%%%%%%%%%%%%%%%%%
\section{Framework}
\label{sec:framework}
In this section starting with the interaction Lagrangians of the two scalar leptoquarks, we obtain the benchmark scenarios which can explain any of the two types of tensions observed in neutral and charged current $B$-decays while being consistent with other data. Then we describe the basic set up used in our analysis to study the collider phenomenology at the LHC, FCC as well as at the proposed multi-TeV muon collider. 

\subsection{Theory and benchmark points}
\label{sec:theory}
We consider two scalar leptoquarks $S_1(\bar{\bf 3}, {\bf 1},1/3)$ and $S_3(\bar{\bf 3}, {\bf 3}, 1/3)$ separately, and write the interaction Lagrangians for them with the SM fermions as
\begin{align}
\label{eq:LS1}
\mathcal{L}_{S_1}\; &=\;  \overline{Q^c}^i\, i \tau_2\, Y_{\tiny S_1}^{i\alpha} L^\alpha\; S_1 + \overline{ u^c_R}^i \,Z_{\tiny S_1}^{i\alpha} \ell_R^\alpha\; S_1+ \rm{h.c.}\,, \\
\label{eq:LS3}
\mathcal{L}_{S_3}\; &=\overline{Q^c}^i\, Y_{\tiny S_3}^{i\alpha} \,i\tau_2  \, {\boldsymbol \tau\bf \cdot S_3} \, L^\alpha+ \rm{h.c.}  \,,
\end{align}
respectively, where we denote the left-handed SM quark (lepton) doublets as $Q$ ($L$), while $u_R$ ($d_R$) and $\ell_R$ are the right-handed up (down)-type quark and lepton singlets, respectively. The notation $f^c\equiv \mathcal{C}\bar f^{\, T}$ indicates the charge-conjugated field of the fermion $f$.
Here $Y_{\tiny\rm LQ}$ and $Z_{\tiny \rm LQ}$ are completely arbitrary Yukawa-type matrices in flavour space and $\tau_k,~k\in \{1,2,3\}$ are the Pauli matrices. Expanding the interaction terms in the mass-eigenstate basis we get
\begin{align}
\label{eq:LS12}
\mathcal{L}_{S_1}\; &= \left[ {\overline{u^c_{L\!}}}^{\, i} (V^* Y_{\tiny S_1})^{ij}  \ell_L^j - {\overline{d^c_{L\!}}}^{\, i}\, Y_{\tiny S_1}^{ij}\,  \nu_L^j + {\overline{ u^c_{R\!}}}^{\, i}\,  Z_{\tiny S_1}^{ij} \ell_R^j  \right] S_1+ \rm{h.c.}\,, \\
\label{eq:LS32}
\mathcal{L}_{S_3}\; &=  -{\overline{d^c_{L\!}}}^{\, i}  Y_{\tiny S_3}^{ij} \nu_L^j\, S_3^{1/3}  \!-\! \sqrt{2}\, {\overline{d^c_{L\!}}}^{\, i}  Y_{\tiny S_3}^{ij}\, \ell_L^j\, S_3^{4/3} + \sqrt{2}\, {\overline{u^c_{L\!}}}^{\, i}  (V^*Y_{\tiny S_3})^{ij} \nu_L^j\, S_3^{-2/3} - {\overline{u^c_{L\!}}}^{\, i}  (V^* Y_{\tiny S_3})^{ij} \ell_L^j\, S_3^{1/3} + \rm{h.c.}\,.
\end{align}
The transformation from the fermion interaction eigenstates to mass eigenstates is simply given by $u_L \to V^\dagger u_L$, where $V$ is the quark Cabibbo-Kobayashi-Maskawa (CKM) matrix \cite{Cabibbo:1963yz,Kobayashi:1973fv} and we have neglected the unitary matrix in the neutrino sector. Note that, being a triplet under $SU(2)_L$, $S_3$ has three components differing in electric charges which are shown in the superscripts.

It is apparent from \autoref{eq:LS12} that at tree level, $S_1$ contributes to the $b\to c \tau \bar{\nu}$ transition, whereas, $S_3$ in \autoref{eq:LS32} promotes both $b \to s \mu \mu$ and $b\to c \tau \bar{\nu}$ modes. Considering one leptoquark at a time,  the minimal set of non-zero couplings required for the above mentioned leptoquarks to explain either of the $b \to s \mu \mu$ or $b \to c \tau \bar\nu$ anomalies are summarized in \autoref{tab:all}. Here the contribution to CC mode is via the following effective Hamiltonian 
\bea
\label{eq:Heffb2c}
{\cal H}_\text{eff}^{\rm CC}\, =\, \frac{ 4 G_F V_{cb}}{\sqrt{2}}\left[ \mathcal{C}_{L}^{S}
\left(\bar{c}\, P_L b\right)\left(\bar{\tau} P_L \nu \right) +\mathcal{C}_{L}^{T} \left(\bar{c}\,\sigma^{\mu \nu} P_L b\right)\left(\bar{\tau} \sigma_{\mu \nu} P_L \nu\right) \right]\,,
\eea
where
\begin{align}
\mathcal{C}_{L}^{S} (M_{\tiny S_1}) &=-4  \mathcal{C}_{L}^{T} (M_{\tiny S_1})  = - \dsp\frac{v^2}{4 M_{\tiny S_1}^2} \frac{1}{V_{cb}} Y_{\tiny S_1}^{33} Z_{\tiny S_1}^{ *23}\,.
\end{align}
The ratios, defined as $R(D^{(*)})\equiv \mathrm{BR}(B \to D^{(*)} \tau \bar{\nu})/ \mathrm{BR}(B \to D^{(*)} \ell \bar{\nu})$, with $\ell=\{e,\,\mu \}$, can then be expressed as~\cite{Mandal:2020htr}
\bea
R(D)/R(D)_{\rm SM} &\!\!\approx&\!\! 1  + 1.504 \,  \Re \left[ \mathcal{C}^{S*}_{L}  \right] + 1.171\, \Re \left[  \mathcal{C}^{T*}_{L} \right]+ 1.037 |\mathcal{C}^S_{L}|^2 +
0.939 |\mathcal{C}^{T}_{L}|^2\,, \\
{\cal R}({D^*}) / {\cal R}({D^*})_\text{SM} &\!\! \approx &\!\!\! 1
- 0.114 \, \Re \left[ \mathcal{C}^{S*}_{L} \right] - 5.130 \, \Re \left[ \mathcal{C}^{T*}_{L} \right] - 0.037 |\mathcal{C}^S_{L}|^2+
17.378 |\mathcal{C}^{T}_{L}|^2 \, ,
\eea
where the Wilson coefficients are evaluated at the $m_b$ scale using renormalization group equations and neglecting electroweak contributions: 
$\mathcal{C}^{S(T)}_{L}(m_b) = 1.67(0.84)\times \mathcal{C}^{S(T)}_{L} (\Lambda=\mathcal{O}({\rm TeV}))\,.$ The latest HFLAV average of $R(D^{(*)})$ data indicates 14\% enhancements~\cite{HFLAV:2019otj} compared to the SM predictions and two  desired benchmark values are quoted in \autoref{tab:all}. We have checked that such benchmark points are allowed by the one-loop induced $Z\to \tau \tau$, $Z\to \nu \nu$ decays~\cite{Arnan:2019olv,Falkowski:2019hvp,ColuccioLeskow:2016dox,Crivellin:2020mjs} and $B_s$ mixing bound~\cite{DiLuzio:2019jyq}. Note that, $S_3$ also generates SM-like V-A operator, however, the required couplings to explain the $R(D^{(*)})$ anomalies are forbidden by $Z\to \tau \tau $ and $B_s$-mixing data.

In case of the NC anomalies we generate the following contribution to the effective Hamiltonian 
\be
\begin{aligned}
	\mathcal{H}_\text{eff}^{\rm NC}=-\frac{4 G_F}{\sqrt{2}}\frac{\alpha_\text{EM}}{4\pi}V_{td}V^*_{ts}\left[ C_9^{\rm NP} \,(\bar{s} \gamma^\mu  P_{L} b)(\bar{\mu}  \gamma_\mu \mu)+  C_{10}^{\rm NP}  (\bar{s} \gamma^\mu  P_{L} b)(\bar{\mu}  \gamma_\mu \gamma_5 \mu) \right ],
\end{aligned}
\label{eq:lagrangian_bsll}
\ee
where
\bea
C_9^{\rm NP}= - C_{10}^{\rm NP}=  \frac{ v^2}{M_{S_3}^2} \frac{\pi}{\alpha_\text{EM} V_{tb}V_{ts}^*} Y_{\tiny S_3}^{*32}Y_{\tiny S_3}^{22}\,.
\eea
The existing tensions observed in this mode can be achieved via $C_9^{\rm NP}=-C_{10}^{\rm NP}=-0.41^{+0.07}_{-0.07}$~\cite{Altmannshofer:2021qrr} and such benchmark cases are shown in \autoref{tab:all}, which are allowed by the most constraining bounds arising from $Z\to \mu \mu$, $Z\to \nu \nu$ decays\cite{Arnan:2019olv,Falkowski:2019hvp,ColuccioLeskow:2016dox,Crivellin:2020mjs} and $B_s$ mixing~\cite{DiLuzio:2019jyq}. The subsequent sections are devoted for detailed collider phenomenology studies of such benchmark scenarios for these two leptoquarks $S_1$ and $S_3$ at the LHC/FCC and at a multi-TeV muon collider. We mention that among all five scalar leptoquarks, the weak doublet $R_2(\mathbf{3},\, \mathbf{2},\,7/6 )$ can also accommodate CC anomalies~\cite{Angelescu:2018tyl,Angelescu:2021lln} and, with the minimal choice of Yukawa-type couplings, it might give rise to the similar phenomenology as of $S_1$. However, the detailed analysis of $R_2$ is left for our future work. Additionally, the collider phenomenology is very much dependent on the flavour structure of the leptoquark Yukawa-type couplings, and we stick to the minimal choice required to explain the observed tensions in $B$-decays. Altering the flavour structure of the entries of these couplings will give rise to completely different phenomenology which is beyond the focus of this work. The minimal choice of couplings are presented in \autoref{tab:all}. In order to understand the relevance of the magnitude of such chosen couplings, the variation of signal significance for the most promising cases will be discussed in \autoref{sec:reachplots}.

\begin{table}[h!]
	\renewcommand{\arraystretch}{1.1}
	\centering
	\begin{tabular}{|c|c|c|c|c|c|}
		\hline \noalign{\vskip 2pt}
		LQ		& Mass & NC & CC &  Couplings  & Benchmark \\
		&  (TeV) & && & points \\ 
		%		\cline{5-6} & & \\
		\hline \hline&&&&&\\[-3mm]
		\multirow{2}{*}{$S_1$}&	1.5&	\multirow{2}{*}{\textcolor{red}{\xmark}}&	\multirow{2}{*}{\textcolor{Darkgreen}{\cmark}} & 	$Y_{\tiny S_1}^{33}= 0.91,\, Z_{\tiny S_1}^{23}= -0.50$ & BP1\\[1ex]
		& 2.0& & &$Y_{\tiny S_1}^{33}= 1.10,\, Z_{\tiny S_1}^{23}= -0.74$ & BP2 \\[1ex] \hline\hline&&&&&\\[-3mm]
		\multirow{3}{*}{ $S_3$}&1.5& \multirow{3}{*}{ \textcolor{Darkgreen}{\cmark}} &\multirow{3}{*}{ \textcolor{red}{\xmark}} &$Y_{\tiny S_3}^{22}=0.50 ,\, Y_{\tiny S_3}^{32}=0.003$ & BP1\\[1ex]
		& 2.0 & & & $Y_{\tiny S_3}^{22}=0.60 ,\, Y_{\tiny S_3}^{32}=0.003$ & BP2  \\[1ex]
		& 1.5 & & & $Y_{\tiny S_3}^{22}=0.008 ,\, Y_{\tiny S_3}^{32}=0.20$ & BP3  \\ \hline
		%		\multirow{2}{*}{$R_2$}&\multirow{2}{*}{1.5}& \multirow{2}{*}{\textcolor{red}{\xmark}}&\multirow{2}{*}{ \textcolor{Darkgreen}{\cmark}} &  \\[2ex]		& 2.0& & & \\
	
	\end{tabular}
	\caption{The benchmark points defined with the minimal set of coupling values required for the CC or NC anomalies for $S_1$ and $S_3$ leptoquarks, respectively. The symbol `\textcolor{Darkgreen}{\cmark}' (`\textcolor{red}{\xmark}') denotes agreement (disagreement) at $\pm1\,\sigma$ level for the corresponding observables.}  \label{tab:all}
\end{table}

\subsection{Set up for the LHC/FCC and muon colliders}
\label{sec:setup_colliders}

In this subsection we summarise the kinematic cuts and definition of the collider set up that are  used in simulations. Implementing  the models in  {\tt SARAH} \cite{sarah}, model files are generated for CalcHEP \cite{calchep}. The ``.lhe"  event  files are then generated and interfaced with PYTHIA6.4.5 \cite{pythia6} for hadronization with initial  state radiation (ISR) and final state radiation (FSR). The jet is formed using {\tt Fastjet-3.0.3}\cite{fastjet} with Cambridge/Aachen jet algorithm with  a jet radius of 0.5. The additional  basic cuts, written below, are also implemented.

\begin{itemize}
	\item The calorimeter coverage is $|\eta|<4.5$.
	\item The minimum jet transverse momentum $p_T=20$ GeV and jets are ordered in $p_T$.
	\item Leptons are selected with $p_T \geq 20$\,GeV and $|\eta|<2.5$.
	\item $\Delta R_{\ell j} \geq 0.4 $ and $\Delta R_{jj} \geq 0.2$, where  $\Delta R_{ij}= \sqrt{\Delta \eta_{ij}^2 +\Delta \Phi_{ij}^2}$ is the angle between the $i$-th and $j$-th particles, with $\Delta \Phi_{ij}$ is the difference of the azimuthal angle and $\Delta \eta_{ij}$ is the difference of the pseudo-rapidities.
	\item  We demand that hadronic activity within a cone of $\Delta R=0.3$ of the leptons should be $\leq 0.15\, p^{\ell}_T$\,GeV in the specified cone.
	
	\item As our benchmark points are with leptoquark masses of  1.5 TeV or 2.0 TeV, a hardness cut evaluated as the scalar sum of lepton, jet and missing transverse momentum, $p^H_T= \Sigma (p^\ell_T + p^j_T +\ptmiss) \geq 1.2$ TeV is implemented at the analysis level for both signal and  backgrounds. For computational convergence and to get events at the high-momentum tail, the SM background events were generated with $\sqrt{\hat{s}} \geq$ 1.2 TeV.

\end{itemize}

Armed with the above mentioned collider set up, in the following sections we analyse the phenomenologies of  the single production of $S_1, \, S_3$ leptoquarks at the LHC/FCC with three different choices of the centre-of-mass energies 14 TeV, 30 TeV and 100 TeV.  In this  article  we focus  on the  single leptoquark production for probing  the leptoquark Yukawa couplings. Finalstates coming from such production processes solely  depend  on the  Yukawa couplings, absence of  which make  the finalstates cease  to exist. However, leptoquark pair production dominated by the strong coupling constant can contaminate such finalstates arising from the single leptoquark productions.  We define such contamination as model backgrounds, that can be estimated once we have the information of the leptoquark mass and excitations for a given choice of Yukawa-type coupling that we already have benchmarked. In \autoref{uncer} we discuss the impact of such effects and the corresponding uncertainties in the signal significance.

%%%%%%%%%%%%%%%%%%%%%%%%%%%%

\section{$S_1$ at the LHC/FCC}
\label{sec:S1}

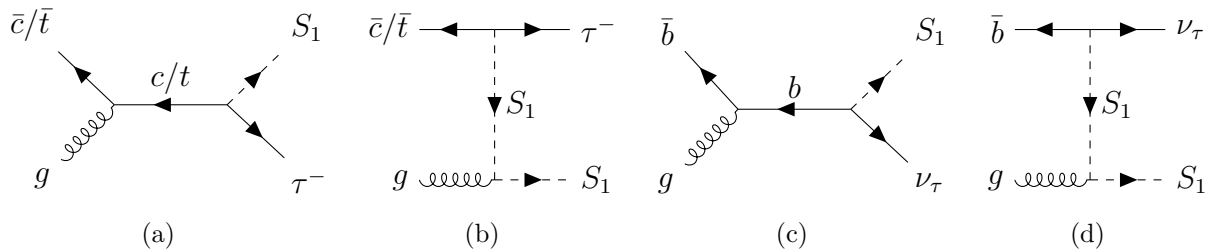
\begin{figure}[h!]
	\subfigure[]{
	\begin{tikzpicture}
		\begin{feynman}
			\vertex (a1);
			\vertex [above left=1cm of a1] (a0){$\bar c/\bar t$};
			\vertex [right=1.5cm of a1] (a2);
			\vertex [above right=1 cm of a2] (a3){$S_1$};
			\vertex [below left=1cm of a1] (b0){$g$};
			\vertex [below right=1cm of a2] (b3){$\tau^-$};
			\diagram {(a0)--[anti fermion](a1)--[anti fermion,edge label=$c/t$](a2)--[charged scalar](a3),
				(b0)--[gluon](a1),(a2)--[fermion](b3)};
		\end{feynman}
		\end{tikzpicture}}
	\subfigure[]{
	\begin{tikzpicture}		
%		\hspace{5.3cm}
		\begin{feynman}
			\vertex (a1);
			\vertex [above=1cm of a1] (z1);
			\vertex [left=10mm of z1] (z0) {$\bar c/\bar t$};
			\vertex [right=10mm of z1] (z2) {$\tau^-$};
			\vertex [below=1cm of a1] (b1);
			\vertex [left=10mm of b1] (b0){$g$};
			\vertex [right=10mm of b1] (b2) {$S_1$};
			\diagram {(z0)--[anti fermion](z1)--[charged scalar, edge label=$S_1$](b1)--[charged scalar](b2),
				(b0)--[gluon](b1),
				(z1)--[fermion](z2)};
		\end{feynman}
		\end{tikzpicture}}		
	\subfigure[]{
			\begin{tikzpicture}
%		\hspace{3.8cm}
		\begin{feynman}
			\vertex (a1);
			\vertex [above left=1cm of a1] (a0){$\bar b$};
			\vertex [right=1.5cm of a1] (a2);
			\vertex [above right=1 cm of a2] (a3){$S_1$};
			\vertex [below left=1cm of a1] (b0){$g$};
			\vertex [below right=1cm of a2] (b3){$\nu_\tau$};
			\diagram {(a0)--[anti fermion](a1)--[anti fermion,edge label=$b$](a2)--[charged scalar](a3),
				(b0)--[gluon](a1),(a2)--[fermion](b3)};
		\end{feynman}
		\end{tikzpicture}}	
	\subfigure[]{
		\begin{tikzpicture}	
%		\hspace{5.3cm}
		\begin{feynman}
			\vertex (a1);
			\vertex [above=1cm of a1] (z1);
			\vertex [left=10mm of z1] (z0) {$\bar b$};
			\vertex [right=10mm of z1] (z2) {$\nu_\tau$};
			\vertex [below=1cm of a1] (b1);
			\vertex [left=10mm of b1] (b0){$g$};
			\vertex [right=10mm of b1] (b2) {$S_1$};
			\diagram {(z0)--[anti fermion](z1)--[charged scalar, edge label=$S_1$](b1)--[charged scalar](b2),
				(b0)--[gluon](b1),
				(z1)--[fermion](z2)};
		\end{feynman}
	\end{tikzpicture}}
	\caption{The tree level Feynman diagrams for $c/t-g$ and $b-g$ fusion producing $S_1$ leptoquark  associated with a lepton.}
	\label{fig:S_1_prod}
\end{figure}

In this section, we first start with  the singlet leptoquark $S_1$. In order to perform a collider analysis at the LHC/FCC, we choose the following set of centre-of-mass energy ($E_{\rm{CM}}$) values: 14\,TeV, 30\,TeV and 100\,TeV and the dominant SM backgrounds are also taken into account accordingly. The benchmark points, quoted in \autoref{tab:all}, for two different $S_1$ masses namely, 1.5\,TeV and 2.0\,TeV are motivated from the explanation to CC anomalies seen in $B$-decays. Such parameter spaces are also  allowed by the  recent searches at the LHC~\cite{CMS:2020gru,Aad:2021rrh}.
The main focus of this article is to probe the Yukawa-type coupling via single leptoquark production and the corresponding quark $-$ gluon ($g$) fusion production modes can be seen from the leading order Feynman diagrams in \autoref{fig:S_1_prod}.  The tree-level cross-sections for the $c-g$ and $b-g$ fusions are presented  in \autoref{crosss1} for three different centre-of-mass energies of 14\,TeV, 30\,TeV and 100\,TeV  respectively, where NNPDF$\_$lo$\_$as$\_$0130$\_$qed \cite{nnpdf} is used as parton distribution function, and $\sqrt{\hat{s}}$, the  parton  level centre-of-mass energy is used as the renormalization/factorization scale. It can be seen that the $t-g$ fusion is not negligible at the $E_{\rm{CM}}$ of  30\,TeV and 100\,TeV due to enhanced parton distribution function contribution in  NNPDF$\_$lo$\_$as$\_$0130$\_$qed \cite{nnpdf}. Additionally, extrapolating the results from refs. \cite{Alves:2002tj, Hammett:2015sea}, we take the NLO QCD $K$-factor of 1.5 for the single scalar leptoquark production processes. For the purpose of the analysis, the SM backgrounds contributions are also quoted at NLO QCD, with the $K$-factors calculated using \texttt{MadGraph5\_aMC@NLO}\cite{Alwall:2011uj}, which are presented in Appendix \autoref{sec:nlobg}. The final event numbers and the signal significance are evaluated with NLO cross-sections and assuming Gaussian distribution the signal significance is calculated as $\rm \sigma=\frac{n_{sig}}{\sqrt{n_{sig}+n_{bg}}}$, where $\rm n_{sig}, \, n_{bg}$ are the  signal and the background events numbers presented at certain integrated luminosity at some centre-of-mass energy. 

The leptoquarks produced from these mentioned channels will decay into $b\nu_\tau,\, t\tau~ \rm{and}~ c\tau$ finalstates with the branching ratios quoted in \autoref{brss1}. Here we find that $b\nu_\tau,\, t\tau $ are the dominant modes which give rise to various  finalstate topologies as discussed later in the subsections. Note that this minimal choice of parameter space forbids a decay to $c\mu$ mode which substantially reduces the SM backgrounds and can also nicely reconstruct the leptoquark invariant mass as found in~\cite{Bandyopadhyay:2018syt}. The following subsections describe kinematical distributions and signal events and background events for several chosen topologies.

%%%%%%%%%%%%%%%%%%%%%%%%%%Feynman diagrams for S_1 %%%%%%%%%%%%%%%%%%%%
%%%%%%%%%%%%%%%%%%%%%%%%%%%%%%%%%%%%%%%%%%%%%%%%%%%%%%%%

%%%%%%%%%%%%%%%%%%%%%%%%Cross-sections%%%%%%%%%%%%%%%%%%%%%%%%%%%%%%%%%%%

\begin{table}[h]
	\renewcommand{\arraystretch}{2}
	\centering
	\begin{tabular}{|c||c|c|c||c|c|c||c|c|c|}
		\hline 
Bench-		&\multicolumn{3}{c||}{$\sigma(c-g \to S_1 \tau)$ in fb  }&\multicolumn{3}{|c||}{$\sigma(b-g \to S_1 \nu_\tau)$ in fb }&\multicolumn{3}{c|}{$\sigma(t-g \to S_1 \tau)$ in fb  }\\
mark&\multicolumn{3}{c||}{with $E_{\rm{CM}}$ in TeV}&\multicolumn{3}{|c||}{with $E_{\rm{CM}}$ in TeV}&\multicolumn{3}{c|}{with $E_{\rm{CM}}$ in TeV}\\
\cline{2-10}
	\makecell{Points \\ ($M_{S_1}$)}	& 14 TeV& 30 TeV &100 TeV & 14 TeV& 30 TeV &100 TeV& 14 TeV& 30 TeV &100 TeV\\
	\hline\hline

\makecell{BP1 \\  (1.5 TeV)} &0.24 &4.07&96.65&0.50&9.09&237.29&0.12&2.60&78.21\\
\hline
\makecell{BP2 \\ (2.0 TeV)} &0.08 &1.86&60.62&0.09&2.73&98.16&0.03&0.80&33.20\\
\hline 
	
\end{tabular}
\caption{The cross-sections at the LHC/FCC via $c-g$, $b-g$ and $t-g$ channels for the two benchmark points of $S_1$ leptoquark at three different centre-of-mass energies of 14\,TeV, 30\,TeV and 100\,TeV. We chose NNPDF$\_$lo$\_$as$\_$0130$\_$qed \cite{nnpdf} as the parton distribution function and $\sqrt{\hat{s}}$ as renormalization/factorization scale, with the NLO QCD $K$-factor of 1.5.} \label{crosss1}
\end{table}
%%%%%%%%%%%%%%%%%%%%%%%%%%%%%%%%%%%%%%%%%%%%%%%%%%%%%%%%%%%%
%%%%%%%%%&%%%%%%%%%%  Branching fractions  %%%%%%%%%%%%%%%%%%%%%%%%%%%%%%
\begin{table}[h]
	\renewcommand{\arraystretch}{1.5}
	\centering
	\begin{tabular}{|c|c|c|}
		\hline 
			Decay	& \multicolumn{2}{|c|}{Branching fractions }\\ 
			\cline{2-3}
		Modes &\makecell{BP1 \\ $M_{S_1}$ = 1.5 TeV}  &\makecell{BP2 \\ $M_{S_1}$ = 2.0 TeV} \\
		\hline
	$S_1 \to b \nu_\tau$ & 43.9 & 41.4\\
	\hline
	$S_1 \to t \tau$ & 42.8& 40.4\\
	\hline
		$S_1 \to c\tau$ & 13.3 & 18.6\\
	\hline

	\end{tabular}
	\caption{Decay branching fractions in \% for the allowed benchmark points of $S_1$ leptoquark.}  \label{brss1}
\end{table}
%%%%%%%%%%%%%%%%%%%%%%%%%%%%%%%%%%%%%%%%%%%%%%%%%%%%%

\subsection{Kinematic distributions and topologies}
\label{sec:kine_S1}

Before going into the details of the collider simulation let us have a look at the different differential distributions to motivate the advanced cuts which will be used later on to reduce the SM backgrounds.
Depending on the decays of $S_1$ some finalstates may have more background than the rest. However, to reduce the light QCD-jet  backgrounds we need more flavour tagging viz. $b-$jet  and/or $\tau-$jet. We first consider the production channel $c-g \to S_1 \tau$ (shown in \autoref{fig:S_1_prod}(a)), where $S_1$ can further decay to either $b \nu $ or $c\tau$ states. Thus, finalstates involving $b-$, $c-$ and $\tau-$jets are possible and we discuss them separately. The dominant SM backgrounds arise from $t\bar{t}$, owing to the high cross-section, which contribute in the finalstates involving $b/c/\tau-$ jets. The demand of only one $b/c$-jets, one or two $\tau-$jets, high cuts on missing transverse momentum ($\ptmiss$), and veto on the number of light jets can help us reduce such background contaminations. Each of such demands and cuts are categorically mentioned when we discuss each individual finalstate.  If we consider the decay of  $S_1 \to t \tau$, the finalstates involving leptons are suppressed due to the lower branching of $W^\pm$ in the leptonic mode. 

For this analysis we considered $b$-jet tagging efficiency of $\sim 70\%$  via the secondary vertex reconstruction mechanism\cite{btagging,btagging2,btagging3}.  For $\tau-$jet we reconstruct the hadronic one-prong ($\pi^\pm$)  jet as $\tau-$jet with momentum dependent efficiencies  as shown in \cite{tautagging,tautagging2}.  The $c$-jet tagging efficiency is taken around 56\%  with a mistagging of 12\%,  which is very conservative considering non-loose tagging mechanism \cite{ctagging}. 

%%%%%%%%%%%%%% Jet and Lepton Multiplicity distributions %%%%%%%%%%%%%
\begin{figure*}[h!]
	\centering
	\mbox{\subfigure[]{\includegraphics[scale=0.44]{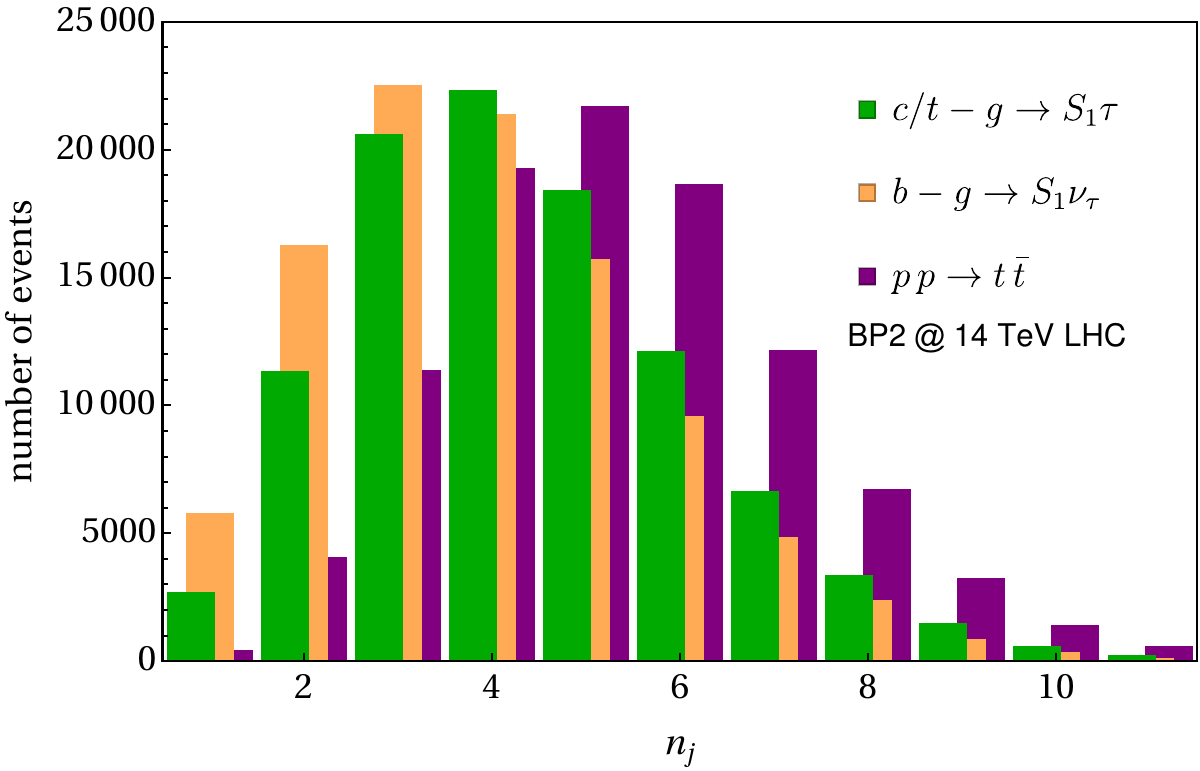}}
		\hfil
		\subfigure[]{	\includegraphics[scale=0.44]{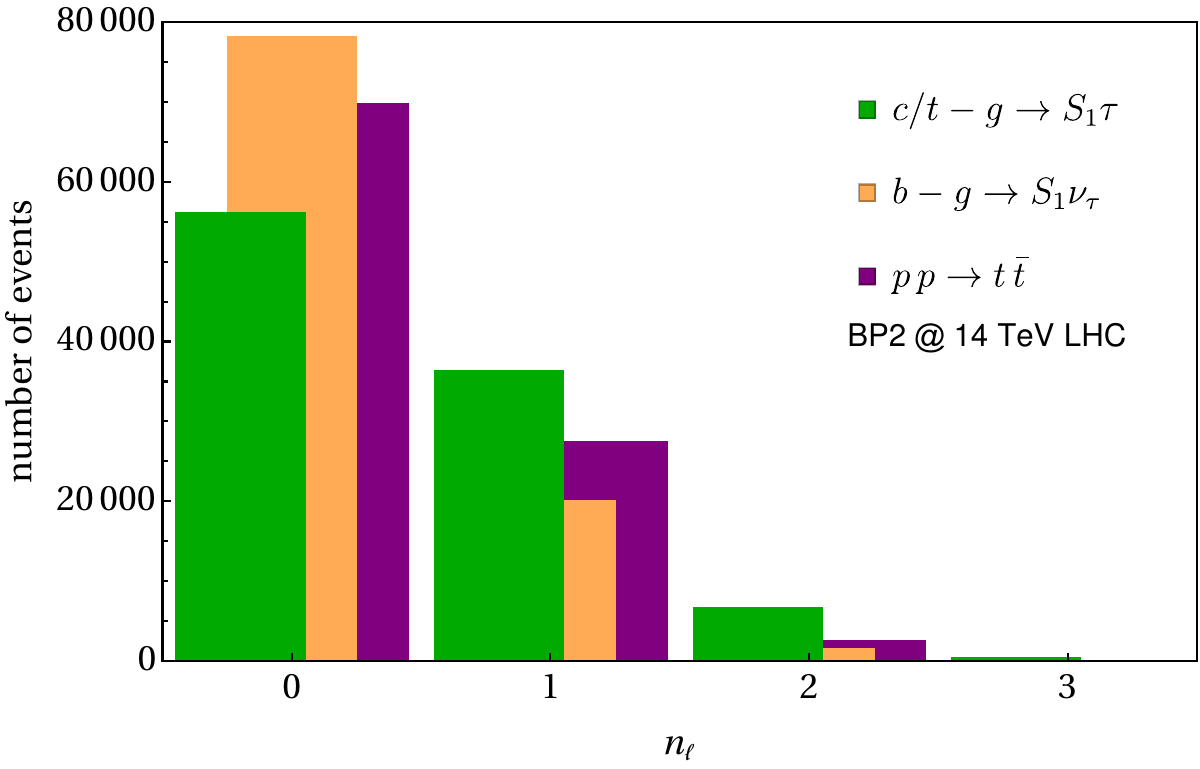}}}

	\caption{The jet multiplicity ($n_j$ in (a)) and lepton multiplicity ($n_\ell $ in (b)) distributions for the BP2 and SM background $t\bar{t}$ at the LHC with centre-of-mass energy of 14\,TeV.}\label{jlml}
\end{figure*}
%%%%%%%%%%%%%%%%%%%%%%%%%%%%%%%%%%%%%%%%%%%%%%%%%%%%

In \autoref{jlml}(a) we display the jet multiplicity distribution ($n_j$) for the two signal processes of $c/t-g \to S_1 \tau$ (green) and $b-g \to S_1 \nu_\tau$ (orange) for BP2 at the 14 TeV LHC, in comparison with the $t\bar{t}$ SM background (purple). The distribution for $b-g \to S_1 \nu_\tau$ peaks at three jets, with the sources of jets being the daughter top quark of $S_1$, as well as the $\tau$-jet in the $S_1 \to t\tau$ decay channel (if tagged). This peak increases to four jets for $c/t-g \to S_1 \tau$, where the additional $\tau$-jet produced with $S_1$ contributes. The $t\bar{t}$ background distribution shows the peak at five jets, as both the top quarks and their daughter $W^\pm$ bosons contribute. The ISR/FSR effects give the tails for these jet multiplicity distributions. \autoref{jlml}(b) shows the distribution of lepton multiplicity ($n_\ell$) for the same processes, following the same colour codes. While both the signal processes and the background peaks at zero leptons, the $b-g \to S_1 \nu_\tau$ process has more events there owing to less sources of leptons in the production and decay products. The hard charged lepton ($e/ \mu$) mainly comes from the decay to top quark which is produced from the $S_1$ decay. The source of the second lepton is mostly from the $\tau$ decay or the semileptonic decays of $b$ quark. On the other hand, $c/t-g \to S_1 \tau$ gives the least number of zero-lepton events, as the leptonic decay of the recoiled $\tau$ can also contribute. The background shows similar behaviour as the signal, as mainly the $W^\pm$ bosons coming from the top quarks can contribute to the lepton multiplicity.

%%%%%%%%%%%%%% Jet p_T distributions %%%%%%%%%%%%%
\begin{figure*}[h!]
	\centering
	\mbox{\subfigure[]{\includegraphics[scale=0.44]{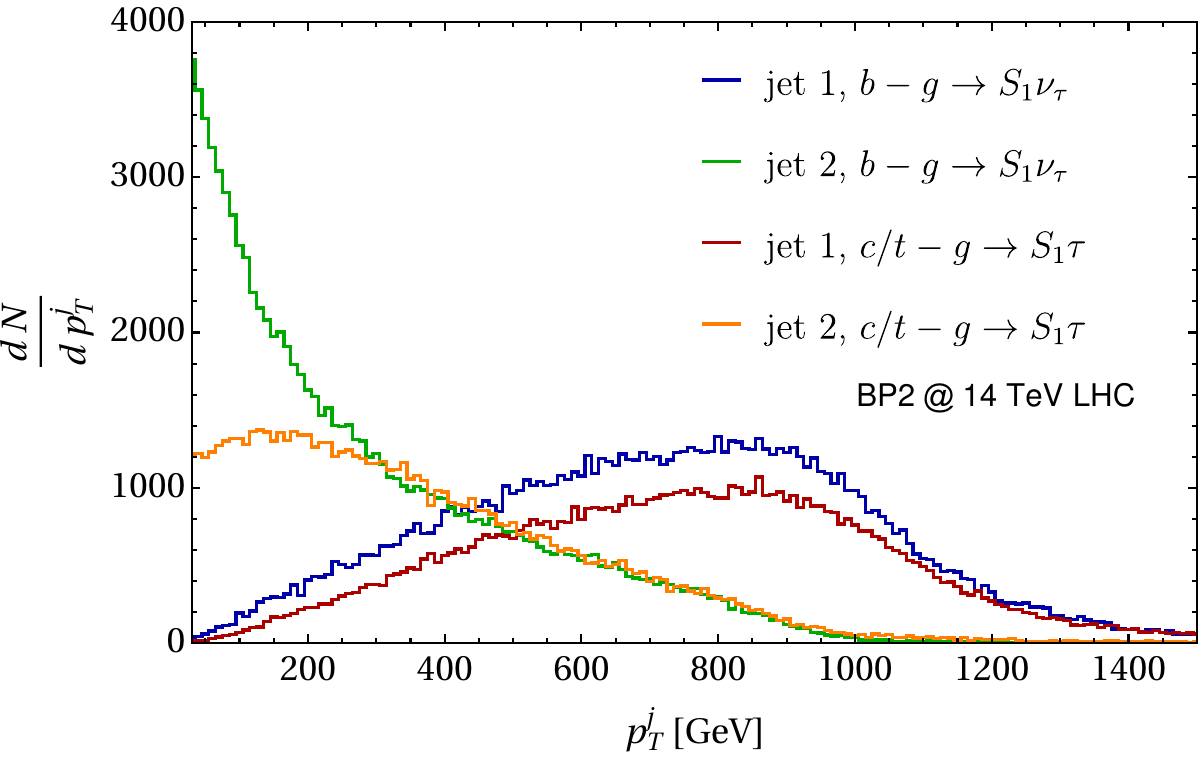}}
		%		\hfil
		%		\subfigure[]{	\includegraphics[scale=0.44]{plots/jet_pT_tt_14}}
	}
	\caption{The $p_T$ distribution of the two hardest jets ($p_T^j$) from each of the production processes $b-g \to S_1 \nu_\tau$ and $c/t-g \to S_1 \tau$, at the 14 TeV LHC, for BP2.}\label{jpt}
\end{figure*}
%%%%%%%%%%%%%%%%%%%%%%%%%%%%%%%%%%%%%%%%%%%%%%%%%%%%

In \autoref{jpt} we depict the jet $p_T$ ($p_T^j$) distributions of the two hardest jets at the 14 TeV LHC, emanating from each of the two production modes considered. The jets from $b-g \to S_1 \nu_\tau$ process are shown in blue and green, while those from $c/t-g \to S_1 \tau$ are shown in red and orange. In each case, it is evident that the hardest jets (blue and red) peak at $\sim$850 GeV, which lies roughly around half of the leptoquark mass, as expected. In case of $b-g \to S_1 \nu_\tau$, the second hardest jet's source is the daughter $W$-boson of the top quark from $S_1 \to t\tau$ decay, and so the $p_T$ peak is observed at around 40 GeV. However, for $c/t-g \to S_1 \tau$, the hadronic $\tau$-jet produced alongside the leptoquark accounts for the second hardest jet, showing a wide peak at $\sim 150$ GeV.

%%%%%%%%%%%%%% Lepton pT distributions %%%%%%%%%%%%%
\begin{figure*}[h!]
	\centering

\mbox{\subfigure[]{\includegraphics[scale=0.44]{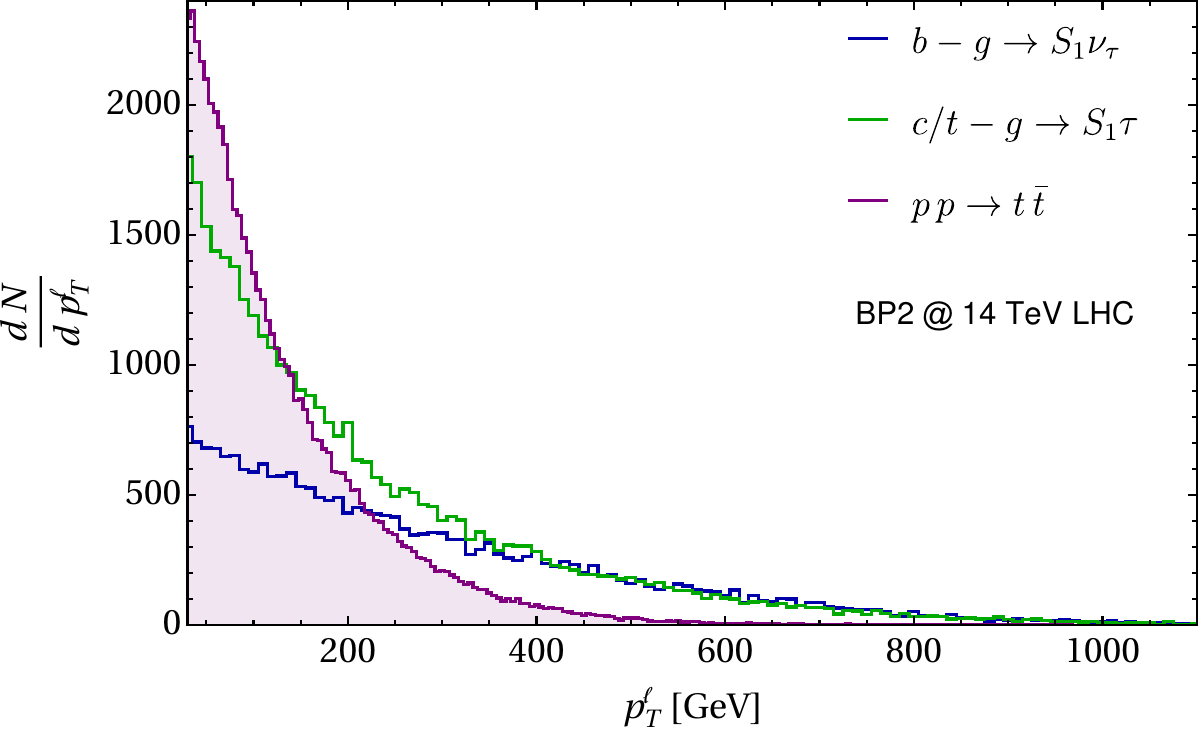}}
		\hfil
		\subfigure[]{	\includegraphics[scale=0.44]{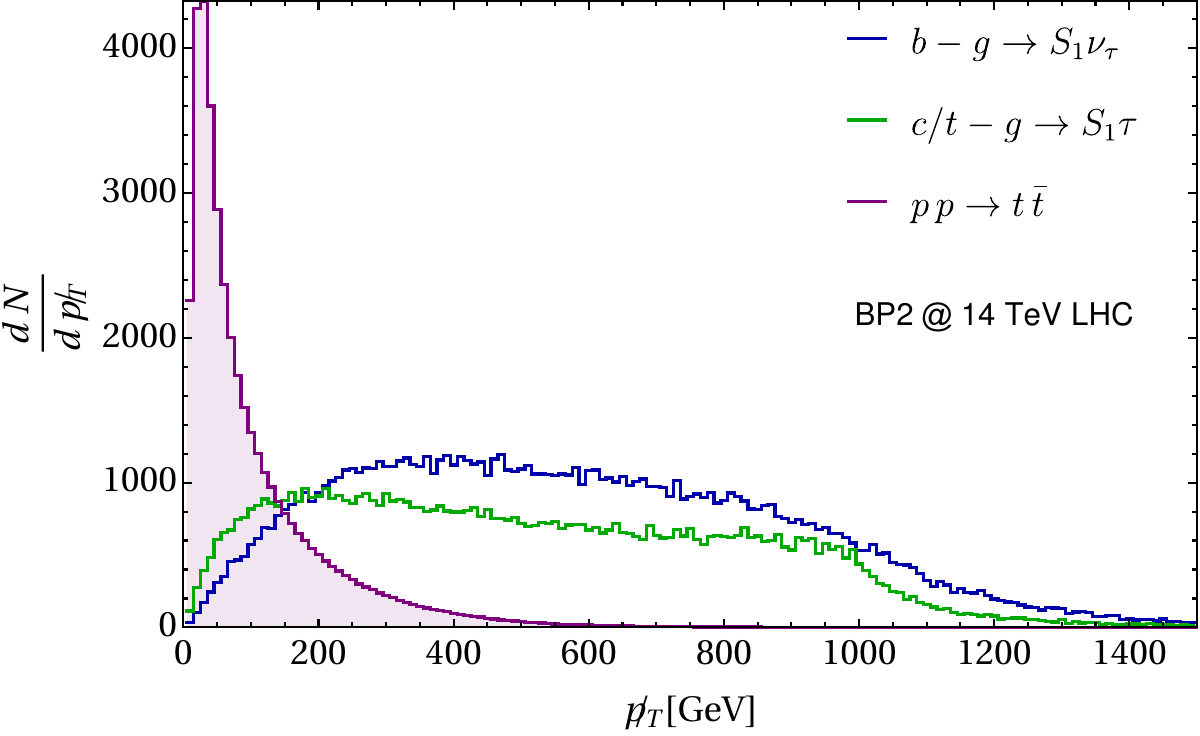}}}
	\caption{The lepton $p_T$ ( $p_T^\ell$ in (a)) and missing transverse momentum ($\ptmiss$ in (b)) distributions for BP2 and the SM background $t\bar{t}$ at the LHC with centre-of-mass energy of 14 TeV.
	The background in (a) and (b) are scaled with 1/2 and 1/100 respectively for convenience.}\label{lmpt}
\end{figure*}
 
We now move to the lepton  $p_T$ distributions as depicted in \autoref{lmpt}(a), showing the $p_T$ distribution of the hardest lepton $p_T^\ell$ obtained from the two production processes $b-g \to S_1 \nu_\tau$ (blue) and $c/t-g \to S_1 \tau$ (green), for BP2 at the 14 TeV LHC. For comparison, the same distribution is shown in shaded purple for the $t\bar{t}$ background, scaled down by 1/2 for illustrative purposes. The lack of a lepton as a direct decay product of the leptoquark means that in either of the signal production processes, the lepton $p_T$ peaks at $\sim30$ GeV, same as the $t\bar{t}$ case. However, the distributions in case of signal processes have more events at the tail, which can help us put advanced cut of $p_T^{\ell} \geq 200$ GeV, to reduce the background contamination, later in our analysis.  \autoref{lmpt}(b)  shows the missing transverse momentum $\ptmiss$ distribution for the three aforementioned signal and background processes with the same colour coding. it is evident that a large missing transverse momentum is observed which arises from the recoiled neutrino coming either at the production level for $b-g \to S_1\,  \nu_\tau $ or  at the later stage from $c-g \to S_1 (\to b\nu_\tau) \tau $. On the contrary, the missing transverse momentum $\ptmiss$ due to neutrinos in case of $t\bar{t}$ peaks near $\sim 50$\,GeV, and the tail is much shorter. We can thus apply missing transverse momentum cut $\ptmiss \geq$ 500 GeV for the considered finalstates later in our analysis which reduce the SM backgrounds substantially.

\subsection{Finalstates including $b$ and $\tau$ jets }
\label{sec:S1_btau}
In this subsection we describe the finalstate topologies comprising $b-$ and $\tau-$jets  for $S_1$ leptoquark production mainly via $c-g$ and $b-g$ fusions as well as with $t-g$ fusion which contributes at high energy.  Once produced in association with $\tau-$jet in  $c-g$ fusion, the $S_1$ leptoquark further decays  to $b \nu_\tau, \,  t \tau$ states governed by the decay  branching given in \autoref{brss1} giving rise to the following topologies composed of at least one $b-$ and $\tau-$jet.
\bea
\hspace*{-1cm}\text{BP1, BP2:}\qquad\qquad c/t-g &\to& S_1 \tau ,\nn\\
&\to& (b \, \nu_\tau) + \tau \to 1b-\rm{jet} +  1  \tau-\rm{jet} + \ptmiss,\\
&\to& (t \,\tau)+\tau \to 1b-\rm{jet} +  2  \tau-\rm{jet} + 1\ell + \ptmiss, \label{1b1l2t}\\
&\to& (t \,\tau)+\tau \to 1b-\rm{jet} +  2  \tau-\rm{jet} + 2-\rm{jets}.
\label{1b2t2j}\eea
Similarly, $b-g \to S_1 \nu_\tau$ can give rise to the following topologies  with $b-$ and $\tau-$jet.
\bea
\hspace*{-0.4cm}\text{BP1, BP2:}\qquad\qquad b-g &\to& S_1 \, \nu_\tau , \label{bmis}\nn\\
&\to& (b \, \nu_\tau) + \nu \to 1b-\rm{jet} + \ptmiss,\\
&\to& (t \,\tau)+\nu \to 1b-\rm{jet} +  1 \tau-\rm{jet} + 1\ell + \ptmiss, \label{1b1l1t}\\
&\to& (t \,\tau)+\nu \to 1b-\rm{jet} +  1  \tau-\rm{jet} + 2-\rm{jets} +\ptmiss. \label{1b1t2j}
\eea
Note that, unlike the $c/t-g$ fusion, for $b-g$ fusion we can have  mono $b-$jet plus missing energy as an unique signature  (\autoref{bmis}). The $b-$jet and $\tau-$jet tagging are followed with the corresponding efficiencies \cite{btagging,btagging2,btagging3,tautagging, tautagging2}, as mentioned earlier. From now onward, in the rest of the analysis, the light-jets are denoted as `jets' ensuring no flavour tagging has been implemented. In the subsequent subsections we discuss all these finalstate signatures involving at least one $b-$ and $\tau-$jet at the LHC/FCC with two different centre-of-mass energies namely $30\,$TeV and $100\,$TeV and we leave 14 TeV results as the signal significances are lower than $3\sigma$ even with the integrated luminosity  of  3000 fb$^{-1}$.

%%%%%%%%%%$1b-\rm{jet} + 1  \tau-\rm{jet} +1l + \ptmiss  %%%%%%%%%%%%
\begin{table*}[h!]
	\renewcommand{\arraystretch}{1.2}
	\centering
	\begin{tabular}{|c|c||c|c||c|c|c|c|c|}
		\hline
		\multirow{2}{*}{ $E_{\rm{CM}}$ in}&&\multicolumn{7}{c|}{$1b-\rm{jet} + \geq 1  \tau-\rm{jet}+ \geq 1\ell + \ptmiss\geq 500$ GeV}\\
		\cline{3-9}
		\multirow{2}{*}{TeV}& Mode &\multicolumn{2}{c||}{Signal}&\multicolumn{5}{c|}{Backgrounds}\\
		\cline{3-9}
		&&BP1&BP2 &$t\bar t$&$VV$&$VVV$&$t\bar{t}V$&$tVV$\\
		\hline
		\hline
		
		\multirow{2}{*}{30} &$c/t-g \to S_1\, \tau$&56.08&39.76&\multirow{2}{*}{643.35}&\multirow{2}{*}{1.36}&\multirow{2}{*}{0.00}&\multirow{2}{*}{30.66}&\multirow{2}{*}{7.84}\\
		\cline{2-4}
		&$b-g\to S_1\,\nu$&7.44&3.24&&&&&\\
		\hline
		\multicolumn{2}{|c||}{Total}&63.52&43.00&\multicolumn{5}{c|}{683.21}\\
		\hline
		\multicolumn{2}{|c||}{Significance ($\sigma$)}&2.32&1.60&\multicolumn{5}{c|}{ }\\
		\cline{1-4}
		\multicolumn{2}{|c||}{$\mathcal{L}_{5\sigma}$ (fb$^{-1}$)}&$\gg 3000$&$\gg3000$&\multicolumn{5}{c|}{ }\\
		\hline\hline
		
		\multirow{2}{*}{100} &$c/t-g \to S_1\, \tau$&294.42&232.08&\multirow{2}{*}{1313.83}&\multirow{2}{*}{7.67}&\multirow{2}{*}{7.04}&\multirow{2}{*}{91.84}&\multirow{2}{*}{28.02}\\
		\cline{2-4}
		&$b-g\to S_1\,\nu$&23.96&18.53&&&&&\\
		\hline
		\multicolumn{2}{|c||}{Total}&318.36&250.61&\multicolumn{5}{c|}{1448.40}\\
		\hline
		\multicolumn{2}{|c||}{Significance ($\sigma$)}&7.57&6.08&\multicolumn{5}{c|}{ }\\
		\cline{1-4}
		\multicolumn{2}{|c||}{$\mathcal{L}_{5\,\sigma}$ (fb$^{-1}$)}&43.58&67.63&\multicolumn{5}{c|}{ }\\
		\hline
	\end{tabular}
	\caption{The number of events for  $1b-\rm{jet} + \geq 1  \tau-\rm{jet} + \geq 1\ell + \ptmiss \geq 500$ GeV finalstate for the benchmark points and dominant SM backgrounds at the LHC/FCC with the centre-of-mass energies of   30\,TeV  and 100\,TeV with   integrated luminosities  at 1000\,fb$^{-1}$ and 100\,fb$^{-1}$, respectively. The required luminosities to achieve a $5\sigma$ signal ($\mathcal{L}_{5\sigma}$) are also shown for both the cases.}\label{btau1}
\end{table*}
%%%%%%%%%%%%%%%%%%%%%%%%%%%%%%%%%%%%%%%%%%%%%%%%%%%%%%%%%
\subsubsection{$1b-\rm{jet} + 1  \tau-\rm{jet} +1\ell + \ptmiss $}\label{321}
Here we consider \autoref{1b1l1t} and \autoref{1b1l2t} which lead to $1b-\rm{jet} + 1  \tau-\rm{jet} +1\ell + \ptmiss $ finalstate topology. The complete finalstate including the advanced cuts and veto are given below.

\begin{center}
	$n_{b-\rm{jet}} = 1$, $n_{\tau-\text{jet}} \geq 1$, $n_{j} \geq 2$, $n_\ell \geq 1$ \& \\ $\ptmiss \geq 500$ GeV, $p_T^{j_1 , j_2} \geq 200$ GeV, $p_T^{\ell} \geq 200$ GeV, $p_T^H \geq 1200$ GeV.
\end{center}
The event numbers at the centre-of-mass energies of  30\,TeV and 100\,TeV at the LHC/FCC with the respective integrated luminosities of  1000\,fb$^{-1}$ and 100\,fb$^{-1}$ are presented in \autoref{btau1}. 
It can be seen that $1b-\rm{jet} + 1  \tau-\rm{jet} +1\ell + \ptmiss $ finalstate  arises from both $c/t$-gluon and $b$-gluon  fusion, where $S_1$ decays to $t\, \tau$ states. The top quark then provides the $b$-jet and the charged lepton via subsequent decays. When any of the two $\tau$-jets in \autoref{1b1l2t} is tagged we obtain the mentioned finalstate from $c/t-g$ fusion. However, for $b-g$ fusion we have only one $\tau$-jet finalstate making the contribution significantly reduced in this case. The missing energy for the signal is relatively higher as can be seen from \autoref{lmpt}(b) and we apply a cut of $\ptmiss >500$ GeV. The benchmark points are with leptoquark masses of 1.5 and 2.0 TeV, so we apply a hardness cut of $1.2$ TeV to reduce the background number of the events substantially, where the transverse variable total hardness defined as $p^H_T= \Sigma (p^\ell_T + p^j_T +\ptmiss)$, is the scalar sum of lepton, jet and missing transverse momentum. The first two hard jets $p_T$  and the charged lepton $p_T$ are demanded to be $\geq 200$\,GeV in order to reduce the SM backgrounds further as demonstrated in \autoref{jpt}. Although 14  TeV  numbers are  not encouraging, the numbers presented in \autoref{btau1} at centre-of-mass energy of  30\,TeV give rise to the signal significances of $2.32 \,\sigma$ and $1.60\, \sigma$ at  1000\,fb$^{-1}$ of integrated luminosity for BP1 and BP2, respectively, which again is not a very pleasant scenario. However, at the 100 TeV centre-of-mass energy of the LHC/FCC, we see promising numbers, with the signal significances of $7.57\sigma$ and $6.08\sigma$ obtained at 100 \fbi of integrated luminosity, for BP1 and BP2, respectively. In all cases, $t\bar{t}$ remains the dominant background owing to the availability of a $b$-jet and the high cross-section, while $t\bar{t}V$ also contributing significantly.

\subsubsection{$1b-\rm{jet} + 1  \tau-\rm{jet} +2-{\rm jets} + \ptmiss $}

%%%%%%%%%%%%%$1b-\rm{jet} + 1  \tau-\rm{jet} +2j + \ptmiss $ %%%%%%%%%%%%%%%%
\begin{table*}[h]
	\renewcommand{\arraystretch}{1.2}
	\centering
	\begin{tabular}{|c|c||c|c||c|c|c|c|c|}
		\hline
		\multirow{2}{*}{ $E_{\rm{CM}}$ in}&&\multicolumn{7}{c|}{$1b-\rm{jet} + \geq 1  \tau-\rm{jet}+ \geq 2-{\rm jets} + \ptmiss \geq 500$ GeV}\\
		\cline{3-9}
		\multirow{2}{*}{TeV}& Mode &\multicolumn{2}{c||}{Signal}&\multicolumn{5}{c|}{Backgrounds}\\
		\cline{3-9}
		&&BP1&BP2 &$t\bar t$&$VV$&$VVV$&$t\bar{t}V$&$tVV$\\
	\hline
		\multirow{2}{*}{30} &$c/t-g \to S_1\, \tau$&426.62&256.89&\multirow{2}{*}{28097.22}&\multirow{2}{*}{128.64}&\multirow{2}{*}{28.06}&\multirow{2}{*}{737.05}&\multirow{2}{*}{94.04}\\
		\cline{2-4}
		&$b-g\to S_1\,\nu$&239.59&123.33&&&&&\\
		\hline
		\multicolumn{2}{|c||}{Total}&666.21&380.22&\multicolumn{5}{c|}{29085.01}\\
		\hline
		\multicolumn{2}{|c||}{Significance ($\sigma$)}&3.86 &2.21&\multicolumn{5}{c|}{}\\
		\cline{1-4}
		\multicolumn{2}{|c||}{$\mathcal{L}_{5\sigma}$ (fb$^{-1}$)}&1675.80&$\gg 3000$&\multicolumn{5}{c|}{ }\\
		\hline\hline
		\multirow{2}{*}{100} &$c/t-g \to S_1\, \tau$&1575.29&1105.74&\multirow{2}{*}{59677.63}&\multirow{2}{*}{151.75}&\multirow{2}{*}{75.81}&\multirow{2}{*}{1682.02}&\multirow{2}{*}{320.77}\\
		\cline{2-4}
		&$b-g\to S_1\,\nu$&830.63&546.74&&&&&\\
		\hline
		\multicolumn{2}{|c||}{Total}&2405.92&1664.48&\multicolumn{5}{c|}{61907.98}\\
		\hline
		\multicolumn{2}{|c||}{Significance ($\sigma$)}&9.49&6.60&\multicolumn{5}{c|}{ }\\
		\cline{1-4}
		\multicolumn{2}{|c||}{$\mathcal{L}_{5\sigma}$ (fb$^{-1}$)}&27.78&57.37&\multicolumn{5}{c|}{ }\\
		\hline
	\end{tabular}
	\caption{The number of events for  $1b-\rm{jet} + \geq 1  \tau-\rm{jet} + \geq 2-{\rm jets} + \ptmiss \geq 500$ GeV finalstate for the benchmark points and dominant SM backgrounds at the LHC/FCC with the centre-of-mass energies of  30\,TeV  and 100\,TeV for  the integrated luminosities  of 1000\,fb$^{-1}$ and 100\,fb$^{-1}$ for 100\,TeV, respectively. The required luminosities to achieve a $5\sigma$ signal ($\mathcal{L}_{5\sigma}$) are also shown for both the cases.}\label{btau2}
\end{table*}
%%%%%%%%%%%%%%%%%%%%%%%%%%%%%%%%%%%%%%%%%%%%%%%%%%%

Now we consider the $1b-\rm{jet} + 1  \tau-\rm{jet} +2-{\rm jets}  + \ptmiss $ finalstate, which is almost similar to the previous decay topologies with only exception of the $W^\pm$, coming  from the top quark, decays hadronically (\autoref{1b2t2j}, \autoref{1b1t2j}). Certainly, due to higher branching fraction  in the hadronic mode, the event numbers for this finalstate are expected to increase substantially as compared to $1b-\rm{jet} + 1  \tau-\rm{jet} +1\ell + \ptmiss $ in \autoref{btau1}.  The complete finalstate with the advanced cuts is given as

\begin{center}
	$n_{b-\rm{jet}} = 1$, $n_{\tau-\rm{jet}} \geq 1$, $n_{j} \geq 4$, $n_\ell = 0$ \& \\ $\ptmiss \geq 500$ GeV, $p_T^{j_1 , j_2} \geq 200$ GeV, $p_T^H \geq 1200$ GeV.
\end{center}

Similar to $1b-\rm{jet} + 1  \tau-\rm{jet} +1\ell + \ptmiss $ , here also the 14 TeV numbers are insignificant. Therefore, in \autoref{btau2} we list only the  number of events for the benchmark points as well as the dominant SM backgrounds for the centre-of-mass energies of 30\,TeV and 100\, TeV  at integrated luminosities  of 1000 fb$^{-1}$, 100 fb$^{-1}$, respectively.

 In this case, while the signal events increase as expected, we observe an overwhelming rise of the background contribution, owing to the abundance of zero-lepton events. $t\bar{t}$ and $t\bar{t}V$ remain the most dominant backgrounds, contributing to the demand of a $b$-jet. Such high backgrounds reduce our chances of obtaining a good signal strength at the 14 TeV LHC for both benchmark points and we do not list them here. The situation improves for BP1 when we move to  the centre-of-mass energy of 30 TeV, where $3.86\sigma$ signal significance can be obtained at  1000 \fbi of integrated luminosity. The required $5\sigma$ discovery can be predicted to be made with a luminosity of 1675.80 \fbi for BP1. The BP2 signal however remains weak with $2.21\sigma$ significance. The most promising scenario again is the 100 TeV LHC/FCC, where $9.49\sigma$ and $6.60\sigma$ significance can be obtained for BP1 and BP2 respectively, with an integrated luminosity of 100 \fbi.

%%%%%%%%%%%%%%%%%%%  $1b-\rm{jet} + \ptmiss $ %%%%%%%%%%%%%%%%%%%%%%%%%%%%
\begin{table*}[h!]
	\renewcommand{\arraystretch}{1.2}
	\centering
	\begin{tabular}{|c|c||c|c||c|c|c|c|c|}
		\hline
		\multirow{2}{*}{ $E_{\rm{CM}}$ in}&&\multicolumn{7}{c|}{$1b-\rm{jet} +\ptmiss >500$ GeV}\\
		\cline{3-9}
		\multirow{2}{*}{TeV}& Mode &\multicolumn{2}{c||}{Signal}&\multicolumn{5}{c|}{Backgrounds}\\
		\cline{3-9}
		&&BP1&BP2 &$t\bar t$&$VV$&$VVV$&$t\bar{t}V$&$tVV$\\
		\hline

		\multirow{2}{*}{30} &$c/t-g \to S_1\, \tau$&27.98&13.76&\multirow{2}{*}{6439.58}&\multirow{2}{*}{650.04}&\multirow{2}{*}{50.14}&\multirow{2}{*}{74.10}&\multirow{2}{*}{25.86}\\
		\cline{2-4}
		&$b-g\to S_1\,\nu$&401.73&146.02&&&&&\\
		\hline
		\multicolumn{2}{|c||}{Total}&429.71&159.78&\multicolumn{5}{c|}{7239.72}\\
		\hline
		\multicolumn{2}{|c||}{Significance ($\sigma$)}&4.90&1.86&\multicolumn{5}{c|}{ }\\
		\cline{1-4}
		\multicolumn{2}{|c||}{$\mathcal{L}_{5\,\sigma}$ (fb$^{-1}$)}&1038.39&$\gg$3000&\multicolumn{5}{c|}{ }\\
		\hline\hline
		
		\multirow{2}{*}{100} &$c/t-g \to S_1\, \tau$&66.33&39.59&\multirow{2}{*}{11196.26}&\multirow{2}{*}{432.21}&\multirow{2}{*}{33.51}&\multirow{2}{*}{96.66}&\multirow{2}{*}{57.61}\\
		\cline{2-4}
		&$b-g\to S_1\,\nu$&827.77&294.36&&&&&\\
		\hline
		\multicolumn{2}{|c||}{Total}&894.10&433.95&\multicolumn{5}{c|}{11816.25}\\
		\hline
		\multicolumn{2}{|c||}{Significance ($\sigma$)}&7.93&3.92&\multicolumn{5}{c|}{ }\\
		\cline{1-4}
		\multicolumn{2}{|c||}{$\mathcal{L}_{5\,\sigma}$ (fb$^{-1}$)}&39.75&162.63&\multicolumn{5}{c|}{ }\\
		\hline
	\end{tabular}
	\caption{The number of events for  $1b-\rm{jet} +\ptmiss >500$ GeV finalstate for the benchmark points and dominant SM backgrounds at the LHC/FCC with the centre-of-mass energies of  30\,TeV  and 100\,TeV  for  the integrated luminosities of1000\,fb$^{-1}$ and 100\,fb$^{-1}$, respectively. The required luminosities to achieve a $5\,\sigma$ signal ($\mathcal{L}_{5\,\sigma}$) are also shown for both the cases.}\label{btau3}
\end{table*}
%%%%%%%%%%%%%%%%%%%%%%%%%%%%%%%%%%%%%%%%%%%%%%%%%%%%%%

\subsubsection{$1b-\rm{jet} + \ptmiss $}
\label{sec:1b1jpt}
In this case we consider the mode where $S_1$ decays to $b\nu_\tau$ states and this finalstate may only be composed of mono $b$-jet and missing energy, when $S_1$ being produced from $b-g$ fusion (\autoref{bmis}). In order to obtain a cleaner signal and elimination of the SM background, further advanced cuts are applied on this finalstate. The full finalstate is given as follows:

\begin{center}
	$n_{b-\rm{jet}} = 1$, $n_{j} \leq 2$, $n_{\tau-\rm{jet}} = 0 $, $n_\ell = 0$ \& \\
	$\ptmiss \geq 500$ GeV, $p_T^{j_1} \geq 400$ GeV, $ p_T^H \geq 1.2$ TeV \& \\ $\abs{M_{\ell\ell}-M_Z} \geq 5\, \text{GeV} \,+ \,\abs{M_{jj}-M_W} \geq 10 \,\text{GeV}$. 
\end{center}

While we keep the cut on missing energy to be $\geq 500$ GeV accounting for the recoiled neutrino or the neutrino coming from the $S_1$ decay, we increase the leading jet $p_T$ cut to 400 GeV. The hardness cut remains the same as the previous two cases. However, the absence of a top quark-induced contribution to this decay topology, we can put a veto on the di-jet invariant mass $M_{jj}$, demanding it to be at least 10 GeV away from the $W$-boson mass peak. Similar veto is applied to the di-lepton invariant mass $M_{\ell\ell}$, demanding a 5 GeV minimum separation from the $Z$-boson mass. This helps us reduce the background contribution further. Such a decay topology has a very unique signature, and we show the number of events and the SM backgrounds in \autoref{btau3} for the two centre-of-mass energies at the LHC/FCC. Similar to the previous cases the 14\,TeV signal numbers are not very encouraging and we do not list them here. The 30\, TeV and 100\, TeV event numbers are given in \autoref{btau3} at integrated luminosities of  1000, 100\,fb$^{-1}$, respectively. At the 30 TeV LHC, results are a bit more promising for BP1 with a healthy $4.9\sigma$ of signal significance, while the BP2 signal remains weak with a $1.86\sigma$ significance. The $5\sigma$ reach for BP1 can be achieved at a luminosity of 1038.39 \fbi. At 100 TeV, the situation  improves  for BP2, as we reach a strength of $3.92\sigma$ with 100 \fbi luminosity, with a requirement of 162.63 \fbi for the desired $5\sigma$ strength. For BP1, we achieve $7.93\sigma$ significance 100\, fb$^{-1}$ luminosity, with the $5\sigma$ strength predicted to be obtainable at 39.75 \fbi of luminosity. In all the cases, $t\bar{t}$ remains the dominant background due to availability of a $b$-jet and higher cross-section, and the diboson ($VV$) background is the next dominant one.

\subsection{Finalstates including $c$ and $\tau$ jets  }
\label{sec:S1_ctau}

After studying the finalstates involving $b$-jets, we now aim to probe the other decay mode of $S_1$, namely to $c\, \tau$ states. With the two dominant production channels via $c-g$ and $b-g$ fusions, we look for topologies composed of at least one $c-$jet and $\tau-$jet.  The entire decay chain prompt us the following signals.
\bea
c/t-g &\to& S_1 \tau \nn\\
&\to& (c \, \tau) + \tau \to 1c-\rm{jet} + 2  \tau-\rm{jet}\,,\\
b-g &\to& S_1 \nu_\tau \nn\\
&\to& (c \, \tau) + \nu \to 1c-\rm{jet} + 1  \tau-\rm{jet}  + \ptmiss .
\eea
 Below we list the event numbers for the benchmark points (defined in \autoref{tab:all}) for the above mentioned finalstates along with the dominant SM backgrounds. Just to remind, the $c$-jet tagging efficiency is taken around 56\%  with a mistagging of 12\%,  which is very conservative considering non-loose tagging mechanism \cite{ctagging}.

\subsubsection{$1c-\rm{jet} + 1  \tau-\rm{jet} + \ptmiss$}

%%%%%%%%%%%%$1c-\rm{jet} + 1  \tau-\rm{jet} + \ptmiss$%%%%%%%%%%%%%%%%%%%%%
\begin{table*}[h]
	\renewcommand{\arraystretch}{1.2}
	\centering
	\begin{tabular}{|c|c||c|c||c|c|c|c|c|}
		\hline
		\multirow{2}{*}{ $E_{\rm{CM}}$ in}&&\multicolumn{7}{c|}{$1c-\rm{jet} + \geq 1  \tau-\rm{jet} + \ptmiss \geq 500$ GeV}\\
		\cline{3-9}
		\multirow{2}{*}{TeV}& Mode &\multicolumn{2}{c||}{Signal}&\multicolumn{5}{c|}{Backgrounds}\\
		\cline{3-9}
		&&BP1&BP2 &$t\bar t$&$VV$&$VVV$&$t\bar{t}V$&$tVV$\\

\hline
		
		\multirow{2}{*}{30} &$c/t-g \to S_1\, \tau$& 355.17&192.99&\multirow{2}{*}{21345.00}&\multirow{2}{*}{136.84}&\multirow{2}{*}{30.07}&\multirow{2}{*}{356.38}&\multirow{2}{*}{29.77}\\
		\cline{2-4}
		&$b-g\to S_1\,\nu$&172.02&109.77&&&&&\\
		\hline
		\multicolumn{2}{|c||}{Total}&527.19&302.76&\multicolumn{5}{c|}{21898.07}\\
		\hline
		\multicolumn{2}{|c||}{Significance($\sigma$)}&3.52&2.03&\multicolumn{5}{c|}{ }\\
		\cline{1-4}
		\multicolumn{2}{|c||}{$\mathcal{L}_{5\sigma}$ (fb$^{-1}$)}&2017.17&$\gg3000$&\multicolumn{5}{c|}{ }\\
		\hline\hline
		
		\multirow{2}{*}{100} &$c/t-g \to S_1\, \tau$&1395.93&929.10&\multirow{2}{*}{42636.18}&\multirow{2}{*}{199.77}&\multirow{2}{*}{54.66}&\multirow{2}{*}{953.79}&\multirow{2}{*}{129.25}\\
		\cline{2-4}
		&$b-g\to S_1\,\nu$&619.95&479.67&&&&&\\
		\hline
		\multicolumn{2}{|c||}{Total}&2015.88&1408.77&\multicolumn{5}{c|}{43973.66}\\
		\hline
		\multicolumn{2}{|c||}{Significance ($\sigma$)}&9.40&6.61&\multicolumn{5}{c|}{ }\\
		\cline{1-4}
		\multicolumn{2}{|c||}{$\mathcal{L}_{5\sigma}$ (fb$^{-1}$)}&28.29&57.17&\multicolumn{5}{c|}{ }\\
		\hline
	\end{tabular}
	\caption{The number of events for  $1c-\rm{jet} + \geq 1\tau-\rm{jet} +   \ptmiss \geq$ 500 GeV finalstate for the benchmark points and dominant SM backgrounds at the LHC/FCC, with centre-of-mass energies of   30\,TeV  and 100\,TeV, for the integrated luminosities  of 1000\,fb$^{-1}$ and 100\,fb$^{-1}$, respectively. The required luminosities to achieve a $5\,\sigma$ signal ($\mathcal{L}_{5\,\sigma}$) are also shown for both cases.}\label{ctau1}
\end{table*}
%%%%%%%%%%%%%%%%%%%%%%%%%%%%%%%%%%%%%%%%%%%%%%%%%%%%%%%%%%%%%

\autoref{ctau1} presents the results for the $1c-\rm{jet} + 1\tau-\rm{jet} +   \ptmiss $  finalstate, where $c-$gluon, $b-$gluon as well as $t-$gluon contribute. The complete finalstate comprised of the advanced cuts and veto is given as follows:
\begin{center}
	$n_{c-\rm{jet}} = 1$, $n_{\tau-\rm{jet}} \geq 1$, $n_{j} \geq 2$, $n_\ell = 0$ \& \\
	$\ptmiss \geq 500$ GeV, $p_T^{j_1 , j_2} \geq 200$ GeV, $p_T^{\tau-\text{jet, } c-\rm{jet}} \geq 200$ GeV, $p_T^H \geq 1.2$ TeV.
\end{center}

Since the $c-$jet originates directly from the leptoquark decay we demand a relatively hard cut of $p_T > 200$ GeV for the $c$-jet. The $\tau-$jet however, can either come  directly from the production channel (for $c-g$ fusion) or from the leptoquark decay. Hence we also demand  $p_T > 200$ GeV for the $\tau$-jet. This almost implied that the first two $p_T$ ordered jets are with  $p_T > 200$ GeV.  The missing transverse momentum $\ptmiss \geq 500$ GeV is demanded as well  since the relatively boosted neutrino arise at the production level.  The number of events listed for the benchmark points and dominant SM backgrounds in \autoref{ctau1} for 30\,TeV and 100\, TeV centre-of-mass energies at integrated luminosities of  1000, 100 fb$^{-1}$, respectively at the LHC/FCC. Once again, at the 14 TeV LHC, we do not even reach $1\sigma$ for either benchmark point and we do not list them. Moving to the 30 TeV LHC, we get a $3.52\sigma$ significance for BP1, with the $5\sigma$ strength being obtainable with $\sim2020$ \fbi luminosity. At the highest centre-of-mass energy of 100 TeV, both the BPs cross $5\sigma$ significance, with $9.40\sigma$ for BP1, and $6.61\sigma$ for BP2. High cross-section and more number of jets keep $t\bar{t}$ as the substantially dominant background here.

\subsubsection{$1c-\rm{jet} + 2 \tau-\rm{jet} + \ptmiss$}

%%%%%%%%%%%%%%%%%%$1c-\rm{jet} + 2  \tau-\rm{jet} + \ptmiss$%%%%%%%%%%%%%%%%%
\begin{table*}[h!]
	\renewcommand{\arraystretch}{1.2}
	\centering
	\begin{tabular}{|c|c||c|c||c|c|c|c|c|}
		\hline
		\multirow{2}{*}{ $E_{\rm{CM}}$ in}&&\multicolumn{7}{c|}{$1c-\rm{jet} + \geq 2  \tau-\rm{jet} + \ptmiss$}\\
		\cline{3-9}
		\multirow{2}{*}{TeV}& Mode &\multicolumn{2}{c||}{Signal}&\multicolumn{5}{c|}{Backgrounds}\\
		\cline{3-9}
		&&BP1&BP2 &$t\bar t$&$VV$&$VVV$&$t\bar{t}V$&$tVV$\\
		\hline

		\multirow{2}{*}{30} &$c/t-g \to S_1\, \tau$&55.28&35.07&\multirow{2}{*}{619.31}&\multirow{2}{*}{10.93}&\multirow{2}{*}{4.00}&\multirow{2}{*}{30.65}&\multirow{2}{*}{2.34}\\
		\cline{2-4}
		&$b-g\to S_1\,\nu$&4.35&3.21&&&&&\\
		\hline
		\multicolumn{2}{|c||}{Total}&59.64&38.28&\multicolumn{5}{c|}{667.23}\\
		\hline
		\multicolumn{2}{|c||}{Significance($\sigma$)}&2.21&1.44&\multicolumn{5}{c|}{ }\\
		\cline{1-4}
		\multicolumn{2}{|c||}{$\mathcal{L}_{5\sigma}$ (fb$^{-1}$)}&$\gg3000$&$\gg3000$&\multicolumn{5}{c|}{ }\\
		\hline\hline
		
		\multirow{2}{*}{100} &$c/t-g \to S_1\, \tau$&329.07&222.62&\multirow{2}{*}{1118.48}&\multirow{2}{*}{9.60}&\multirow{2}{*}{8.82}&\multirow{2}{*}{103.10}&\multirow{2}{*}{15.56}\\
		\cline{2-4}
		&$b-g\to S_1\,\nu$&17.33&14.02&&&&&\\
		\hline
		\multicolumn{2}{|c||}{Total}&346.40&236.64&\multicolumn{5}{c|}{1255.56}\\
		\hline
		\multicolumn{2}{|c||}{Significance($\sigma$)}&8.65&6.13&\multicolumn{5}{c|}{ }\\
		\cline{1-4}
		\multicolumn{2}{|c||}{$\mathcal{L}_{5\sigma}$ (fb$^{-1}$)}&33.38&66.62&\multicolumn{5}{c|}{ }\\
		\hline
	\end{tabular}
	\caption{The number of events for  $1c-\rm{jet} + 2\tau-\rm{jet} +   \ptmiss \geq  200$ GeV finalstate for the benchmark points and dominant SM backgrounds at the LHC/FCC with centre-of-mass energies of   30\,TeV  and 100\,TeV for  the integrated luminosities of 1000, 100\,fb$^{-1}$, respectively. The required luminosities to achieve a $5\sigma$ signal ($\mathcal{L}_{5\sigma}$) are also shown for both cases.}\label{ctau2}
\end{table*}
%%%%%%%%%%%%%%%%%%%%%%%%%%%%%%%%%%%%%%%%%%%%%%%%%%%%%%%%%%%%%
In \autoref{ctau2} we now tag one  more $\tau$-jet compared to the previous case and  present $1c-\rm{jet} + 2\tau-\rm{jet} +   \ptmiss \geq  500$ GeV finalstates for the benchmark points and dominant SM backgrounds for the two different centre-of-mass energies. Here, the complete finalstate including the advanced cuts and veto is described as follows:
	\begin{center}
		$n_{c-\rm{jet}} = 1$, $n_{\tau-\rm{jet}} \geq 2$, $n_{j} \geq 3$, $n_\ell = 0$ \& \\
		$\ptmiss \geq 500$ GeV, $p_T^{j_1 , j_2} \geq 200$ GeV, $p_T^{\tau-\text{jet, } c-\rm{jet}} \geq 200$ GeV, $p_T^H \geq 1.2$ TeV.
	\end{center}

Tagging  one more $\tau$-jet and demanding high momentum for both of them definitely reduces the events numbers both for the signal as well as for the backgrounds. We see a overall drop in the significance. The signal remains very weak with $<1\sigma$ significance for both BPs, at the 14 TeV LHC, which are not listed. At the centre-of-mass energy of 30 TeV, the signal for BP1 shows a $2.21\sigma$ strength, while BP2 stays weaker with $1.44\sigma$ significance. Moving to 100 TeV, both benchmark points show promising outcomes, with $8.65\sigma$ and $6.13\sigma$ significance for BP1 and BP2, respectively at 100 \fbi of integrated luminosity. Nonetheless, this set up will help us in reconstructing the  invariant mass edge of  $c\,\tau$  which  we discuss in the next in \autoref{sec:edge}. 

\subsection{Invariant mass edge distribution}
\label{sec:edge}

%%%%%%%%%%%%%%%Feynman diag of  S_1 decay  %%%%%%%%%%%%
\begin{figure*}[h!]
	\begin{center}
		\subfigure[]{$\vcenter{\hbox{
					\begin{tabular}{|c|}
						\hline\\[5mm]
						\begin{tikzpicture}
							\begin{feynman}
								\vertex (a1);
								\vertex [left=1cm of a1] (a0){$S_1$};
								\vertex [above right=1.2cm of a1] (a2) {$\bar c$};
								\vertex [below right=1.2 cm of a1] (a3){$\tau^+$};
								\vertex [above right=2cm of a3] (a4);
								\vertex [right=2cm of a3] (a5){$\bar\nu_\tau$};
								\vertex [right = 1.5cm of a4] (a6);
								\vertex [above = 0.2cm of a6] (a7){$u$};
								\vertex [below = 0.2cm of a6] (a8){$\bar d$};
								
								\diagram {(a0)--[charged scalar](a1)--[anti fermion](a2),
									(a1)--[anti fermion](a3),
									(a5)--[fermion](a3),
									(a3)--[charged boson,quarter left,edge label'=\(W^+\)](a4),
									(a4)--[fermion,out=80,in=180](a7),
									(a8)--[fermion,in=270,out=180](a4)};
								
								\draw [decoration={brace},thick, decorate] (a7.north east) -- (a8.south east)
								node [pos=0.5, right] {\(\,\pi^+\)};
							\end{feynman}
						\end{tikzpicture}\\[8mm]
						\hline
					\end{tabular}
					\vspace*{7mm}
			}}$}
		\hfil
		\subfigure[]{$\vcenter{\hbox{\includegraphics[height=0.245\textheight]{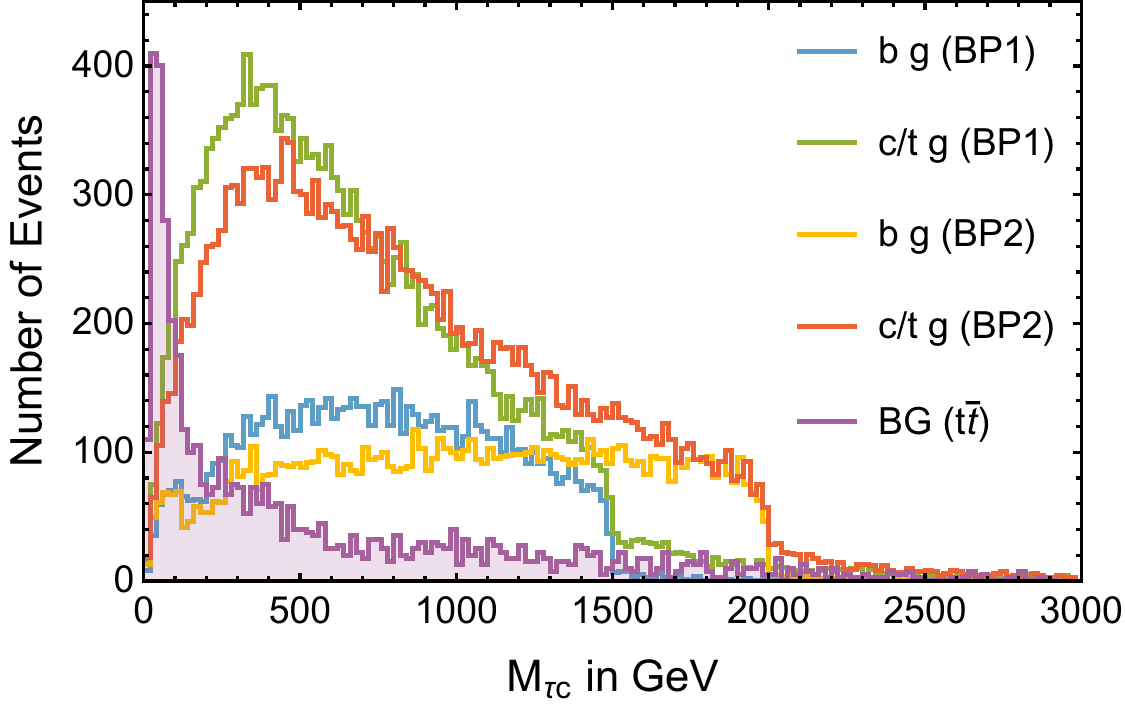}}}$}
		\caption{ Panel (a) presents the Feynman diagram of $S_1\to \bar{c} \, \bar{\tau} \to \bar{c}\,  \pi^+ \bar{\nu}_\tau$ and panel (b) shows the invariant  mass distribution of $c$-jet and $\tau$-jet ($M_{\tau c}\equiv m_{\pi c}$) for the chosen scenarios BP1, BP2 and the SM background $t\bar{t}$ (scaled by 5) at 14 TeV centre-of-mass energy  at the LHC. Invariant mass edges of $m^{\rm max}_{c\pi^+}$ at the leptoquark masses $M_{S_1}$ are clearly identifiable for both the benchmark scenarios.}\label{LQdcy}
	\end{center}
\end{figure*}
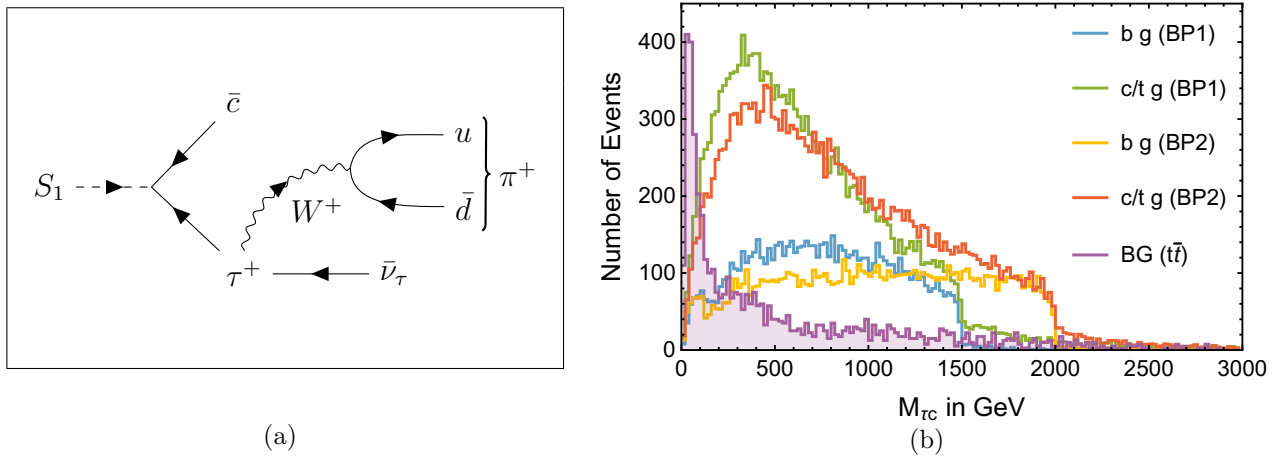
%%%%%%%%%%%%%%%%%%%%%%%%%%%%%%%%%%%%%%%%%%%%%%%%%%%%%%

Ensuring the finalstates with excess events, we now look for invariant mass distributions for the resonance discovery of the leptoquark. The decay branching fractions quoted in \autoref{brss1} show that the leptoquark $S_1$ decays mostly to third generation fermions. It has been demonstrated in \cite{Bandyopadhyay:2018syt} that the third generation fermions give rise to a very rich finalstate; however, in the presence of a large number of jets, and specially the missing momentum from neutrino, the peaks are smeared. In case of a decay to $c\mu$ finalstate a very clear invariant mass peak can be constructed~\cite{Bandyopadhyay:2018syt}. In this paper due the absence of such mode we demonstrate how invariant mass edge can be constructed, which is similar to a situation arises in supersymmetric theories with neutralino decays~\cite{Mohr:2009jd,CMS:2009twa}. 

%Unlike $c-\mu$ invariant mass peak the $c-\tau$ invariant mass distributions does not give a peak rather it produces an edge similarly to supersymmetric neutralino decays \cite{susyneutralino}. 
As schematically shown in \autoref{LQdcy}(a), $S_1$ decays into a $c-$jet and a $\tau$, which is detected as hadronic $\tau-$jet \cite{tautagging,tautagging2}. The neutrino in the finalstate contributes to missing energy but not to the $\tau-$jet energy, which  is identified as hadronic one-prong ($\pi^\pm$) jet. This results in a mass edge rather than a mass peak at the $S_1$ mass in the  $c$-jet$-\tau$-jet invariant mass distribution as given in \autoref{invedge}.
\bea\label{invedge}
M_{\tau c}^{\rm max}\equiv\, m^{\rm max}_{\pi c}&=& \frac{1}{m_\tau}[(m^2_{S_1}-m^2_{\tau})(m^2_{\tau} -m^2_{\nu})]^{1/2}\nn\\
&\simeq& \frac{1}{m_\tau}(m^2_{S_1}-m^2_{\tau})^{1/2} m_\tau %\nn \\
%&\simeq& m_{S_1}\left[1 - \frac{m^2_{\tau}}{m^2_{S_1}}\right]^{1/2} \nn \\
%&\simeq&m_{S_1} - \frac{1}{2}\frac{m^2_{\tau}}{m^2_{S_1}}\nn \\
\simeq m_{S_1}\,.
\eea

The mass edge can be calculated from a three-body decay $S_1$, where the $\tau-$jet accumulates the energy of the pion ($\pi^\pm$).  This can be expressed in terms of the mass of the leptoquarks, mass of  $\tau$ and the neutrino. As the leptoquark is at the TeV scale, from the collider perspective we can consider the last two particles as massless and this leaves us with the mass edge at $\sim m_{S_1}$ as shown in \autoref{invedge}. In \autoref{LQdcy}(b) we show the distributions at 14\,TeV LHC for the two benchmark points BP1, BP2 and the dominant SM background $t\bar{t}$. It is clear that the invariant mass of $c$ and $\tau$ (rather the $\pi^+$)  $m^{\rm max}_{c\pi^+}$, gives mass edge at $\sim $ 1.5\,TeV and 2.0\,TeV for the respective benchmark points, where the contributions are coming from all three production modes $b-g$ and $c/t-g$ fusions. The dominant SM background $t\bar{t}$ does not show any mass edge at these two regions.

%%%%%%%%%%%%Invariant mass edge %%%%%%%%%%%%%%%%%%%%%%%%%%%%%%%%
\begin{table*}[h!]
	\renewcommand{\arraystretch}{1.2}
	\centering
	\begin{tabular}{|c|c||c|c||c|c|}
			\hline
			\multicolumn{6}{|c|}{$(m_{\pi c} <m_{\rm edge})- (m_{\pi c} \geq m_{\rm edge}$)}\\
		\hline
	$E_{\rm{CM}}$&Mode&BP1 & Background & BP2 & Background \\
	\hline
		
		\multirow{2}{*}{30\,TeV} &$c/t-g \to S_1\, \tau$&358.97&\multirow{3}{*}{31077.18}&307.53&\multirow{3}{*}{24119.70}\\
		\cline{2-3} \cline{5-5}
		&$b-g\to S_1\,\nu$&315.35&&287.93&\\
			\cline{1-3} \cline{5-5}
		\multicolumn{2}{|c||}{Total}&674.32&&595.46&\\
		\hline
		\multicolumn{2}{|c||}{Significance($\sigma$)}&3.78&&3.79&\\
		\cline{1-3} \cline{5-5}
		\multicolumn{2}{|c||}{$\mathcal{L}_{5\sigma}$ (fb$^{-1}$)}&1745.76&&1742.63&\\
		\hline\hline
		
		\multirow{2}{*}{100\,TeV} &$c/t-g \to S_1\, \tau$&868.43&\multirow{3}{*}{56479.22}&871.02&\multirow{3}{*}{49375.35} \\
		\cline{2-3} \cline{5-5}
		&$b-g\to S_1\,\nu$& 840.59&&791.48&\\
			\cline{1-3} \cline{5-5}
		\multicolumn{2}{|c||}{Total}&1709.02&&1662.50&\\
		\hline
		\multicolumn{2}{|c||}{Significance($\sigma$)}&7.08&&7.36&\\
		\cline{1-3} \cline{5-5}
		\multicolumn{2}{|c||}{$\mathcal{L}_{5\sigma}$ (fb$^{-1}$)}&49.80&&46.16&\\
		\hline
	\end{tabular}
\caption{the number of event combinations for  $(m_{\pi c}<m_{\rm edge})- (m_{\pi c} \geq m_{\rm edge})$  with reconstructed invariant mass of $\tau-$ and $c-$ jets as $M_{\tau c}\equiv m_{\pi c}$ for the benchmark points and the total SM background at the LHC/FCC with centre-of-mass energies of  30\,TeV  and 100\,TeV for the integrated luminosities of 1000, 100\,fb$^{-1}$, respectively. The required luminosities to achieve a $5\,\sigma$ signal ($\mathcal{L}_{5\,\sigma}$) are also shown for both the cases.} \label{medge}
\end{table*}
%%%%%%%%%%%%%%%%%%%%%%%%%%%%%%%%%%%%%%%%%%%%%%%%%%%%%

We also present, in \autoref{medge}, the number of events for the interval $(m_{\pi c} <m_{\rm edge})- (m_{\pi c} \geq m_{\rm edge})$  with reconstructed invariant mass of $\tau-$ and $c-$ jets, denoted as $M_{\tau c}\equiv m_{\pi c}$, for the benchmark points and the total SM background at the LHC/FCC at two different centre-of-mass energies by identifying the $\tau$-jet as hadronic one prong ($\pi^\pm$) jet. Additionally, we implement the $W$- and $Z$-boson vetoes on the di-jet and di-lepton invariant masses, and put the hardness cut of $p_T^H \geq 1200$ GeV to obtain these numbers. The top quark backgrounds are further reduced by demanding $n_{b-\rm{jet}} = 0$. Similar to the previous analysis, we present the numbers at the 30\,TeV and 100\,TeV results are for 1000, 100 fb$^{-1}$ integrated luminosities, respectively. Here $m_{\rm edge}$ represents  the  mass-edge (or mass-wall) that we see for BP1 and BP2 in \autoref{LQdcy}(b) and thus a asymmetry around it is constructed by selecting events in the interval $(m_{\pi c} <m_{\rm edge})- (m_{\pi c} \geq m_{\rm edge})$. In both benchmark points, we achieve a $\sim3.8\sigma$ significance at the centre-of-mass energy of 30 TeV. This increases to $7.08\sigma$ for BP1, and $7.36\sigma$ for BP2, when we move to the 100 TeV LHC/FCC. The background numbers in  \autoref{medge}  includes contributions from all possible backgrounds i.e. $t\bar{t}, \, VV,\, VVV,\, t\bar{t}V,$ and $tVV$. 

\section{$S_3$ at the LHC/FCC}
\label{sec:S3}

%%%%%%%%%%%%%%%Feynman diagrams of different production modes of S_3 %%%%%%%%%%%%%%
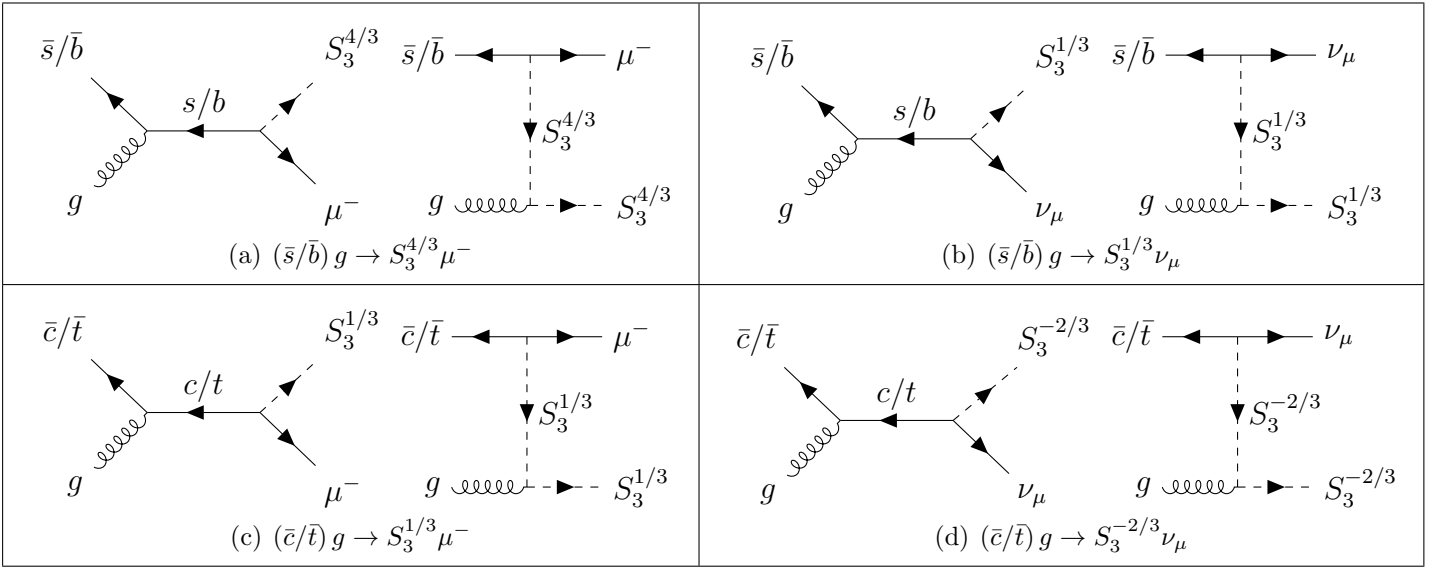
\begin{figure}[t!]
	\hspace*{-10mm}
	\begin{tabular}{|c|c|}
		\hline
		\subfigure[$(\bar s/\bar b)\,g\to S_3^{4/3} \mu^-$]{
			\begin{tikzpicture}
			\begin{feynman}
			\vertex (a1);
			\vertex [above left=1cm of a1] (a0){$\bar s /\bar b$};
			\vertex [right=1.5cm of a1] (a2);
			\vertex [above right=1 cm of a2] (a3){$S_3^{4/3}$};
			\vertex [below left=1cm of a1] (b0){$g$};
			\vertex [below right=1cm of a2] (b3){$\mu^-$};
			\diagram {(a0)--[anti fermion](a1)--[anti fermion,edge label=$s/b$](a2)--[charged scalar](a3),
				(b0)--[gluon](a1),(a2)--[fermion](b3)};
			\end{feynman}
			\end{tikzpicture}\hfill
			\begin{tikzpicture}
			\begin{feynman}
			\vertex (a1);
			\vertex [above=1cm of a1] (z1);
			\vertex [left=10mm of z1] (z0) {$\bar s/\bar b$};
			\vertex [right=10mm of z1] (z2) {$\mu^-$};
			\vertex [below=1cm of a1] (b1);
			\vertex [left=10mm of b1] (b0){$g$};
			\vertex [right=10mm of b1] (b2) {$S_3^{4/3}$};
			\diagram {(z0)--[anti fermion](z1)--[charged scalar, edge label=$S_3^{4/3}$](b1)--[charged scalar](b2),
				(b0)--[gluon](b1),
				(z1)--[fermion](z2)};
			\end{feynman}
			\end{tikzpicture}}&
		\subfigure[$(\bar s/\bar b)\,g\to S_3^{1/3} \nu_\mu$]{
			\begin{tikzpicture}
			\begin{feynman}
			\vertex (a1);
			\vertex [above left=1cm of a1] (a0){$\bar s/\bar b$};
			\vertex [right=1.5cm of a1] (a2);
			\vertex [above right=1 cm of a2] (a3){$S_3^{1/3}$};
			\vertex [below left=1cm of a1] (b0){$g$};
			\vertex [below right=1cm of a2] (b3){$\nu_\mu$};
			\diagram {(a0)--[anti fermion](a1)--[anti fermion,edge label=$s/b$](a2)--[charged scalar](a3),
				(b0)--[gluon](a1),(a2)--[fermion](b3)};
			\end{feynman}
			\end{tikzpicture}\hfill
			\begin{tikzpicture}
			\begin{feynman}
			\vertex (a1);
			\vertex [above=1cm of a1] (z1);
			\vertex [left=10mm of z1] (z0) {$\bar s/\bar b$};
			\vertex [right=10mm of z1] (z2) {$\nu_\mu$};
			\vertex [below=1cm of a1] (b1);
			\vertex [left=10mm of b1] (b0){$g$};
			\vertex [right=10mm of b1] (b2) {$S_3^{1/3}$};
			\diagram {(z0)--[anti fermion](z1)--[charged scalar, edge label=$S_3^{1/3}$](b1)--[charged scalar](b2),
				(b0)--[gluon](b1),
				(z1)--[fermion](z2)};
			\end{feynman}
			\end{tikzpicture}}\\
		\hline
		\subfigure[$(\bar c/\bar t)\,g\to S_3^{1/3} \mu^-$]{	
			\begin{tikzpicture}
			\begin{feynman}
			\vertex (a1);
			\vertex [above left=1cm of a1] (a0){$\bar c/\bar t$};
			\vertex [right=1.5cm of a1] (a2);
			\vertex [above right=1 cm of a2] (a3){$S_3^{1/3}$};
			\vertex [below left=1cm of a1] (b0){$g$};
			\vertex [below right=1cm of a2] (b3){$\mu^-$};
			\diagram {(a0)--[anti fermion](a1)--[anti fermion,edge label=$c/t$](a2)--[charged scalar](a3),
				(b0)--[gluon](a1),(a2)--[fermion](b3)};
			\end{feynman}
			\end{tikzpicture}\hfill
			\begin{tikzpicture}
			\begin{feynman}
			\vertex (a1);
			\vertex [above=1cm of a1] (z1);
			\vertex [left=10mm of z1] (z0) {$\bar c/\bar t$};
			\vertex [right=10mm of z1] (z2) {$\mu^-$};
			\vertex [below=1cm of a1] (b1);
			\vertex [left=10mm of b1] (b0){$g$};
			\vertex [right=10mm of b1] (b2) {$S_3^{1/3}$};
			\diagram {(z0)--[anti fermion](z1)--[charged scalar, edge label=$S_3^{1/3}$](b1)--[charged scalar](b2),
				(b0)--[gluon](b1),
				(z1)--[fermion](z2)};
			\end{feynman}
			\end{tikzpicture}}&
		\subfigure[$(\bar c/\bar t)\,g\to S_3^{-2/3} \nu_\mu$]{
			\begin{tikzpicture}
			\begin{feynman}
			\vertex (a1);
			\vertex [above left=1cm of a1] (a0){$\bar c/\bar t$};
			\vertex [right=1.5cm of a1] (a2);
			\vertex [above right=1 cm of a2] (a3){$S_3^{-2/3}$};
			\vertex [below left=1cm of a1] (b0){$g$};
			\vertex [below right=1cm of a2] (b3){$\nu_\mu$};
			\diagram {(a0)--[anti fermion](a1)--[anti fermion,edge label=$c/t$](a2)--[charged scalar](a3),
				(b0)--[gluon](a1),(a2)--[fermion](b3)};
			\end{feynman}
			\end{tikzpicture}\hfill
			\begin{tikzpicture}
			\begin{feynman}
			\vertex (a1);
			\vertex [above=1cm of a1] (z1);
			\vertex [left=10mm of z1] (z0) {$\bar c/\bar t$};
			\vertex [right=10mm of z1] (z2) {$\nu_\mu$};
			\vertex [below=1cm of a1] (b1);
			\vertex [left=10mm of b1] (b0){$g$};
			\vertex [right=10mm of b1] (b2) {$S_3^{-2/3}$};
			\diagram {(z0)--[anti fermion](z1)--[charged scalar, edge label=$S_3^{-2/3}$](b1)--[charged scalar](b2),
				(b0)--[gluon](b1),
				(z1)--[fermion](z2)};
			\end{feynman}
			\end{tikzpicture}}\\
		\hline
		
	\end{tabular}
	\caption{The tree level Feynman diagrams for $(s/b)-g$ and $(c/t)-g$ fusions producing different components of the leptoquark $S_3$ in association with a lepton. }
	\label{fig:s3prod}
\end{figure}
%%%%%%%%%%%%%%%%%%%%%%%%%%%%%%%%%%%%%%%%%%%%%%%%%%%%%%%

In this section we discuss the collider phenomenology of the $S_3$ leptoquark. Unlike $S_1$, the $SU(2)_L$ triplet leptoquark $S_3$ has three components namely $S^{4/3}_3,\, S^{1/3}_3, \, S^{2/3}_3$ which are degenerate at the tree-level (see \autoref{eq:LS32}). Finding distinguishable signatures for these different excitations can be challenging. In this article we illustrate how  production modes vary depending on the leptoquark excitations.  In \autoref{tab:all} we present three benchmark scenarios corresponding to two different mass scales 1.5\,TeV and 2\,TeV for the leptoquark, and three different Yukawa-type coupling combinations. Note that, BP3 has largest $Y_{S_3}^{32}$ value which can lead to sizable interactions between the second and third generation of fermions. For this reason we separately discuss BP3 as lepton flavour violating (LFV) signatures in decay in \autoref{sec:lfv}.

%%%%%%%%%%%%%%%Cross-section  S gluon fusion %%%%%%%%%%%%%%%%%%%%%%%%%%%%%
\begin{table}[h!]
	\renewcommand{\arraystretch}{2}
	\centering
	\begin{tabular}{|c||c|c|c|c|c|c|}
		\hline
		\multirow{3}{*}{\makecell{Benchmark \\ Points \\ ($M_{S_3}$)}}&\multicolumn{3}{c|}{$\sigma(s-g \to S_3^{4/3} \mu)$ in fb  }&\multicolumn{3}{c|}{$\sigma(s-g \to S_3^{1/3} \nu_\mu)$ in fb }\\
		&\multicolumn{3}{c|}{with center  of mass energies in TeV}&\multicolumn{3}{c|}{with center  of mass energies in TeV}\\
		\cline{2-7}
		&  $E_{\rm{CM}}$ =14&$E_{\rm{CM}}$ =30  &$E_{\rm{CM}}$ =100  &   $E_{\rm{CM}}$  =14 &$E_{\rm{CM}}$ =30  &$E_{\rm{CM}}$ =100 \\
		\hline
		\makecell{BP1 \\ (1.5 TeV)}& 0.33 &5.15&115.38& 0.17&2.58&57.69\\
		\hline
		%		BP2 &0.13&2.94&88.9&0.06&1.47&44.45 \\ 
		%		\hline
		\makecell{BP2 \\ (2.0 TeV)}&0.08&1.59&48.03&0.03&0.80&24.02\\
		\hline
	\end{tabular}
	\caption{The cross-sections (in fb) at the LHC via $s-g$ fusion of $S^{4/3}_3$ and $S^{1/3}_3$, for the two benchmark points, at three different centre-of-mass energies 14\,TeV, 30\,TeV and 100\,TeV, respectively. NNPDF$\_$lo$\_$as$\_$0130$\_$qed~\cite{nnpdf} is considered as the parton distribution function with $\sqrt{\hat{s}}$ as renormalization/factorization scale with the NLO QCD $K$-factor of 1.5.}  \label{crosss3s}
\end{table}
%%%%%%%%%%%%%%%%%%%%%%%%%%%%%%%%%%%%%%%%%%%%%%%%%%%%%%%%%%%%%%%

We list the Feynman diagrams for dominant single production processes of $S^{4/3}_3,\, S^{1/3}_3$ and $S^{2/3}_3$ via quark-gluon fusions in \autoref{fig:s3prod}. The cross-sections  at the LHC for the centre-of-mass energies of 14\,TeV, 30\,TeV and 100\,TeV are listed in \autoref{crosss3s} for $s-g$ fusion and in \autoref{crosss3c} for $c-g$ fusion for the benchmark points BP1 and BP2. Similarly, the production cross-sections for BP3 from $b-g$ and $t-g$ fusion at the same three values of centre-of-mass energies are presented in \autoref{crosss3b} and \autoref{crosss3t}. Here NNPDF$\_$lo$\_$as$\_$0130$\_$qed~\cite{nnpdf} has been used as parton distribution function where top quark is also included. The parton-level centre-of-mass energy, i.e. $\sqrt{\hat{s}}$  is used as renormalization/factorization scale. Again, similar to \autoref{sec:S1}, these cross-sections are enhanced with the NLO QCD $K$-factor of 1.5 \cite{Alves:2002tj, Hammett:2015sea}.  It is interesting to note that  $S^{4/3}_3$ can only be produced via $s-g$ fusion, whereas, $S^{2/3}_3$ is  produced through $c-g$ fusion for the chosen BP1 and BP2 scenarios. Due to different choices of couplings, in the case of BP3, the only production process for $S_3^{4/3}$ ($S_3^{2/3}$) is via $b-g$ ($t-g$) fusion. We notice that $S^{1/3}_3$ has contributions from both the production processes for the considered benchmark points. It is noteworthy that the production cross-section of $S_3^{1/3}$ in any particular fusion process is almost half of the production cross-section of $S_3^{4/3}$ and $S_3^{2/3}$ leptoquarks. This is due to the reason that the interaction vertex of $S_3^{4/3}$ and $S_3^{2/3}$ with quarks and leptons carry an additional $\sqrt2$ factor as can be observed from \autoref{eq:LS32}. However, due to larger mass scale leptoquark (in TeV range), the cross-sections at 14\,TeV centre-of-mass energy is not quite promising and we need to depend on the collisions at 30\,TeV and 100\,TeV centre-of-mass energies at the LHC/FCC.

%%%%%%%%%%%%%%%%%%%%%cross-section C-gluon fusion %%%%%%%%%%%%%%%%%%%%%%%%%
\begin{table}[h!]
	\renewcommand{\arraystretch}{2}
	\centering
	\begin{tabular}{|c||c|c|c|c|c|c|}
		\hline
			\multirow{3}{*}{\makecell{Benchmark \\ Points \\ ($M_{S_3}$)}}&\multicolumn{3}{c|}{$\sigma(c-g \to S_3^{1/3} \mu)$ in fb  }&\multicolumn{3}{c|}{$\sigma(c-g \to S_3^{2/3} \nu_\mu)$ in fb }\\
		&\multicolumn{3}{c|}{with center  of mass energies in TeV}&\multicolumn{3}{c|}{with center  of mass energies in TeV}\\
		\cline{2-7}
		&  $E_{\rm{CM}}$ =14&$E_{\rm{CM}}$ =30  &$E_{\rm{CM}}$ =100  &   $E_{\rm{CM}}$  =14 &$E_{\rm{CM}}$ =30  &$E_{\rm{CM}}$ =100 \\
		\hline
		\makecell{BP1 \\ (1.5 TeV)} & 0.12&2.09&49.76&0.26&4.19&99.47\\
		\hline
		%		BP2&0.05&1.17&37.99&0.09&2.34&75.99\\ \hline
		\makecell{BP2 \\ (1.5 TeV)}  &0.03&0.63&20.51&0.05&1.26&41.06 \\
		\hline
	\end{tabular}
	\caption{The cross-sections (in fb) at the LHC via $c-g$ fusion of $S^{1/3}_3$ and $S^{2/3}_3$, for the two benchmark points, at three different centre-of-mass energies 14\,TeV, 30\,TeV and 100\,TeV. NNPDF$\_$lo$\_$as$\_$0130$\_$qed~\cite{nnpdf} is considered as the parton distribution function  with $\sqrt{\hat{s}}$ as renormalization/factorization scale with the NLO QCD $K$-factor of 1.5.}  \label{crosss3c}
\end{table}
%%%%%%%%%%%%%%%%%%%%%%%%%%%%%%%%%%%%%%%%%%%%%%%%%%%%%%%%

%%%%%%%%%%%%%%%Cross-section  S gluon fusion %%%%%%%%%%%%%%%%%%%%%%%%%%%%%
\begin{table}[h!]
	\renewcommand{\arraystretch}{2}
	\centering
	\begin{tabular}{|c||c|c|c|c|c|c|}
		\hline
		\multirow{3}{*}{\makecell{Benchmark \\ Points \\ ($M_{S_3}$)}}&\multicolumn{3}{c|}{$\sigma(b-g \to S_3^{4/3} \mu)$ in fb  }&\multicolumn{3}{c|}{$\sigma(b-g \to S_3^{1/3} \nu_\mu)$ in fb }\\
		&\multicolumn{3}{c|}{with center  of mass energies in TeV}&\multicolumn{3}{c|}{with center  of mass energies in TeV}\\
		\cline{2-7}
			&  $E_{\rm{CM}}$ =14&$E_{\rm{CM}}$ =30  &$E_{\rm{CM}}$ =100  &   $E_{\rm{CM}}$  =14 &$E_{\rm{CM}}$ =30  &$E_{\rm{CM}}$ =100 \\
		\hline
		\makecell{BP3 \\ (1.5 TeV)} &0.05&0.90&23.56& 0.03 &0.45&11.81\\
		\hline
	\end{tabular}
	\caption{The cross-sections (in fb) at the LHC via $b-g$ fusion of $S^{4/3}_3$ and $S^{1/3}_3$, for the benchmark point BP3, at three different centre-of-mass energies 14\,TeV, 30\,TeV and 100\,TeV. NNPDF$\_$lo$\_$as$\_$0130$\_$qed~\cite{nnpdf} is considered as parton distribution function  with $\sqrt{\hat{s}}$ as renormalization/factorization scale, with the NLO QCD $K$-factor of 1.5.}  \label{crosss3b}
\end{table}
%%%%%%%%%%%%%%%%%%%%%%%%%%%%%%%%%%%%%%%%%%%%%%%%%%%%%%%%%%%%%%%

%%%%%%%%%%%%%%%%%%%%%cross-section C-gluon fusion %%%%%%%%%%%%%%%%%%%%%%%%%
\begin{table}[h!]
	\renewcommand{\arraystretch}{2}
	\centering
	\begin{tabular}{|c||c|c|c|c|c|c|}
		\hline
		\multirow{3}{*}{\makecell{Benchmark \\ Points \\ ($M_{S_3}$)}}&\multicolumn{3}{c|}{$\sigma(t-g \to S_3^{1/3} \mu)$ in fb  }&\multicolumn{3}{c|}{$\sigma(t-g \to S_3^{2/3} \nu_\mu)$ in fb }\\
		&\multicolumn{3}{c|}{with center  of mass energies in TeV}&\multicolumn{3}{c|}{with center  of mass energies in TeV}\\
		\cline{2-7}
		&  $E_{\rm{CM}}$ =14&$E_{\rm{CM}}$ =30  &$E_{\rm{CM}}$ =100  &   $E_{\rm{CM}}$  =14 &$E_{\rm{CM}}$ =30  &$E_{\rm{CM}}$ =100 \\
		\hline
		\makecell{BP3 \\ (1.5 TeV)} & 0.006&0.14&3.90&0.015&0.26&7.77\\
		\hline
	\end{tabular}
	\caption{The cross-sections (in fb) at the LHC via $t-g$ fusion of $S^{4/3}_3$ and $S^{1/3}_3$, for the benchmark point BP3, at three different centre-of-mass energies 14\,TeV, 30\,TeV and 100\,TeV. NNPDF$\_$lo$\_$as$\_$0130$\_$qed~\cite{nnpdf} is considered as parton distribution function  with $\sqrt{\hat{s}}$ as renormalization/factorization scale with the NLO QCD $K$-factor of 1.5.}  \label{crosss3t}
\end{table}
%%%%%%%%%%%%%%%%%%%%%%%%%%%%%%%%%%%%%%%%%%%%%%%%%%%%%%%%

%%%%%%%%%%%%%%%%Decay Branching fraction of S_3 %%%%%%%%%%%%%
\begin{table}[h!]
	\renewcommand{\arraystretch}{2}
	\centering
	\begin{tabular}{|c|c|c|}
		\hline
		Decay	& \multicolumn{2}{|c|}{Branching ratios (\%)}\\ 
		\cline{2-3}
		Modes & \makecell{BP1 \\ $M_{S_3}$ = 1.5 TeV} &\makecell{BP2 \\ $M_{S_3}$ = 2.0 TeV}\\
		\hline
		$S_3^{-4/3} \to s \mu$ & 100  & 100 \\
		\hline
		$S_3^{-1/3} \to c \mu$ & 50  & 50 \\
		\hline
		$S_3^{-1/3} \to s \nu_\mu$ & 50  & 50 \\
		\hline
		$S_3^{2/3} \to c \nu_\mu$ & 100  & 100  \\
		\hline
	\end{tabular}
	\caption{The decay branching ratios (in percentage) of $S_3$ for the chosen benchmark points BP1 and BP2.}  \label{brss3}
\end{table}

%%%%%%%%%%%%%%%%Decay Branching fraction of S_3 %%%%%%%%%%%%%

Next, in \autoref{brss3} we list the  decay branching fractions of the different excitations of $S_3$ for the first two benchmark points. We find that $S^{4/3}_3$ decays to $s\, \mu$ with 100\% branching ratio for BP1 and BP2. Again, in both the BPs, the modes $c\, \mu$  and $s\nu_\mu$ share  50\% branching ratios for $S^{1/3}_3$. The component $S^{2/3}_3$ decays completely (100\% branching fraction) to $c\nu_\mu$ for BP1 and BP2 as well. The decay branching ratios in BP3 for the lepton flavour violating  decays will be separately discussed in \autoref{sec:lfv}.

\subsection{Kinematic distributions and topologies}
\label{sec:kine_S3}

%%%%%%%%%%%%%%%%%%%%%%%%%%%   multiplicity  %%%%%%%%%%%%%%%%%%%%%%%%%%%%%%%%
\begin{figure}[h!]
	\subfigure[]{\includegraphics[width=0.47\textwidth]{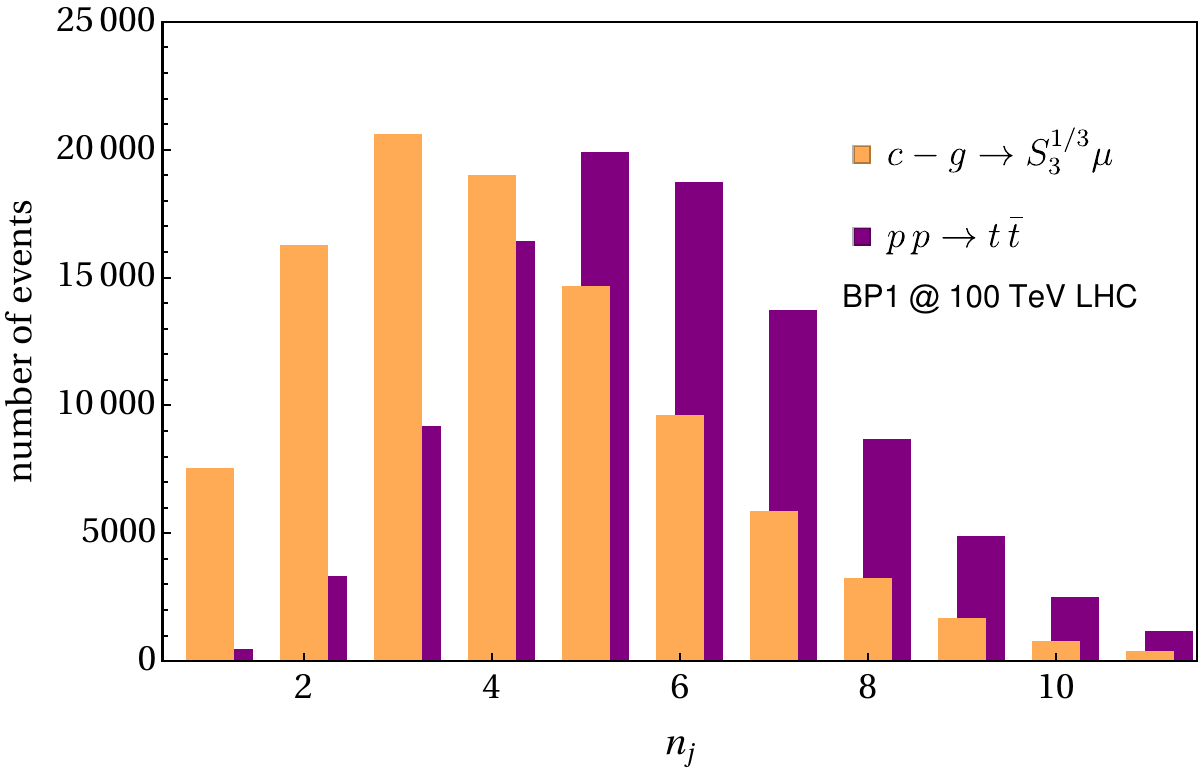}}
	\hfil
	\subfigure[]{\includegraphics[width=0.47\textwidth]{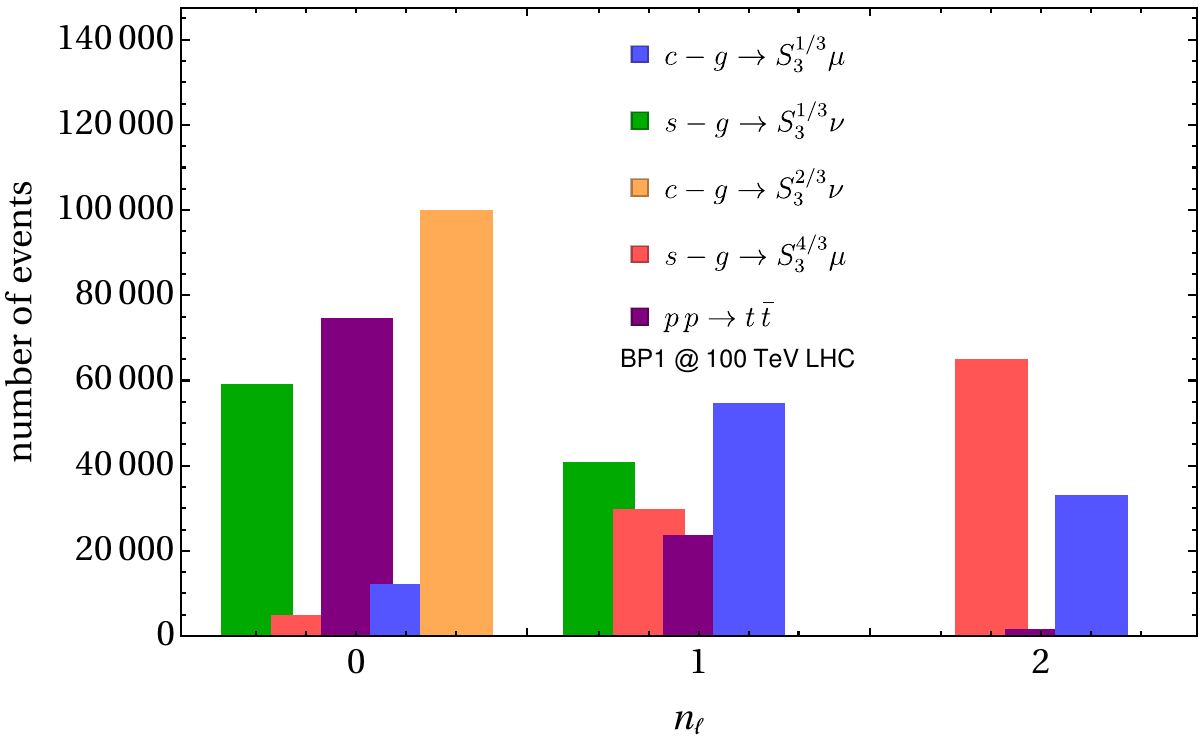}}
	\caption{The jet multiplicity ($n_j$ in (a)) and lepton multiplicity ($n_\ell$ in (b)) distributions of $S_3$ (for BP1) and the SM background $t\bar{t}$ at the LHC/FCC with the centre-of-mass energy of 100\,TeV. The jet multiplicity for signal, shown in (a), represents $c-g\to S_3^{1/3}\mu$ channel only. Since the other single-production channels of $S_3$ show the same jet multiplicity distribution, they are not depicted in (a). }\label{jlml_S3}
\end{figure}

We compare various kinematic distributions for $S_3$ leptoquark with the dominant SM background arising from $t\bar t $ channel. For illustration we choose to discuss these distributions for BP1 at 100\,TeV centre-of-mass energy. The jet multiplicity distribution ($n_j$) for the signal (in orange) and $t\bar t $ background (in purple) are displayed in \autoref{jlml_S3}(a) for the channel $c-g\to S_3^{1/3}\mu$. All the four production channels for different components of $S_3$, as shown in the Feynman diagrams in \autoref{fig:s3prod}, exhibit similar jet multiplicity distribution peaking around three, whereas, the SM background $t\bar t$ peaks at five jets, with more events in the higher multiplicity regions due to large ISR/FSR effects at the 100 TeV centre-of-mass energy. Similarly, \autoref{jlml_S3}(b) illustrates the lepton multiplicity distributions ($n_\ell$) for signal and $t\bar t$ background for BP1 at 100\,TeV centre-of-mass energy. As discussed in the case for $S_1$, the light charged leptons $(e^\pm,\,\mu^\pm)$ for $t\bar t$ essentially come from $W^\pm$ bosons which are produced with the decay of the top quarks to bottom quarks. As the branching fraction of $W^\pm$ to light charged leptons ($e^\pm,\,\mu^\pm$) is only about $22\%$, most of the $W^\pm$ decay hadronically producing no-lepton (dominant) and mono-lepton signatures for the background (in purple). In BP1, $S_3^{1/3}$ couples to both muon and $\nu_\mu$ and thus, $c-g \to S_3^{1/3}\mu$ (in blue) shows mono-lepton and di-lepton signatures, whereas, $s-g \to S_3^{1/3}\nu_\mu$ (in green) shows non-leptonic and mono-leptonic signatures. Now, the component $S_3^{2/3}$ does not couple to any charged lepton, giving almost always zero-lepton events (in orange) in the finalstate. Lastly, it is easy to see that the leptoquark $S_3^{4/3}$ couples to muon only, and hence the process $s-g \to S_3^{4/3} \mu$ has maximum di-lepton events out of all the signal processes considered (in red).

\begin{figure}[h!]
	\centering
	\includegraphics[width=0.5\textwidth]{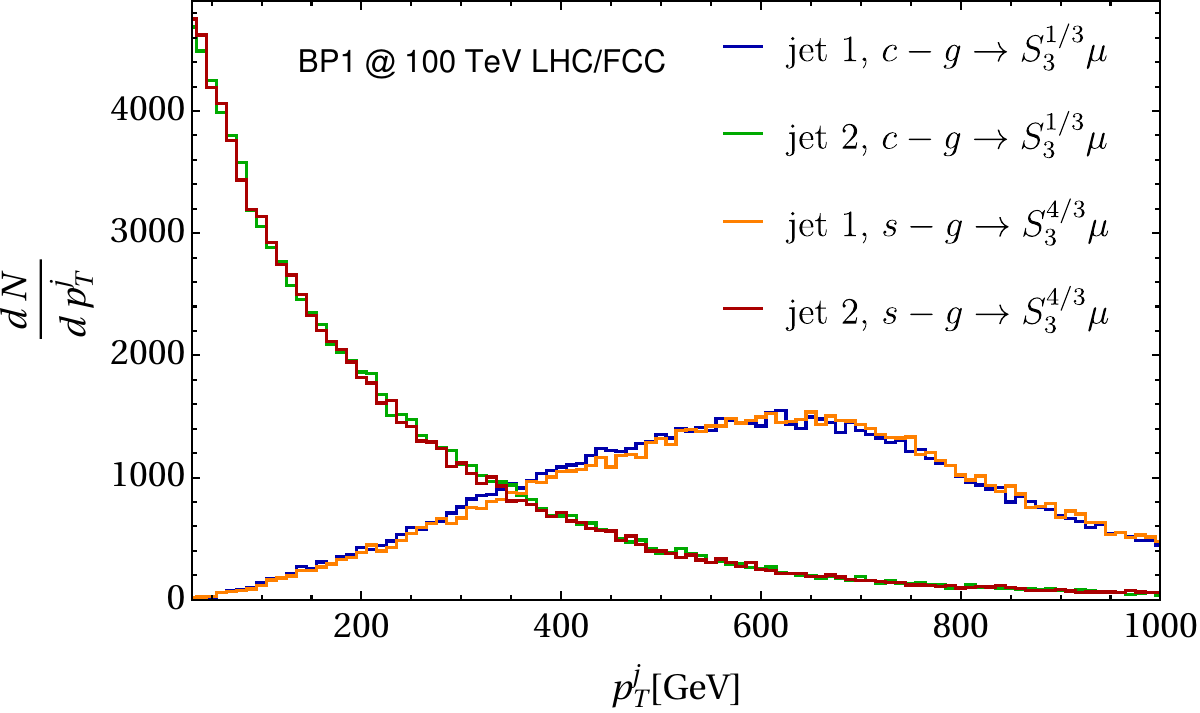}
	\caption{The jet $p_T$ ($p_T^j$) distributions from $S_3^{1/3}$ and $S_3^{4/3}$ production processes at the LHC/FCC with the centre-of-mass energy of 100\,TeV for BP1. $S_3^{2/3}$ shows the same distribution, so it is not shown in the plot.}\label{jpt_S3}
\end{figure}

In \autoref{jpt_S3}, the $p_T$ distribution of the two hardest jets emanating from two different fusion processes are shown for BP1 at the 100 TeV LHC/FCC. The hardest jet ($j_1$) each from the processes $c-g \to S_3^{1/3} \mu$ (blue) and $s-g \to S_3^{4/3} \mu$ (orange) both follow an almost identical distribution, peaking at around half of the leptoquark mass, as expected ($\sim750$) GeV. As there is no recoiled $\tau$-jet in the production processes, the only source of the second hardest jets in each case (green for $S_3^{1/3}$, red for $S_3^{4/3}$) are the ISR/FSR, and they are much softer. Distributions for $S_3^{2/3}$ are not shown here to avoid repetition and overlapping, as they also follow the very same pattern.

%%%%%%%%%%%%%%%%%%%%%%%%%%%%%  jet pT %%%%%%%%%%%%%%%%%%%%%%%%%%%%%%%%%%%

\begin{figure}
	\subfigure[]{\includegraphics[width=0.47\textwidth]{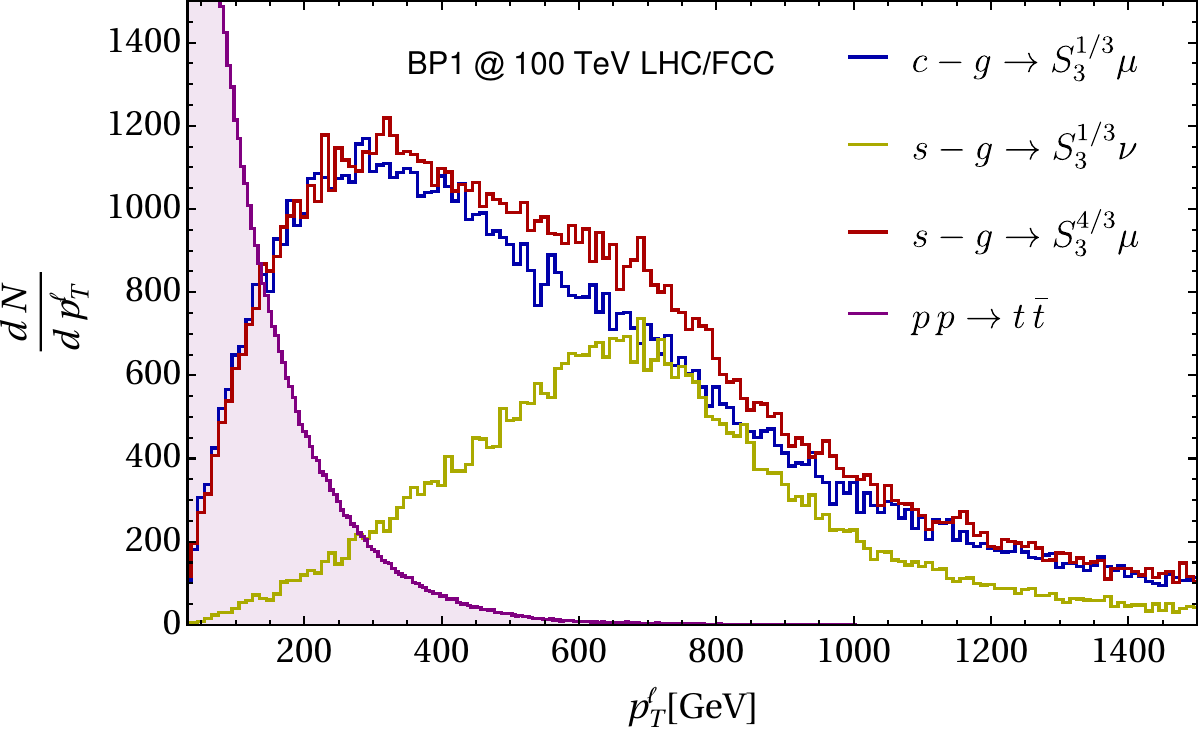}}
	\hfil
	\subfigure[]{\includegraphics[width=0.47\textwidth]{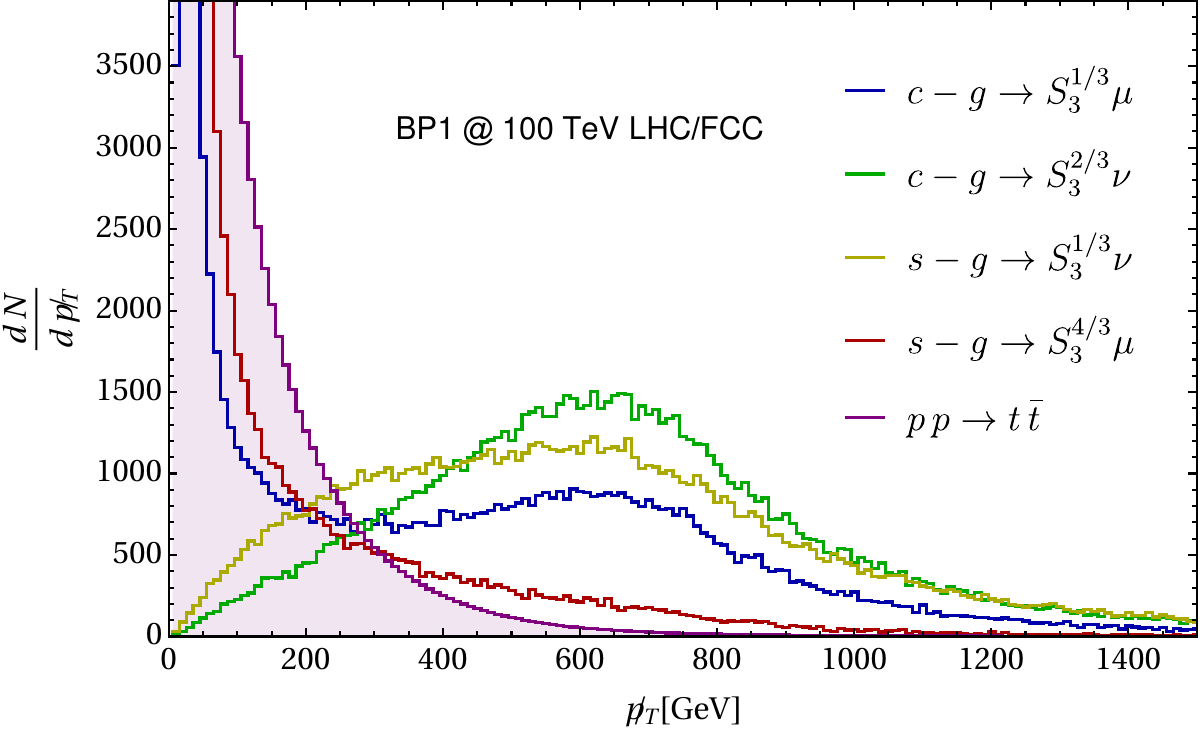}}
	\caption{(a) distributions of lepton $p_T$ ($p_T^\ell$) and (b) missing $p_T$ ($\ptmiss$) of $S_3$ (for BP1) with the SM background $t\bar{t}$ at the LHC with centre-of-mass energy of 100\,TeV. In (a), background has been scaled by 1/10, while in (b) it has been scaled as 1/300. $S_3^{2/3}$ is not shown in (a) as it does not couple to charged leptons.} \label{lmpt_S3}
\end{figure}

We show the transverse momentum distributions for light charged leptons $(e^\pm,\,\mu^\pm)$  for all the production channels of $S_3$ and $t\bar t$ background in \autoref{lmpt_S3}(a), for BP1 at the 100 TeV LHC/FCC. As discussed above, the light charged leptons for $t\bar t$ background can only come from the $W^\pm$, produced from the decay of top quark, the lepton $p_T$ distribution (in purple) exhibits a peak around the half mass of the $W$-boson and becomes insignificant for higher $p_T^{\rm lep}$. The component $S_3^{2/3}$ does not couple to charged leptons and hence does not contribute here. Now for the mode $s-g \to S_3^{1/3}\nu$ (in yellow), as the charged lepton arises only from the decay of $S_3^{1/3}$, the lepton $p_T$ distribution peaks around half the mass of leptoquark (i.e. 750\,GeV). In the other two modes (in blue and red), muons are produced at two stages: firstly, during the production of the leptoquark, and secondly, during its decay. So, the distributions show quite similar behaviour \pp{for $S_3^{4/3}$ and $S_3^{1/3}$.} However, $S_3^{1/3}$ can decay to muon or neutrino, whereas, $S_3^{4/3}$ has channel only to muon (see \autoref{brss3}).  For this reason the $p_T^{\rm lep}$ distribution for $s-g \to S_3^{4/3}\mu$ (in red) remains above the  mode $c-g \to S_3^{1/3}\mu$ (in blue). 

%%%%%%%%%%%%%%%%%%%%%%%%%%%%% lepton pT & pTmiss %%%%%%%%%%%%%%%%%%%%%%%%%%%%%%%

The missing $p_T$ distributions for signals and dominant SM background have been presented in \autoref{lmpt_S3}(b), again for BP1 at the centre-of-mass energy of 100 TeV. During the production and decay of $S_3^{4/3}$ (in red), no neutrino is involved, and thus the $\ptmiss$ peaks at around 30 GeV only, owing to neutrinos from SM sources. The production and decay of $S_3^{2/3}$ (in green) create two neutrinos, moving nearly opposite to each other with different momentum. However, the first neutrino at the production carries most of the missing transverse momentum and we observe a nice bell-shaped curve peaking around half of the leptoquark mass (i.e. 750\,GeV). During the production of $S_3^{1/3}$ through $s-g$ fusion (in yellow), neutrino appears at production level and again, there is 50\% probability for $S_3^{1/3}$ to decay to neutrino as well. Therefore, the $\rm\ptmiss$ distribution curve becomes a bit flat and resembles with in $S_1$ scenario. Finally, for $c\, g\to S_3^{1/3}\mu$ mode (in blue), when $S_3^{1/3}$ decays to muon the missing $p_T$ shows a peak in low $\rm\ptmiss$ region similar to the $t\bar{t}$ background. Although, there is also a possibility for $S_3^{1/3}$ to decay to a neutrino exhibiting a local maximum about the half of the leptoquark mass. Like muons, the neutrinos in $t \bar t$ scenario also arise from leptonic decay of $W^\pm$, consequently the missing $p_T$ distribution (in purple) peaks at lower $\rm\ptmiss$ and decreases gradually with a longer tail, enabling us to demand  large $\rm\ptmiss$ to reduce the $t\bar{t}$ background contamination.

We now focus on decay topologies arising from the single production channels for the two benchmark points BP1 and BP2. Our aim is to identify specific decay finalstates which can distinguish different components of the $S_3$ leptoquark. Due to particular gauge structure of the Lagrangian (in \autoref{eq:LS32}),  only $S_3^{4/3}$ and $S_3^{1/3}$ components of $S_3$ will be produced in $s-g$ fusion. Similarly, $c-g$ fusion produces $S_3^{2/3}$ and $S_3^{1/3}$ components of $S_3$ only. However, upon production, all these leptoquarks will decay to quarks and leptons. While $S_3^{4/3}$ and $S_3^{2/3}$ components decay to $s\mu$ and $c\nu_\mu$ respectively,  $S_3^{1/3}$ decays to both the  $s\nu_\mu$ and $c\mu$ topologies giving rise to the following finalstates:
\begin{align}
\label{eq:s43_prod}
	\hspace*{-1cm}\text{BP1, BP2:}\qquad\qquad s-g \;&\to\; S_3^{4/3} \mu \;\to\; (s\,\mu) + \mu \to 2 \mu+\rm{1-jet}\,,\\[3mm]
\label{eq:s13_prod_sg1}
	s-g \;&\to S_3^{1/3} \nu_\mu \;\to\; (s\,\nu_\mu) + \nu_\mu \to \rm{1-jet} +\ptmiss\,,\\
\label{eq:s13_prod_sg2}
	&\qquad\qquad\;\;\to \;  (c\,\mu) + \nu_\mu \to  \to1\mu+1c-\rm{jet}+\ptmiss\,,\\[3mm]
\label{eq:s13_prod_cg1}
	\qquad\qquad\qquad c-g \;&\to S_3^{1/3} \mu \;\to\; (c\,\mu) + \mu \to 2 \mu+1c-\rm{jet}\,,\\
\label{eq:s13_prod_cg2}
	&\qquad\qquad\;\to \; (s\,\nu_\mu) + \mu \to 1\mu+\rm{1-jet}+\ptmiss\,,\\[3mm]
\label{eq:s23_prod}
	c-g \;&\to S_3^{2/3} \nu_\mu \to\; (c\,\nu_\mu) + \nu_\mu \to 1c-\rm{jet}+\ptmiss\,.
\end{align}
As already mentioned, here `jet' implies light-jets unless the flavour is mentioned. We note that the complete decay chain of  the leptoquark $S_3^{4/3}$ provides a unique finalstate of di-muon plus mono light jet. Similarly, we have unique signature for $S_3^{2/3}$ through the finalstate consisting of mono $c-$jet with missing energy. On the other hand, four different finalstates are possible involving the production of $S_3^{1/3}$ in quark-gluon fusion at LHC/FCC with BP1 and BP2. In the  succeeding few subsections we describe the signal-background analyses for these six finalstates at centre-of-mass energies of 14 TeV, 30 TeV and 100 TeV.

\subsection{$S^{4/3}_3$ component of $S_3$: $1-\rm{jet} +2\mu + \ptmiss$}

\label{sec:s43}

As discussed earlier, leptoquark $S^{4/3}_3$ gets produced in association with muon from $s-g$ fusion via the Feynman diagram shown in \autoref{fig:s3prod}(a) and eventually decays into $s \mu$ with 100\% branching fraction as presented in \autoref{brss3}. This leads to the finalstate of  mono-jet plus di-muon with suitable additional cuts as given below:

\begin{center}
	$n_{j} \geq 1$, $n_\mu \geq 2$, $n_{\tau-\rm{jet}} = 0$ \& \\
	$p_T^{\ell_1} \geq 200$ GeV, $p_T^{j_1} \geq 200$ GeV, $\ptmiss \leq 30$ GeV, $p_T^H \geq 1200$ GeV.
\end{center}

%%%%%%%%%%%%%%%%%%%%%%%%%%%%%%%  1j+ 2\mu+ pTmiss %%%%%%%%%%%%%%%%%%%%%%%%%%%%%%%%%%%%%%%%%%

\begin{table*}[h!]
	\renewcommand{\arraystretch}{1.1}
	\centering
	\begin{tabular}{|c|c|c||c|c||c|c|c|c|c|}
		\hline
		\multirow{2}{*}{ $\sqrt s$ in}&\multirow{3}{*}{Fusion}&&\multicolumn{7}{c|}{$\geq 1-\rm{jet} + \geq2\mu + \ptmiss$}\\
		\cline{4-10}
		\multirow{2}{*}{TeV}	&& Mode &\multicolumn{2}{c||}{Signal}&\multicolumn{5}{c|}{Backgrounds}\\
		\cline{4-10}
		&&&BP1&BP2 &$t\bar t$&$VV$&$VVV$&$t\bar{t}V$&$tVV$\\
		\hline
		\hline
		\multirow{4}{*}{14} &\multirow{2}{*}{$s-g$}&$S_3^{4/3} \mu$& 174.08& 46.53&\multirow{4}{*}{14.47}&\multirow{4}{*}{171.22}&\multirow{4}{*}{20.83}&\multirow{4}{*}{53.75}&\multirow{4}{*}{11.57}\\
		\cline{3-5}
		&&$S_3^{1/3} \nu$&0.00&0.00&&&&&\\
		\cline{2-5}
		&\multirow{2}{*}{$c-g$}&$S_3^{1/3} \mu$&26.22&6.41&&&&&\\
		\cline{3-5}
		&&$S_3^{2/3} \nu$&0.00&0.00&&&&&\\
		\hline
		\multicolumn{3}{|c||}{Total}&200.30&52.94&\multicolumn{5}{c|}{271.84}\\
		\hline
		\multicolumn{3}{|c||}{Significance  ($\sigma$)}&9.21&2.94&\multicolumn{5}{c|}{ }\\
		\cline{1-5}
		\multicolumn{3}{|c||}{$\mathcal{L}_{5\,\sigma}$ (fb$^{-1}$)}&294.22&2897.62&\multicolumn{5}{c|}{ }\\
		\hline\hline
		
		\multirow{4}{*}{30} & \multirow{2}{*}{$s-g$}&$S_3^{4/3} \mu$& 2736.75&1032.44&\multirow{4}{*}{210.43}&\multirow{4}{*}{1145.45}&\multirow{4}{*}{150.41}&\multirow{4}{*}{588.86}&\multirow{4}{*}{74.46}\\
		\cline{3-5}		&&$S_3^{1/3} \nu$&0.00&0.00&&&&&\\
		\cline{2-5}
		&\multirow{2}{*}{$c-g$}&$S_3^{1/3} \mu$&443.89&163.39&&&&&\\
		\cline{3-5}
		&&$S_3^{2/3} \nu$&0.00&0.00&&&&&\\
		\hline
		\multicolumn{3}{|c||}{Total}&3180.64&1195.83&\multicolumn{5}{c|}{2169.62}\\
		\hline
		\multicolumn{3}{|c||}{Significance ($\sigma$)}&43.48&20.61&\multicolumn{5}{c|}{ }\\
		\cline{1-5}
		\multicolumn{3}{|c||}{$\mathcal{L}_{5\,\sigma}$ (fb$^{-1}$)}&13.22&58.84&\multicolumn{5}{c|}{ }\\
		\hline\hline
		
		\multirow{4}{*}{100}& \multirow{2}{*}{$s-g$}&$S_3^{4/3} \mu$& 5360.33&2664.26&\multirow{4}{*}{429.01}&\multirow{4}{*}{699.22}&\multirow{4}{*}{134.04}&\multirow{4}{*}{916.72}&\multirow{4}{*}{121.45}\\
		\cline{3-5}
		&&$S_3^{1/3} \nu$&0.00&0.00&&&&&\\
		\cline{2-5}
		&\multirow{2}{*}{$c-g$}&$S_3^{1/3} \mu$&935.26&440.97&&&&&\\
		\cline{3-5}
		&&$S_3^{2/3} \nu$&0.00&0.00&&&&&\\
		\hline
		\multicolumn{3}{|c||}{Total}&6295.59&3105.23&\multicolumn{5}{c|}{2300.44}\\
		\hline
		\multicolumn{3}{|c||}{Significance ($\sigma$)}&67.90&42.23&\multicolumn{5}{c|}{ }\\
		\cline{1-5}
		\multicolumn{3}{|c||}{$\mathcal{L}_{5\,\sigma}$ (fb$^{-1}$)}&0.54&1.40&\multicolumn{5}{c|}{ }\\
		\hline
	\end{tabular}
	\caption{The number of events for $ \geq 1-\rm{jet} + \geq 2\mu + \ptmiss \le 30$ GeV finalstate (\autoref{eq:s43_prod}) for the benchmark points and dominant SM backgrounds at the LHC/FCC with centre-of-mass energy of  14\,TeV, 30\,TeV and 100 TeV at an integrated luminosity of 1000  fb$^{-1}$  for the first two and 100 fb$^{-1}$ for 100 TeV. The required luminosities to achieve a $5\,\sigma$ signal ($\mathcal{L}_{5\,\sigma}$) are also shown for all three cases.}\label{tab:s343}
\end{table*}

The event numbers for the benchmark points BP1 and BP2 along with the dominant SM backgrounds for this finalstate are given in \autoref{tab:s343}. The numbers are presented for three different centre-of-mass energies viz. 14\,TeV, 30\,TeV and 100\,TeV. The integrated luminosity are taken to be 1000\,fb$^{-1}$ for the first two and 100\,fb$^{-1}$ for  the last one. Since the leptoquark masses in the considered benchmark points (BP1 and BP2) are taken to be 1.5 TeV and 2.0 TeV, a hardness cut of 1.2 TeV  has also been implemented here to reduce the background. Both the hardest jet and lepton $p_T$ cut are demanded to be $\geq 200$ GeV. Moreover, as there is no neutrino involved in this finalstate, we put an upper limit on the missing energy, namely $\ptmiss \leq 30$ GeV. As expected, the dominant contribution to this mode comes from $s-g\to S_3^{4/3}\mu$. Nevertheless, a small contribution arises from $c-g\to S_3^{1/3}\mu$ as well, since it can also provide di-muon finalstate. The dominant background contribution at 14 and 30 TeV LHC/FCC comes from the $VV$ process, which has higher chance of getting us a pair of muons in the finalstate. At 100 TeV, $t\bar{t}V$ becomes dominant due to the higher jump in cross-section, while contributing towards the criteria of di-muons and no upper limit on light jets. The demand of di-muons, accompanied by the hardness cut and the small window of missing energy keeps the background numbers comparatively lower than our previous discussions on $S_1$ in \autoref{sec:S1}, which leads to encouraging signal strengths in all the three centre-of-mass energies. At the 14 TeV LHC, a $9.21 \sigma$ significance is obtained for BP1 with 1000 \fbi of luminosity, whereas for BP2 the strength is $2.94\sigma$. Moving to the higher centre-of-mass energy of 30 TeV, both the BPs give us promising outcomes, with $43.48\sigma$ and $20.61\sigma$ significances for BP1 and BP2, respectively. The strength is further enhanced at the 100 TeV LHC/FCC, where with 100 \fbi luminosity we can obtain $67.90\sigma$ significance for BP1, and $42.23\sigma$ significance for BP2. In both 30 TeV and 100 TeV energies, the required $5\sigma$ discovery is predicted with much earlier data.

\subsection{$S^{2/3}_3$ component of $S_3$: $1c-\rm{jet} +\ptmiss$ }
\label{sec:s23}

%%%%%%%%%%%%%%%%%%%%%%%%%%%%%%%  1c + pTmiss %%%%%%%%%%%%%%%%%%%%%%%%%%%%%%%%%%%%%%%%%%

\begin{table*}[h!]
	\renewcommand{\arraystretch}{1.1}
	\centering
	\begin{tabular}{|c|c|c||c|c||c|c|c|c|c|}
		\hline
		\multirow{2}{*}{ $\sqrt s$ in}&\multirow{3}{*}{Fusion}&&\multicolumn{7}{c|}{$1c-\rm{jet} +\ptmiss \geq 200$ GeV }\\
		\cline{4-10}
		\multirow{2}{*}{TeV}	&& Mode &\multicolumn{2}{c||}{Signal}&\multicolumn{5}{c|}{Backgrounds}\\
		\cline{4-10}
		&&&BP1&BP2 &$t\bar t$&$VV$&$VVV$&$t\bar{t}V$&$tVV$\\
		\hline
		\hline
		\multirow{4}{*}{14} &\multirow{2}{*}{$s-g$}&$S_3^{4/3} \mu$& 0.00& 0.00&\multirow{4}{*}{2.89}&\multirow{4}{*}{31.86}&\multirow{4}{*}{0.96}&\multirow{4}{*}{0.00}&\multirow{4}{*}{0.25}\\
		\cline{3-5}
		&&$S_3^{1/3} \nu$&0.12&0.05&&&&&\\
		\cline{2-5}
		&\multirow{2}{*}{$c-g$}&$S_3^{1/3} \mu$&0.00&0.00&&&&&\\
		\cline{3-5}
		&&$S_3^{2/3} \nu$&27.51&6.64&&&&&\\
		\hline
		\multicolumn{3}{|c||}{Total}&27.63&6.69&\multicolumn{5}{c|}{35.96}\\
		\hline
		\multicolumn{3}{|c||}{Significance($\sigma$)}&3.47&1.02&\multicolumn{5}{c|}{ }\\
		\cline{1-5}
		\multicolumn{3}{|c||}{$\mathcal{L}_{5\,\sigma}$ (fb$^{-1}$)}&2082.10&$\gg$5000&\multicolumn{5}{c|}{ }\\
		\hline\hline
		
		\multirow{4}{*}{30} & \multirow{2}{*}{$s-g$}&$S_3^{4/3} \mu$& 0.00&0.00&\multirow{4}{*}{0.00}&\multirow{4}{*}{173.93}&\multirow{4}{*}{6.01}&\multirow{4}{*}{1.27}&\multirow{4}{*}{0.00}\\
		\cline{3-5}
		&&$S_3^{1/3} \nu$&2.12&0.80&&&&&\\
		\cline{2-5}
		&\multirow{2}{*}{$c-g$}&$S_3^{1/3} \mu$&0.00&0.02&&&&&\\
		\cline{3-5}
		&&$S_3^{2/3} \nu$&366.42&131.13&&&&&\\
		\hline
		\multicolumn{3}{|c||}{Total}&368.54&131.95&\multicolumn{5}{c|}{181.21}\\
		\hline
		\multicolumn{3}{|c||}{Significance($\sigma$)}&15.72&7.46&\multicolumn{5}{c|}{ }\\
		\cline{1-5}
		\multicolumn{3}{|c||}{$\mathcal{L}_{5\,\sigma}$ (fb$^{-1}$)}&101.19&449.71&\multicolumn{5}{c|}{ }\\
		\hline\hline
		
		\multirow{4}{*}{100}& \multirow{2}{*}{$s-g$}&$S_3^{4/3} \mu$& 0.00&0.00&\multirow{4}{*}{19.15}&\multirow{4}{*}{170.96}&\multirow{4}{*}{1.75}&\multirow{4}{*}{0.00}&\multirow{4}{*}{0.00}\\
		\cline{3-5}
		&&$S_3^{1/3} \nu$&8.42&5.33&&&&&\\
		\cline{2-5}
		&\multirow{2}{*}{$c-g$}&$S_3^{1/3} \mu$&0.19&0.16&&&&&\\
		\cline{3-5}
		&&$S_3^{2/3} \nu$&634.05&295.82&&&&&\\
		\hline
		\multicolumn{3}{|c||}{Total}&642.66&301.31&\multicolumn{5}{c|}{191.86}\\
		\hline
		\multicolumn{3}{|c||}{Significance($\sigma$)}&22.25&13.57&\multicolumn{5}{c|}{ }\\
		\cline{1-5}
		\multicolumn{3}{|c||}{$\mathcal{L}_{5\,\sigma}$ (fb$^{-1}$)}&5.05&13.58&\multicolumn{5}{c|}{ }\\
		\hline
	\end{tabular}
	\caption{The number of events for $1c-\rm{jet} +\ptmiss \geq 200$ GeV  finalstate (\autoref{eq:s23_prod}) for the benchmark points and dominant SM backgrounds at the LHC/FCC with centre-of-mass energies of  14\,TeV, 30\,TeV and 100 TeV at an integrated luminosity of 1000  fb$^{-1}$  for the first two and 100 fb$^{-1}$ for 100 TeV. The required luminosities to achieve a $5\,\sigma$ signal ($\mathcal{L}_{5\,\sigma}$) are also shown for all three cases.}\label{tab:s323}
\end{table*}

As we have pointed out earlier, $S^{2/3}_3$ can be produced only via $c-g$ fusion in association with a neutrino (Feynman diagram in \autoref{fig:s3prod}(d)) and then decays to $c\, \nu$ with 100\% branching ratio. This leaves us with mono $c-$jet plus missing energy signature, which is very unique. The recoil of $\nu$ against  $S^{2/3}_3$ leads to larger missing energy as already shown in \autoref{lmpt_S3}(b). The complete finalstate demanded in this case is written as follows:
\begin{center}
	$n_{c-\rm{jet}} = 1$, $n_{j} = 1$, $n_\ell = n_{\tau-\rm{jet}} = 0$ \& \\
	$p_T^{c-\rm{jet}} \geq 200$ GeV, $\ptmiss \geq 200$ GeV, $ p_T^H \geq 1200$ GeV.
\end{center}

In \autoref{tab:s323}, the events for signal and the SM backgrounds are quoted again for the three different center mass energies with the same choices for integrated luminosity as of all other cases discussed in this work. We put veto on the charged lepton as well as on the $\tau$-jet, and demand only one $c$-jet with $p_T \geq 200$ GeV along with $\ptmiss\geq200$ GeV can be present. Besides, we do not allow any light jets, keeping the total number of jets equal to one, which results into a significant drop in all the background events. The results for 14 TeV are not very heartening for BP2, as the signal significance of just $1.02\sigma$ can be reached. Meanwhile, BP1 shows a healthy signal of $3.47\sigma$ significance at this energy, with an integrated luminosity of $\sim 2082$ \fbi being enough to probe the required $5\sigma$ significance. At the 30 TeV energy, both these BPs cross $5\sigma$ significance at 1000 \fbi luminosity, with  $15.72\,\sigma$ and $ 7.46\,\sigma$ strengths being reached by BP1 and BP2, respectively. At the highest energy of 100 TeV, these significances enhance to $22.25\sigma$ for BP1, and $13.57\sigma$ for BP2 with 100 \fbi of integrated luminosity, while the required $5\sigma$ strength can be obtained with very early data. For this finalstate, $VV$ remains the strongest background, with fully  invisible  decay of  $Z$ and/or hadronic decays of $Z,W^\pm$ with a $c$-jet.

%%%%%%%%%%%%%%%%%%%%%%%%%%%%%%%%%%%%%%%%%%%%%%%%%%%%%%%%%%%%%%%%%%%%%%%%%%%%%%%%%%%%%

\subsection{$S^{1/3}_3$ component of $S_3$}
\label{sec:s13}

The component $S^{1/3}_3$ of $S_3$ leptoquark can be produced in association with a muon or a neutrino in $c-g$ and $s-g$ fusions. The produced leptoquark then disintegrates into either $c\,\mu$ or $s\,\nu_\mu$ with equal probability as shown in \autoref{brss3}. Consequently, four different finalstates are possible in this scenario and we investigate them all sequentially.

\subsubsection{$1c-\rm{jet} + 2\mu + \ptmiss$}
In this case, we consider $S_3^{1/3}$ to be produced in accompany with a muon through $c-g$ fusion and eventually decays into a $c-$quark and a muon as shown in \autoref{eq:s13_prod_cg1}. The complete finalstate with the advanced cuts is described below:
\begin{center}
	$n_{c-\rm{jet}} \geq 1$, $n_{j} \geq 1$, $n_\mu \geq 2$, $n_{\tau-\rm{jet}} = 0$ \& \\
	$p_T^{\ell_1} \geq 200$ GeV, $p_T^{j_1} \geq 200$ GeV, $\ptmiss \leq 30$ GeV, $p_T^H \geq 1200$ GeV.
\end{center}

%%%%%%%%%%%%%%%%%%%%%%%%%%%%%% 1c + 2\mu + pTmiss %%%%%%%%%%%%%%%%%%%%%%%%%%%%%%%%%%%%%%

\begin{table*}[h!]
	\renewcommand{\arraystretch}{1.1}
	\centering
	\begin{tabular}{|c|c|c||c|c||c|c|c|c|c|}
		\hline
		\multirow{2}{*}{ $\sqrt s$ in}&\multirow{3}{*}{Fusion}&&\multicolumn{7}{c|}{$ \geq 1c-\rm{jet} + \geq 2\mu + \ptmiss \leq 30$ GeV }\\
		\cline{4-10}
		\multirow{2}{*}{TeV}	&& Mode &\multicolumn{2}{c||}{Signal}&\multicolumn{5}{c|}{Backgrounds}\\
		\cline{4-10}
		&&&BP1&BP2 &$t\bar t$&$VV$&$VVV$&$t\bar{t}V$&$tVV$\\
		\hline
		\hline
		\multirow{4}{*}{14} &\multirow{2}{*}{$s-g$}&$S_3^{4/3} \mu$& 1.17& 0.36&\multirow{4}{*}{11.57}&\multirow{4}{*}{3.98}&\multirow{4}{*}{0.64}&\multirow{4}{*}{10.38}&\multirow{4}{*}{1.75}\\
		\cline{3-5}
		&&$S_3^{1/3} \nu$&0.00&0.00&&&&&\\
		\cline{2-5}
		&\multirow{2}{*}{$c-g$}&$S_3^{1/3} \mu$&16.55&4.17&&&&&\\
		\cline{3-5}
		&&$S_3^{2/3} \nu$&0.00&0.00&&&&&\\
		\hline
		\multicolumn{3}{|c||}{Total}&17.72&5.53&\multicolumn{5}{c|}{28.32}\\
		\hline
		\multicolumn{3}{|c||}{Significance($\sigma$)}&2.61&0.79&\multicolumn{5}{c|}{ }\\
		\cline{1-5}
		\multicolumn{3}{|c||}{$\mathcal{L}_{5\,\sigma}$ (fb$^{-1}$)}&3667.04&$\gg$5000&\multicolumn{5}{c|}{ }\\
		\hline\hline
		
		\multirow{4}{*}{30} & \multirow{2}{*}{$s-g$}&$S_3^{4/3} \mu$& 29.06&11.97&\multirow{4}{*}{90.19}&\multirow{4}{*}{60.26}&\multirow{4}{*}{8.01}&\multirow{4}{*}{94.53}&\multirow{4}{*}{8.62}\\
		\cline{3-5}
		&&$S_3^{1/3} \nu$&0.00&0.00&&&&&\\
		\cline{2-5}
		&\multirow{2}{*}{$c-g$}&$S_3^{1/3} \mu$&284.26&106.83&&&&&\\
		\cline{3-5}
		&&$S_3^{2/3} \nu$&0.00&0.00&&&&&\\
		\hline
		\multicolumn{3}{|c||}{Total}&313.32&118.80&\multicolumn{5}{c|}{261.61}\\
		\hline
		\multicolumn{3}{|c||}{Significance($\sigma$)}&13.07&6.09&\multicolumn{5}{c|}{ }\\
		\cline{1-5}
		\multicolumn{3}{|c||}{$\mathcal{L}_{5\,\sigma}$ (fb$^{-1}$)}&146.41&673.85&\multicolumn{5}{c|}{ }\\
		\hline\hline
		
		\multirow{4}{*}{100}& \multirow{2}{*}{$s-g$}&$S_3^{4/3} \mu$& 91.65&48.14&\multirow{4}{*}{233.65}&\multirow{4}{*}{53.79}&\multirow{4}{*}{14.11}&\multirow{4}{*}{141.78}&\multirow{4}{*}{32.70}\\
		\cline{3-5}
		&&$S_3^{1/3} \nu$&0.00&0.00&&&&&\\
		\cline{2-5}
		&\multirow{2}{*}{$c-g$}&$S_3^{1/3} \mu$&600.77&291.15&&&&&\\
		\cline{3-5}
		&&$S_3^{2/3} \nu$&0.00&0.00&&&&&\\
		\hline
		\multicolumn{3}{|c||}{Total}&692.42&339.29&\multicolumn{5}{c|}{476.03}\\
		\hline
		\multicolumn{3}{|c||}{Significance($\sigma$)}&20.26&11.88&\multicolumn{5}{c|}{ }\\
		\cline{1-5}
		\multicolumn{3}{|c||}{$\mathcal{L}_{5\,\sigma}$ (fb$^{-1}$)}&6.09&17.71&\multicolumn{5}{c|}{ }\\
		\hline
	\end{tabular}
	\caption{The number of events for $2\mu + 1c-\rm{jet} +\ptmiss \le 30$ GeV finalstate (\autoref{eq:s13_prod_cg1}) for the benchmark points and dominant SM backgrounds at the LHC/FCC with centre-of-mass energy of  14\,TeV, 30\,TeV and 100 TeV at an integrated luminosity of 1000  fb$^{-1}$  for the first two and 100 fb$^{-1}$ for 100 TeV. The required luminosities to achieve a $5\,\sigma$ signal ($\mathcal{L}_{5\,\sigma}$) are also shown for all three cases.}\label{s3132mu1c}
\end{table*}

 The signal-background analysis for this finalstate topology at LHC/FCC is illustrated in \autoref{s3132mu1c}. Due to the absence of neutrinos in the entire decay chain ideally there should not be any missing energy and  we impose the missing transverse momentum upper limit $\ptmiss\leq30$ GeV. We demand at least one $c$-jet, and one muon of the two having $p_T\geq 200$ GeV, along with $\tau$-jet veto for the finalstate. Apart from the mentioned process, this finalstate gets very small contribution arising from $s-g\to S_3^{4/3}\mu$ channel (discussed in \autoref{sec:s43}) as well, due to the mistagging of  light-jet as $c$-jet. The applied cut on the total hardness, as well as the specific demand for di-muons keep the backgrounds relatively low, with the dominant contributions coming from $t\bar{t}$ and $t\bar{t}V$. Now, about the outcomes, the 14 TeV scenario is not very inspiring since significances of $2.61\,\sigma$ and $0.79 \,\sigma$ can only be reached with 1000 \fbi of integrated luminosity for the two respective benchmark points which implies the necessity of very high luminosity to attain $5\,\sigma$ reach. However, the situation improves with 30 TeV of centre-of-mass energy where the signal significances of $13.07\,\sigma$ and $6.09\,\sigma$ can be obtained with 1000 \fbi luminosity for BP1 and BP2 respectively which indicates requirement of only $\sim150$ \fbi and $\sim 675$ \fbi integrated luminosities for $5\,\sigma$ reach. On the other hand, the results are very uplifting for 100 TeV centre-of-mass energy as $20.26\,\sigma$ and $11.88\,\sigma$ of signal significances could be gained for BP1 and BP2, respectively, at 100 \fbi of integrated luminosity only. Therefore significance of $5\,\sigma$ is reachable with very early data.

\subsubsection{$1c-\rm{jet} + 1\mu + \ptmiss$}

%%%%%%%%%%%%%%%%  1c + 1\mu + pTmiss %%%%%%%%%%%%%%%%%%%%%%%%%%
\begin{table*}[h!]
	\renewcommand{\arraystretch}{1.1}
	\centering
	\begin{tabular}{|c|c|c||c|c||c|c|c|c|c|}
		\hline
		\multirow{2}{*}{ $\sqrt s$ in}&\multirow{3}{*}{Fusion}&&\multicolumn{7}{c|}{$1c-\rm{jet}  +1\mu +\ptmiss \geq 500$ GeV }\\
		\cline{4-10}
		\multirow{2}{*}{TeV}	&& Mode &\multicolumn{2}{c||}{Signal}&\multicolumn{5}{c|}{Backgrounds}\\
		\cline{4-10}
		&&&BP1&BP2 &$t\bar t$&$VV$&$VVV$&$t\bar{t}V$&$tVV$\\
		\hline
		\hline
		\multirow{4}{*}{14} &\multirow{2}{*}{$s-g$}&$S_3^{4/3} \mu$& 0.09& 0.00&\multirow{4}{*}{212.72}&\multirow{4}{*}{12.95}&\multirow{4}{*}{0.64}&\multirow{4}{*}{8.48}&\multirow{4}{*}{2.27}\\
		\cline{3-5}
		&&$S_3^{1/3} \nu$&21.39&6.68&&&&&\\
		\cline{2-5}
		&\multirow{2}{*}{$c-g$}&$S_3^{1/3} \mu$&2.10&0.72&&&&&\\
		\cline{3-5}
		&&$S_3^{2/3} \nu$&0.00&0.00&&&&&\\
		\hline
		\multicolumn{3}{|c||}{Total}&23.58&7.40&\multicolumn{5}{c|}{237.07}\\
		\hline
		\multicolumn{3}{|c||}{Significance($\sigma$)}&1.46&0.47&\multicolumn{5}{c|}{ }\\
		\cline{1-5}
		\multicolumn{3}{|c||}{$\mathcal{L}_{5\,\sigma}$ (fb$^{-1}$)}&$\gg$5000&$\gg$5000&\multicolumn{5}{c|}{ }\\
		\hline\hline
		
		\multirow{4}{*}{30} & \multirow{2}{*}{$s-g$}&$S_3^{4/3} \mu$& 3.71&1.71&\multirow{4}{*}{4311.08}&\multirow{4}{*}{179.28}&\multirow{4}{*}{24.07}&\multirow{4}{*}{114.96}&\multirow{4}{*}{21.17}\\
		\cline{3-5}
		&&$S_3^{1/3} \nu$&440.00&192.28&&&&&\\
		\cline{2-5}
		&\multirow{2}{*}{$c-g$}&$S_3^{1/3} \mu$&54.55&28.94&&&&&\\
		\cline{3-5}
		&&$S_3^{2/3} \nu$&0.00&0.00&&&&&\\
		\hline
		\multicolumn{3}{|c||}{Total}&498.26&222.93&\multicolumn{5}{c|}{4650.56}\\
		\hline
		\multicolumn{3}{|c||}{Significance($\sigma$)}&6.94&3.19&\multicolumn{5}{c|}{ }\\
		\cline{1-5}
		\multicolumn{3}{|c||}{$\mathcal{L}_{5\,\sigma}$ (fb$^{-1}$)}&518.49&2451.56&\multicolumn{5}{c|}{ }\\
		\hline\hline
		
		\multirow{4}{*}{100}& \multirow{2}{*}{$s-g$}&$S_3^{4/3} \mu$& 21.71&19.88&\multirow{4}{*}{9127.84}&\multirow{4}{*}{199.77}&\multirow{4}{*}{47.62}&\multirow{4}{*}{291.62}&\multirow{4}{*}{66.96}\\
		\cline{3-5}
		&&$S_3^{1/3} \nu$&1155.23&662.25&&&&&\\
		\cline{2-5}
		&\multirow{2}{*}{$c-g$}&$S_3^{1/3} \mu$&256.21&157.54&&&&&\\
		\cline{3-5}
		&&$S_3^{2/3} \nu$&0.00&0.00&&&&&\\
		\hline
		\multicolumn{3}{|c||}{Total}&1433.15&839.67&\multicolumn{5}{c|}{9733.81}\\
		\hline
		\multicolumn{3}{|c||}{Significance($\sigma$)}&13.56&8.17&\multicolumn{5}{c|}{ }\\
		\cline{1-5}
		\multicolumn{3}{|c||}{$\mathcal{L}_{5\,\sigma}$ (fb$^{-1}$)}&13.59&37.49&\multicolumn{5}{c|}{ }\\
		\hline
	\end{tabular}
	\caption{The number of events for $1\mu + 1c-\rm{jet} +\ptmiss \ge 500$ GeV finalstate (\autoref{eq:s13_prod_sg2}) for the benchmark points and dominant SM backgrounds at the LHC/FCC with centre-of-mass energy of  14\, TeV, 30\,TeV and 100 TeV at an integrated luminosity of 1000  fb$^{-1}$  for the first two and 100 fb$^{-1}$ for 100 TeV. The required luminosities to achieve a $5\,\sigma$ signal ($\mathcal{L}_{5\,\sigma}$) are also shown for all three cases.}\label{s3131mu1c}
\end{table*} 
%%%%%%%%%%%%%%%%%%%%%%%%%%%%%%%%%%%%%%%%%%%%%%%%%%%%%%%%%%
While considering the production of $S_3^{1/3}$ along with a neutrino via $s-g$ fusion and its disintegration into $c-$quark and a muon, the finalstate $1c-\rm{jet} + 1\mu + \ptmiss$ arises (see \autoref{eq:s13_prod_sg2}). The demands are almost the same as the previous one except we have only one muon in this finalstate and due to the presence of a high energetic neutrino here, we put a lower bound on the missing transverse momentum as $\ptmiss\geq500$ GeV.  The complete finalstate is given as:
\begin{center}
	$n_{c-\rm{jet}} = 1$, $n_{j} \geq 1$, $n_\mu = 1$, $n_{\tau-\rm{jet}} = 0$ \& \\
	$p_T^{\ell_1} \geq 200$ GeV, $p_T^{c-{\rm{jet}}} \geq 200$ GeV, $\ptmiss \geq 500$ GeV, $p_T^H \geq 1200$ GeV.
\end{center}

The signal and background analysis at the LHC/FCC for this decay topology  is presented in \autoref{s3131mu1c}. The demand for only one muon keeps the background numbers higher than the previous case with two muons. With 14 TeV centre-of-mass energy and 1000 \fbi of luminosity, both the BPs give very weak signals, with strengths of $1.46\sigma$ and $0.47\sigma$ respectively for BP1 and BP2. The situation improves for BP1 at the centre-of-mass energy of 30 TeV, where $6.94\sigma$ signal significance can be achieved with a luminosity of 1000 \fbi. BP2 here shows a $3.19\sigma$ significance, with $\sim2450$ \fbi luminosity required to reach the desired $5\sigma$. Promising outcomes are obtained at the 100 TeV LHC/FCC, where $13.56\sigma$ and $8.17\sigma$ significances are predicted for BP1 and BP2, respectively with 100 \fbi luminosity. Owing to the high cross-section and no upper limit on jets, $t\bar{t}$ still contributes dominantly as background.

\subsubsection{$2-\rm{jet}+\ptmiss$}
%%%%%%%  2j + pTmiss %%%%%%%%%%%%%%%%
\begin{table*}[h!]
	\renewcommand{\arraystretch}{1.1}
	\centering
	\begin{tabular}{|c|c|c||c|c||c|c|c|c|c|}
		\hline
		\multirow{2}{*}{ $\sqrt s$ in}&\multirow{3}{*}{Fusion}&&\multicolumn{7}{c|}{$2-\rm{jet} +\ptmiss \geq 500$ GeV }\\
		\cline{4-10}
		\multirow{2}{*}{TeV}	&& Mode &\multicolumn{2}{c||}{Signal}&\multicolumn{5}{c|}{Backgrounds}\\
		\cline{4-10}
		&&&BP1&BP2 &$t\bar t$&$VV$&$VVV$&$t\bar{t}V$&$tVV$\\
		\hline
		\hline
		\multirow{4}{*}{14} &\multirow{2}{*}{$s-g$}&$S_3^{4/3} \mu$& 0.11& 0.03&\multirow{4}{*}{289.41}&\multirow{4}{*}{1335.87}&\multirow{4}{*}{57.15}&\multirow{4}{*}{4.39}&\multirow{4}{*}{3.39}\\
		\cline{3-5}
		&&$S_3^{1/3} \nu$&47.01&12.35&&&&&\\
		\cline{2-5}
		&\multirow{2}{*}{$c-g$}&$S_3^{1/3} \mu$&2.25&0.54&&&&&\\
		\cline{3-5}
		&&$S_3^{2/3} \nu$&118.31&29.34&&&&&\\
		\hline
		\multicolumn{3}{|c||}{Total}&167.68&42.26&\multicolumn{5}{c|}{1690.22}\\
		\hline
		\multicolumn{3}{|c||}{Significance($\sigma$)}&3.89&1.02&\multicolumn{5}{c|}{ }\\
		\cline{1-5}
		\multicolumn{3}{|c||}{$\mathcal{L}_{5\,\sigma}$ (fb$^{-1}$)}&1652.15&$\gg$5000&\multicolumn{5}{c|}{ }\\
		\hline\hline
		
		\multirow{4}{*}{30} & \multirow{2}{*}{$s-g$}&$S_3^{4/3} \mu$& 2.16&1.29&\multirow{4}{*}{2441.14}&\multirow{4}{*}{8349.32}&\multirow{4}{*}{435.17}&\multirow{4}{*}{34.49}&\multirow{4}{*}{22.73}\\
		\cline{3-5}
		&&$S_3^{1/3} \nu$&614.52&227.99&&&&&\\
		\cline{2-5}
		&\multirow{2}{*}{$c-g$}&$S_3^{1/3} \mu$&42.86&14.00&&&&&\\
		\cline{3-5}
		&&$S_3^{2/3} \nu$&1648.86&604.69&&&&&\\
		\hline
		\multicolumn{3}{|c||}{Total}&2308.40&847.97&\multicolumn{5}{c|}{11282.85}\\
		\hline
		\multicolumn{3}{|c||}{Significance($\sigma$)}&19.80&7.70&\multicolumn{5}{c|}{ }\\
		\cline{1-5}
		\multicolumn{3}{|c||}{$\mathcal{L}_{5\,\sigma}$ (fb$^{-1}$)}&63.76&421.75&\multicolumn{5}{c|}{ }\\
		\hline\hline
		\multirow{4}{*}{100}& \multirow{2}{*}{$s-g$}&$S_3^{4/3} \mu$& 12.93&8.75&\multirow{4}{*}{6304.84}&\multirow{4}{*}{7095.89}&\multirow{4}{*}{527.31}&\multirow{4}{*}{69.28}&\multirow{4}{*}{46.70}\\
		\cline{3-5}
		&&$S_3^{1/3} \nu$&1094.90&525.45&&&&&\\
		\cline{2-5}
		&\multirow{2}{*}{$c-g$}&$S_3^{1/3} \mu$&133.38&58.56&&&&&\\
		\cline{3-5}
		&&$S_3^{2/3} \nu$&3157.52&1465.86&&&&&\\
		\hline
		\multicolumn{3}{|c||}{Total}&4398.72&2058.62&\multicolumn{5}{c|}{14044.02}\\
		\hline
		\multicolumn{3}{|c||}{Significance($\sigma$)}&32.39&16.22&\multicolumn{5}{c|}{ }\\
		\cline{1-5}
		\multicolumn{3}{|c||}{$\mathcal{L}_{5\,\sigma}$ (fb$^{-1}$)}&2.38&9.50&\multicolumn{5}{c|}{ }\\
		\hline
	\end{tabular}
	\caption{The number of events for $2-\rm{jet} +\ptmiss \geq 500$ GeV finalstate (\autoref{eq:s13_prod_sg1}) for the benchmark points and dominant SM backgrounds at the LHC/FCC with centre-of-mass energy of  14\,TeV, 30\,TeV and 100\,TeV at an integrated luminosity of 1000  fb$^{-1}$  for the first two and 100 fb$^{-1}$ for 100 TeV. The required luminosities to achieve a $5\,\sigma$ signal ($\mathcal{L}_{5\,\sigma}$) are also shown for all three cases.}\label{s3131j}
\end{table*}
%%%%%%%%%%%%%%%%%%%%%%%%%%%%%%%%%%%%%%%%%%%%%%%%%%%%%%%%
The finalstate of $1-\rm{jet}+\ptmiss$ ensues from the production of $S_3^{1/3}$ in association with a neutrino via $s-$gluon fusion followed by its disintegration into a $s-$quark and a neutrino (see \autoref{eq:s13_prod_sg1}).  However, we  cannot avoid a ISR/FSR  jet and to avoid the reduction on  the  signal  cross-section, we allow one  such ISR/FSR  jet  in the finalstate. The complete finalstate with advanced cuts is as follows:
\begin{center}
	$1\leq n_{j} \leq 2$, $n_{b-\rm{jet}} = n_{\tau-\rm{jet}} = n_\ell = 0$ \& \\
	$p_T^{j_1} \geq 400 $ GeV, $\ptmiss \geq 500$ GeV, $p_T^H \geq 1200$ GeV.
\end{center}

 The signal and backgrounds for this finalstate are simulated in \autoref{s3131j}. Due to the fact that this finalstate incorporates two neutrinos,  we have imposed a very high missing energy cut as $\ptmiss\geq500$ GeV. We also impose veto on charged leptons ($e^\pm,\,\mu^\pm$),  $b$-jets, and $\tau$-jet. Apart from the single jet from the leptoquark ($s$ quark), we keep room for one ISR/FSR jet, so that the total number of jets in the finalstate can be maximum of two. 
 
 Demand of lesser jets, veto on $b-$jets, and high $\ptmiss$ cut means $VV$ is the dominant background here, over the subdominant $t\bar{t}$. On the contrary, the signal gets a huge contribution from the mode $c-g\to S_3^{2/3}\nu$ as the $c-$jet mimics the light jet. A tiny contribution from $c-g\to S_3^{1/3}\mu$ arises here as well. The simulation is performed with the centre-of-mass energies of  14\,TeV, 30\,TeV and 100\,TeV at an integrated luminosity of 1000  fb$^{-1}$  for the first two and 100 fb$^{-1}$ for 100 TeV. At the 14 TeV LHC, BP1 gives us a fairly strong $3.89\sigma$ significance, which means $5\sigma$ can be reached with $\sim1650$ \fbi of luminosity. BP2 signal remains very weak with $\sim1\sigma$ significance. The situation becomes hopeful when we move to the 30 TeV LHC, where $19.80\sigma$ and $7.70\sigma$ significances are predicted at 1000 \fbi luminosity, for BP1 and BP2, respectively. These strengths are further enhanced at 100 TeV, with significances of $32.39\sigma$ and $16.22\sigma$ for BP1 and BP2, respectively with 100 \fbi luminosity. The required $5\sigma$ significance here is predicted to be obtained with much earlier data.

\subsubsection{$ 1-\rm{jet} +1\mu +\ptmiss$}

%%%%%%%%%%%%%%%%%%  1j+ 1\mu+ pTmiss %%%%%%%%%%%%%%%%%%%%%%%%%%%%
\begin{table*}[h!]
	\renewcommand{\arraystretch}{1.1}
	\centering
	\begin{tabular}{|c|c|c||c|c||c|c|c|c|c|}
		\hline
		\multirow{2}{*}{ $\sqrt s$ in}&\multirow{3}{*}{Fusion}&&\multicolumn{7}{c|}{$ \geq 1-\rm{jet} + \geq 1\mu +\ptmiss \geq 500$ GeV}\\
		\cline{4-10}
		\multirow{2}{*}{TeV}	&& Mode &\multicolumn{2}{c||}{Signal}&\multicolumn{5}{c|}{Backgrounds}\\
		\cline{4-10}
		&&&BP1&BP2&$t\bar t$&$VV$&$VVV$&$t\bar{t}V$&$tVV$\\
		\hline
		\hline
		\multirow{4}{*}{14} &\multirow{2}{*}{$s-g$}&$S_3^{4/3} \mu$& 6.57& 2.69&\multirow{4}{*}{295.20}&\multirow{4}{*}{226.96}&\multirow{4}{*}{32.51}&\multirow{4}{*}{13.52}&\multirow{4}{*}{6.05}\\
		\cline{3-5}
		&&$S_3^{1/3} \nu$&36.81&10.88&&&&&\\
		\cline{2-5}
		&\multirow{2}{*}{$c-g$}&$S_3^{1/3} \mu$&55.58&14.85&&&&&\\
		\cline{3-5}
		&&$S_3^{2/3} \nu$&0.00&0.00&&&&&\\
		\hline
		\multicolumn{3}{|c||}{Total}&98.96&28.42&\multicolumn{5}{c|}{574.24}\\
		\hline
		\multicolumn{3}{|c||}{Significance($\sigma$)}&3.81&1.16&\multicolumn{5}{c|}{ }\\
		\cline{1-5}
		\multicolumn{3}{|c||}{$\mathcal{L}_{5\,\sigma}$ (fb$^{-1}$)}&1718.72&$\gg$5000&\multicolumn{5}{c|}{ }\\
		\hline\hline
		
		\multirow{4}{*}{30} & \multirow{2}{*}{$s-g$}&$S_3^{4/3} \mu$& 163.71&87.84&\multirow{4}{*}{5549.70}&\multirow{4}{*}{2426.37}&\multirow{4}{*}{419.14}&\multirow{4}{*}{159.67}&\multirow{4}{*}{52.52}\\
		\cline{3-5}
		&&$S_3^{1/3} \nu$&760.50&313.38&&&&&\\
		\cline{2-5}
		&\multirow{2}{*}{$c-g$}&$S_3^{1/3} \mu$&974.61&463.29&&&&&\\
		\cline{3-5}
		&&$S_3^{2/3} \nu$&0.00&0.00&&&&&\\
		\hline
		\multicolumn{3}{|c||}{Total}&1898.82&804.96&\multicolumn{5}{c|}{8607.40}\\
		\hline
		\multicolumn{3}{|c||}{Significance($\sigma$)}&18.52&8.30&\multicolumn{5}{c|}{ }\\
		\cline{1-5}
		\multicolumn{3}{|c||}{$\mathcal{L}_{5\,\sigma}$ (fb$^{-1}$)}&72.85&363.15&\multicolumn{5}{c|}{ }\\
		\hline\hline
		
		\multirow{4}{*}{100}& \multirow{2}{*}{$s-g$}&$S_3^{4/3} \mu$& 970.04&604.41&\multirow{4}{*}{10675.32}&\multirow{4}{*}{2153.36}&\multirow{4}{*}{536.16}&\multirow{4}{*}{349.62}&\multirow{4}{*}{151.04}\\
		\cline{3-5}
		&&$S_3^{1/3} \nu$&1999.71&1076.01&&&&&\\
		\cline{2-5}
		&\multirow{2}{*}{$c-g$}&$S_3^{1/3} \mu$&2396.01&1328.94&&&&&\\
		\cline{3-5}
		&&$S_3^{2/3} \nu$&0.00&0.00&&&&&\\
		\hline
		\multicolumn{3}{|c||}{Total}&5365.76&3009.36&\multicolumn{5}{c|}{13865.50}\\
		\hline
		\multicolumn{3}{|c||}{Significance($\sigma$)}&38.69&23.17&\multicolumn{5}{c|}{ }\\
		\cline{1-5}
		\multicolumn{3}{|c||}{$\mathcal{L}_{5\,\sigma}$ (fb$^{-1}$)}&1.67&4.66&\multicolumn{5}{c|}{ }\\
		\hline
	\end{tabular}
	\caption{The number of events for $ \geq 1-\rm{jet} + \geq 1\mu +\ptmiss \ge 500$ GeV finalstate (\autoref{eq:s13_prod_cg2}) for the benchmark points and dominant SM backgrounds at the LHC/FCC with centre-of-mass energy of  14\,TeV, 30\,TeV and 100 TeV at an integrated luminosity of 1000  fb$^{-1}$  for the first two and 100 fb$^{-1}$ for 100 TeV. The required luminosities to achieve a $5\,\sigma$ signal ($\mathcal{L}_{5\,\sigma}$) are also shown for all three cases.}\label{s3131mu1j}
\end{table*}
%%%%%%%%%%%%%%%%%%%%%%%%%%%%%%%%%%%%%%%%%%%%%%%%%%%%%%%%%

If the leptoquark $S_3^{1/3}$ is produced in $c-$gluon fusion associated with a muon and eventually decays to a $s-$quark and a neutrino, the  finalstate $ 1-\rm{jet} +1\mu +\ptmiss$ appears as quoted in \autoref{eq:s13_prod_cg2}. The complete requirements and cuts for this finalstate are given below:
\begin{center}
	$n_{\text{j}} \geq 1$, $n_\mu \geq 1$, $n_{b-\rm{jet}} = n_{\tau-\rm{jet}} = 0$ \& \\
	$p_T^{\ell_1} \geq 200$ GeV, $p_T^{j_1} \geq 200$ GeV, $\ptmiss \geq 500$ GeV, $p_T^H \geq 1200$ GeV.
\end{center}

 The event numbers along with different SM backgrounds have been shown in \autoref{s3131mu1j}. As this finalstate involves one neutrino, a cut on missing transverse momentum is applied as $\slashed p_T\geq500$ GeV along with $p_T\geq 200$ GeV for both the muon and the light-jet. Additionally, no $b-$jet and $\tau-$jet are demanded to reduce the SM backgrounds, and the hardness cut of $p_T^H \geq 1.2$ TeV comes into play here as well. No upper limit on the number of jets means $t\bar{t}$ still contributes dominantly to the background. At the 14 TeV LHC, the BP1 signal is fairly healthy with a $3.81\sigma$ significance obtainable at 1000 \fbi of integrated luminosity, with the $5\sigma$ being achievable with luminosity of $\sim1720$ \fbi. BP2 however, gives a weak signal of $1.16\sigma$ significance. Moving to the centre-of-mass energy of 30 TeV, we obtain encouraging signals with significances of $18.52\sigma$ for BP1, and $8.30\sigma$ for BP2 with 1000 \fbi luminosity. These are enhanced further at the 100 TeV predictions, where with 100 \fbi luminosity, BP1 and BP2 signals carry significances of $38.69\sigma$ and $23.17\sigma$ respectively, indicating the feasibility of a $5\sigma$ probe with $<5$ \fbi integrated luminosity.

%%%%%%%%%%%%%%%%%%%%%%%%%%%%%%%%%%%%%%%%%%%%%%%%%%%%%%%%%%%%%%%%%%%%%%%%

\section{Lepton flavour violating decay signatures}
\label{sec:lfv}

\begin{table}[h!]
	\renewcommand{\arraystretch}{2}
	\centering
	\begin{tabular}{|c|c|}
		\hline 
		 Decay & Branching  ratios\\ 
		\cline{2-2}
		 Modes & \makecell{BP3 \\ $M_{S_3}$ = 1.5 TeV}\\
		\hline
		$S_3^{-4/3} \to b \mu$  & 100\\
		\hline
		 $S_3^{-1/3} \to t \mu$ & 50 \\
		\hline
		 $S_3^{-1/3} \to b \nu_\mu$ & 50\\
		\hline
		 $S_3^{2/3} \to t \nu_\mu$ & 100 \\
		\hline
	\end{tabular}
	\caption{The decay branching ratios (in percentage) of $S_3$ for BP3.}  \label{brss3_bp3}
\end{table}

In this section we discuss the signatures involving second and third generation fermion decays corresponding to the benchmark choice BP3 as quoted in \autoref{tab:all}. Due to different choice of coupling values, it can be seen from \autoref{brss3_bp3} that, we have different decay channels for the three components of $S_3$ as compared to the two previously investigated cases BP1 and BP2, discussed in \autoref{brss3}. In this case, $S_3$ is produced in association with a muon or a neutrino through via $b-g$ and $t-g$ fusions. Now, the components $S^{4/3}_3$ and $S_3^{2/3}$ decay to $b\mu$ and $t\nu_\mu$ states, respectively, with 100\% probability. Whereas, $S_3^{1/3}$ disintegrates into $t\mu$ and $b\nu_\mu$ with equal probabilities i.e. 50\% each. The further decay of $t-$quark to a $b-$quark and a $W$-boson, and finally the $W$-boson decay modes will give rise to two jets or lepton plus missing energy signatures. The complete decay chains of these processes are as following.
\begin{align}
\label{S43_BP3}
	\text{BP3:}\qquad b-g \;&\to\; S_3^{4/3} \mu \;\to\; (b \,\mu) + \mu \to 1b\rm{ -jet}+2 \mu\,,\\[3mm]
	\label{bgS13_BP3_1}
	b-g \;&\to\; S_3^{1/3} \nu_\mu \;\to\; (t \, \mu) + \nu_\mu \to 1b\rm{-jet} + 1\ell +1\mu +\ptmiss\,,\\
	\label{bgS13_BP3_2}
	&\qquad\qquad\;\;\;\to (t \, \mu) + \nu_\mu \to \;1b-\rm{jet}+2-jet+1\mu+\ptmiss\,,\\
	\label{bgS13_BP3_3}
	&\qquad\qquad\;\;\;\to  (b \, \nu_\mu) + \nu_\mu \to \; 1b\rm{ -jet}+\ptmiss\,,\\[3mm]
	\label{tgS13_BP3_1}
	t-g \;&\to S_3^{1/3} \mu \;\to\; (b \, \nu_\mu) + \mu \to 1b-\rm{jet}+1 \mu+\ptmiss\,,\\
	\label{tgS13_BP3_2}
	&\qquad\qquad\;\to (t \, \mu) + \mu \to \;1b-\rm{jet}+2-jet+2\mu\,,\\
		\label{tgS13_BP3_3}
	&\qquad\qquad\;\to (t \, \mu) + \mu \to \;1b-\rm{jet}+1\ell+2\mu+\ptmiss\,,\\
	\label{tgS23_BP3_1}
	t-g \;&\to S_3^{2/3} \nu_\mu \to\; (t \, \nu_\mu) + \nu_\mu \to 1b-\rm{jet}+1\ell+\ptmiss\,.
%	\label{tgS23_BP3_2}
%	&\qquad\qquad\,\;\to (t \, \mu) + \nu_\mu \to \;1b-\rm{jet}+2-jet+\ptmiss\,.
\end{align}

We can see that, the production channel of $S_3^{4/3}$ provides unique signature as one $b$-jet plus di-muon. Whereas, for $S_3^{2/3}$ we get two finalstates depending on the decay of the top-quark, which arises from $S_3^{2/3}$. %$b-\rm{jet}$ plus single lepton plus missing energy or $b-\rm{jet}$ plus di-jet with missing energy. 
However, six different finalstates are possible for the two production processes of $S_3^{1/3}$. It is interesting to notice that unlike BP1 and BP2 scenarios of $S_3$ leptoquark, some finalstates for BP3 exhibit lepton flavour violating signatures (different lepton flavours in the finalstate) though the Lagrangian (in \autoref{eq:LS3}) does not contain any explicit lepton flavour violating interaction.

Next, we analyze these finalstates at the LHC/FCC adopting the similar procedures described in previous sections at 14 TeV and 30 TeV centre-of-mass energies with an integrated luminosity of 1000 \fbi, also at 100 TeV collision with 100 \fbi of integrated luminosity. The signal numbers for all the above mentioned finalstates are very low for 14\,TeV results and do not list them here. The events at 30\,TeV and 100\,TeV centre-of-mass energies are noticeable, however, in most cases, they fail to attain a $5\,\sigma$ signal strength within the proposed lifetime of LHC/FCC. 

\subsection{$\rm 1b -jet+ 2- jet + 2\mu$}

The only encouraging scenario is the finalstate of $1b -$jet $+ 2-$ jet $+ 2\mu$, which according to the topologies quoted above arises from $S_3^{1/3}$ component (\autoref{tgS13_BP3_2}). However, due to presence of initial state radiations and large production cross-section of $S_3^{4/3}$ (see \autoref{crosss3b}), this component contributes dominantly via \autoref{S43_BP3}. As no neutrino is present in this finalstate, we have applied a cut in the missing transverse momentum $\ptmiss<30\,$GeV. Moreover, a total hardness cut $p_T^H \geq 1200$ GeV is also applied, like the previous analysis for $S_1$ and $S_3$. The 
complete finalstate is written below:
\begin{center}
	$n_{b-\rm{jet}} = 1$, $n_{j} \leq 3$, $n_\mu \geq 2$, $n_\ell \geq 2$, $n_{\tau-\rm{jet}} = 0$ \& \\
	$\ptmiss \leq 30$ GeV, $p_T^H \geq 1200$ GeV \& $\abs{M_{jj} - M_{W}} \geq 10$ GeV, $\abs{M_{\ell\ell} - M_{Z}} \geq 5$ GeV.
\end{center}
%%%%%%%%%%%%    1b + 2j + 2mu + pTmiss   %%%%%%%%%%%%%%%%%%%
\begin{table*}[h!]
	\renewcommand{\arraystretch}{1.1}
	\centering
	\begin{tabular}{|c|c|c||c||c|c|c|c|c|}
		\hline
		\multirow{2}{*}{ $\sqrt s$ in}&\multirow{3}{*}{Fusion}&&\multicolumn{6}{c|}{$1b-\rm{jet}+2-jet+ \geq 2\mu+\ptmiss \leq 30$ GeV   }\\
		\cline{4-9}
		\multirow{2}{*}{TeV}	&& Mode &\multicolumn{1}{c||}{Signal}&\multicolumn{5}{c|}{Backgrounds}\\
		\cline{4-9}
		&&& BP3 &$t\bar t$&$VV$&$VVV$&$t\bar{t}V$&$tVV$\\
		\hline
		\hline
		
		\multirow{4}{*}{30} &\multirow{2}{*}{$b-g$}&$S_3^{4/3}\mu$&55.41 &\multirow{4}{*}{12.02}&\multirow{4}{*}{0.00}&\multirow{4}{*}{0.00}&\multirow{4}{*}{0.00}&\multirow{4}{*}{0.78}\\
		\cline{3-4}
		&&$S_3^{1/3}\nu$&0.00&&&&&\\
		\cline{2-4}
		&\multirow{2}{*}{$t-g$}&$S_3^{1/3}\mu$&0.23&&&&&\\
		\cline{3-4}
		&&$S_3^{2/3}\nu$&0.00&&&&&\\
		\hline
		\multicolumn{3}{|c||}{Total}&55.64&\multicolumn{5}{c|}{12.80}\\
		\hline
		\multicolumn{3}{|c||}{Significance($\sigma$)}&6.73&\multicolumn{5}{c|}{ }\\
		\cline{1-4}
		\multicolumn{3}{|c||}{$\mathcal{L}_{5\,\sigma}$ (fb$^{-1}$)}&552.68&\multicolumn{5}{c|}{ }\\
		\hline\hline
		
		\multirow{4}{*}{100} &\multirow{2}{*}{$b-g$}&$S_3^{4/3}\mu$&100.07 &\multirow{4}{*}{3.83}&\multirow{4}{*}{0.00}&\multirow{4}{*}{0.00}&\multirow{4}{*}{0.00}&\multirow{4}{*}{0.00}\\
		\cline{3-4}
		&&$S_3^{1/3}\nu$&0.00&&&&&\\
		\cline{2-4}
		&\multirow{2}{*}{$t-g$}&$S_3^{1/3}\mu$&0.32&&&&&\\
		\cline{3-4}
		&&$S_3^{2/3}\nu$&0.00&&&&&\\
		\hline
		\multicolumn{3}{|c||}{Total}&100.39&\multicolumn{5}{c|}{3.83}\\
		\hline
		\multicolumn{3}{|c||}{Significance($\sigma$)}&9.83&\multicolumn{5}{c|}{ }\\
		\cline{1-4}
		\multicolumn{3}{|c||}{$\mathcal{L}_{5\,\sigma}$ (fb$^{-1}$)}&25.85&\multicolumn{5}{c|}{ }\\
		\hline
	\end{tabular}
	\caption{The number of events for $1b-\rm{jet}+2-jet+ \geq 2\mu+\ptmiss\leq 30$ GeV for BP3 and dominant SM backgrounds at the LHC/FCC with centre-of-mass energy of 30\,TeV and 100\,TeV at an integrated luminosity of 1000  fb$^{-1}$ and 100 fb$^{-1}$, respectively. The required luminosities to achieve a $5\,\sigma$ signal ($\mathcal{L}_{5\,\sigma}$) are also shown for both the cases. }\label{bgs343_BP3}
\end{table*}
%%%%%%%%%%%%%%%%%%%%%%%%%%%%%%%%%%%%%%%%%%%%%%%%%

While $t\bar{t}$ is the dominant background in this case, the contribution is very low, owing to the stringent cuts on missing energy and hardness. It is interesting to note that, this background contribution decreases when we move from 30 TeV to 100 TeV energies. This is accounted for by the less number of events with $\ptmiss \leq 30$ GeV and $n_{\text{jet}} \leq 2$ at 100 TeV, compared to 30 TeV, due to the increase in jets coming from ISR/FSR. Additionally, we reintroduce the $W$- and $Z$-boson resonance vetoes on the di-jet and di-lepton invariant mass, helping us reduce the background further. The numbers for the signal and the SM background events are given in \autoref{bgs343_BP3}. The signal significances of $6.73\,\sigma$ at 30\,TeV with 1000 \fbi integrated luminosity and $9.83\,\sigma$ at 100 TeV with luminosity of 100 \fbi can be attained for this benchmark point (BP3). The required luminosity for a $5\sigma$ discovery is 552.68 \fbi at 30 TeV, which reduces to 25.85 \fbi at 100 TeV.

\subsection{$1b-\rm{jet}+1\ell+1\mu$}
Instead of the demand of two muons in the finalstate, we have also investigated the situations with one muon, namely, the finalstates quoted in \autoref{bgS13_BP3_1}, \autoref{bgS13_BP3_2} and \autoref{tgS13_BP3_1}. Among these the scenario in \autoref{bgS13_BP3_1} is promising and the results are shown in \autoref{bgs313_2}. In this case, the complete finalstate with the appropriate cuts is described as follows:
\begin{center}
	$n_{b-\rm{jet}} = 1$, $n_{j} \leq 2$, $n_\mu = 1$, $n_e = 1$, $n_\ell = 2$, $n_{\tau-\rm{jet}} = 0$ \& \\ $p_T^H \geq 1200$ GeV \& $\abs{M_{jj} - M_{W}} \geq 10$ GeV, $\abs{M_{\ell\ell} - M_{Z}} \geq 5$ GeV.
\end{center}

%%%%%%%%%%%%%    1b + 1l + 1mu + pTmiss   %%%%%%%%%%%%%%%%%%%%%%%%%%%%%%
\begin{table*}[h!]
	\renewcommand{\arraystretch}{1.1}
	\centering
	\begin{tabular}{|c|c|c||c||c|c|c|c|c|}
		\hline
		\multirow{2}{*}{ $\sqrt s$ in}&\multirow{3}{*}{Fusion}&&\multicolumn{6}{c|}{$1b-\rm{jet}+1\ell+1\mu$}\\
		\cline{4-9}
		\multirow{2}{*}{TeV}	&& Mode &\multicolumn{1}{c||}{Signal}&\multicolumn{5}{c|}{Backgrounds}\\
		\cline{4-9}
		&&& BP3 &$t\bar t$&$VV$&$VVV$&$t\bar{t}V$&$tVV$\\
		\hline
		\hline
		
		\multirow{4}{*}{30} &\multirow{2}{*}{$b-g$}&$S_3^{4/3}\mu$&80.70 &\multirow{4}{*}{625.32}&\multirow{4}{*}{12.32}&\multirow{4}{*}{14.04}&\multirow{4}{*}{65.14}&\multirow{4}{*}{8.62}\\
		\cline{3-4}
		&&$S_3^{1/3}\nu$&2.43&&&&&\\
		\cline{2-4}
		&\multirow{2}{*}{$t-g$}&$S_3^{1/3}\mu$&0.53&&&&&\\
		\cline{3-4}
		&&$S_3^{2/3}\nu$&0.02&&&&&\\
		\hline
		\multicolumn{3}{|c||}{Total}&83.68&\multicolumn{5}{c|}{725.44}\\
		\hline
		\multicolumn{3}{|c||}{Significance($\sigma$)}&2.94&\multicolumn{5}{c|}{ }\\
		\cline{1-4}
		\multicolumn{3}{|c||}{$\mathcal{L}_{5\,\sigma}$ (fb$^{-1}$)}&2889.44&\multicolumn{5}{c|}{ }\\
		\hline\hline
		
		\multirow{4}{*}{100} &\multirow{2}{*}{$b-g$}&$S_3^{4/3}\mu$&148.16 &\multirow{4}{*}{628.19}&\multirow{4}{*}{26.90}&\multirow{4}{*}{10.58}&\multirow{4}{*}{37.06}&\multirow{4}{*}{15.56}\\
		\cline{3-4}
		&&$S_3^{1/3}\nu$&5.24&&&&&\\
		\cline{2-4}
		&\multirow{2}{*}{$t-g$}&$S_3^{1/3}\mu$&0.78&&&&&\\
		\cline{3-4}
		&&$S_3^{2/3}\nu$&0.03&&&&&\\
		\hline
		\multicolumn{3}{|c||}{Total}&154.21&\multicolumn{5}{c|}{718.29}\\
		\hline
		\multicolumn{3}{|c||}{Significance($\sigma$)}&5.22&\multicolumn{5}{c|}{ }\\
		\cline{1-4}
		\multicolumn{3}{|c||}{$\mathcal{L}_{5\,\sigma}$ (fb$^{-1}$)}&91.73&\multicolumn{5}{c|}{ }\\
		\hline
	\end{tabular}
	\caption{The number of events for  $1b-\rm{jet}+1\ell+1\mu$ for BP3 and dominant SM backgrounds at the LHC/FCC with centre-of-mass energy of 30\,TeV and 100\,TeV at an integrated luminosity of 1000  fb$^{-1}$ and 100 fb$^{-1}$, respectively. The required luminosities to achieve a $5\,\sigma$ signal ($\mathcal{L}_{5\,\sigma}$) are also shown for both the cases.}\label{bgs313_2}
\end{table*}
%%%%%%%%%%%%%%%%%%%%%%%%%%%%%%%%%%%%%%%%%%%%%%%%%%%

 Here the $S_3^{1/3}$ is produced in association with a neutrino from $b-g$ fusion and decays into a muon and top quark that further decomposes semi-leptonically into a bottom quark, a light charged lepton and a neutrino. In this finalstate, we demand this accompanying lepton to be an electron. However, due to the higher cross-section and the high probability of having a $b-$jet and at least one muon, we still have dominant contribution from the $b-g \to S_3^{4/3}$ process. In this case, we do not put a cut on the $\ptmiss$ to avoid the risk of losing signal events. The backgrounds are reduced by the $W$- and $Z$-boson vetoes, along with the hardness cut and the demand of $\leq 2$ total jets. In both the centre-of-mass energies of 30 and 100 TeV, $t\bar{t}$ remains the dominant background, contributing to the $b-$jet criteria. The signal strength at this finalstate is feeble compared to \autoref{bgs343_BP3}, as we only obtain a $2.94\sigma$ significance at the 30 TeV LHC with 1000 \fbi of integrated luminosity. However, the situation is more promising at 100 TeV centre-of-mass energy, where we have a $5.22\sigma$ signal strength at 100 \fbi of integrated luminosity, with the requirement of 91.73 \fbi for a $5\sigma$ probe. 
 	
However, for the two other topologies (\autoref{bgS13_BP3_2} and \autoref{tgS13_BP3_1}) the signal numbers are quite low and the SM $t\bar t$ background numbers are significant which in turn reduces the signal strength considerably.  The situations further worsen in the topologies where no muon is present, which are \autoref{bgS13_BP3_3} for the $S_3^{1/3}$ component, and \autoref{tgS23_BP3_1} for the $S_3^{2/3}$ part. In these cases, we do not obtain any significant signal strength due to the overwhelming SM background numbers. Hence we infer that unlike the previous cases with BP1 and BP2, here for BP3, different components of the $S_3$ leptoquark can not be discriminated via looking at distinguishable signatures.

\section{Leptoquarks at muon collider}\label{sec:muon}
%%%%%%%%%%%%%%%%%%%% Feynman diagrams %%%%%%%%%%%%%%%%%%%%%%%%%%%%%
\begin{figure}[h!]
	\subfigure[]{
		\begin{tikzpicture}
			\begin{feynman}
				\vertex (a1);
				\vertex [above left=1cm of a1] (a0){$\mu^+$};
				\vertex [right=1.5cm of a1] (a2);
				\vertex [above right=1 cm of a2] (a3){$S_1/S_3$};
				\vertex [below left=1cm of a1] (b0){$\mu^-$};
				\vertex [below right=1cm of a2] (b3){$S_1^c/S_3^c$};
				\diagram {(a0)--[anti fermion](a1)--[boson,edge label=$\gamma/ Z^0$](a2)--[charged scalar](a3),
					(b0)--[fermion](a1),(b3)--[charged scalar](a2)};
			\end{feynman}
	\end{tikzpicture}}\hfill
	\subfigure[]{\begin{tikzpicture}
			\begin{feynman}
				\vertex (a1);
				\vertex [above=1cm of a1] (z1);
				\vertex [left=10mm of z1] (z0) {$\mu^+$};
				\vertex [right=10mm of z1] (z2) {$S_3^{4/3}$};
				\vertex [left=10mm of b1] (b0){$\mu^-$};
				\vertex [right=10mm of b1] (b2) {$S_3^{-4/3}$};
				\diagram {(z0)--[anti fermion](z1)--[fermion, edge label=$s/b$](b1)--[anti charged scalar](b2),
					(b0)--[fermion](b1),
					(z1)--[charged scalar](z2)};
			\end{feynman}
	\end{tikzpicture}}\hfill
	\subfigure[]{\begin{tikzpicture}
			\begin{feynman}
				\vertex (a1);
				\vertex [above=1cm of a1] (z1);
				\vertex [left=10mm of z1] (z0) {$\mu^+$};
				\vertex [right=10mm of z1] (z2) {$S_3^{1/3}$};
				\vertex [left=10mm of b1] (b0){$\mu^-$};
				\vertex [right=10mm of b1] (b2) {$S_3^{-1/3}$};
				\diagram {(z0)--[anti fermion](z1)--[fermion, edge label=$c/t$](b1)--[anti charged scalar](b2),
					(b0)--[fermion](b1),
					(z1)--[charged scalar](z2)};
			\end{feynman}
	\end{tikzpicture}}
	\caption{The tree-level Feynman diagrams for the pair production of $S_1$ and  ${S_3}$ leptoquarks at a muon collider for the benchmark points specified in \autoref{tab:all}.}
	\label{fig:muon_coll}
\end{figure}
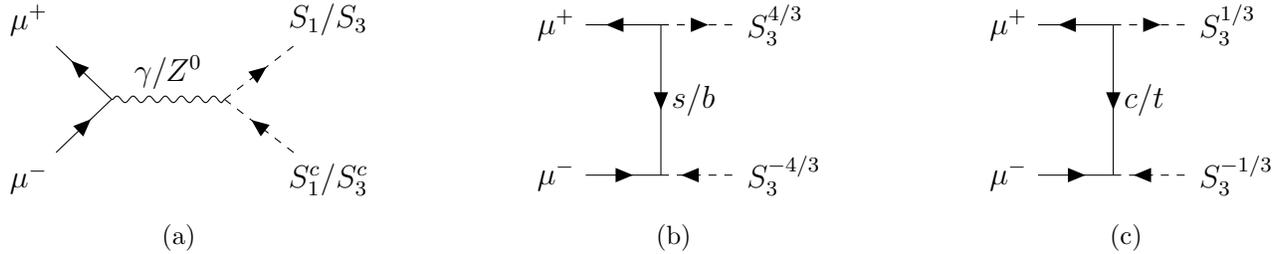
%%%%%%%%%%%%%%%%%%%%%%%%%%%%%%%%%%%%%%%%%%%%%%%%%%%%%%
%%%%%%%%%%%%%%%%%%%%%%%%%%%%%%%%%%%%%%%%%%%%%%%%%%%%%%%%%%%%
\begin{figure}[h!]
	\centering
	\subfigure[]{\includegraphics[height=0.23\textheight,width=0.46\textwidth]{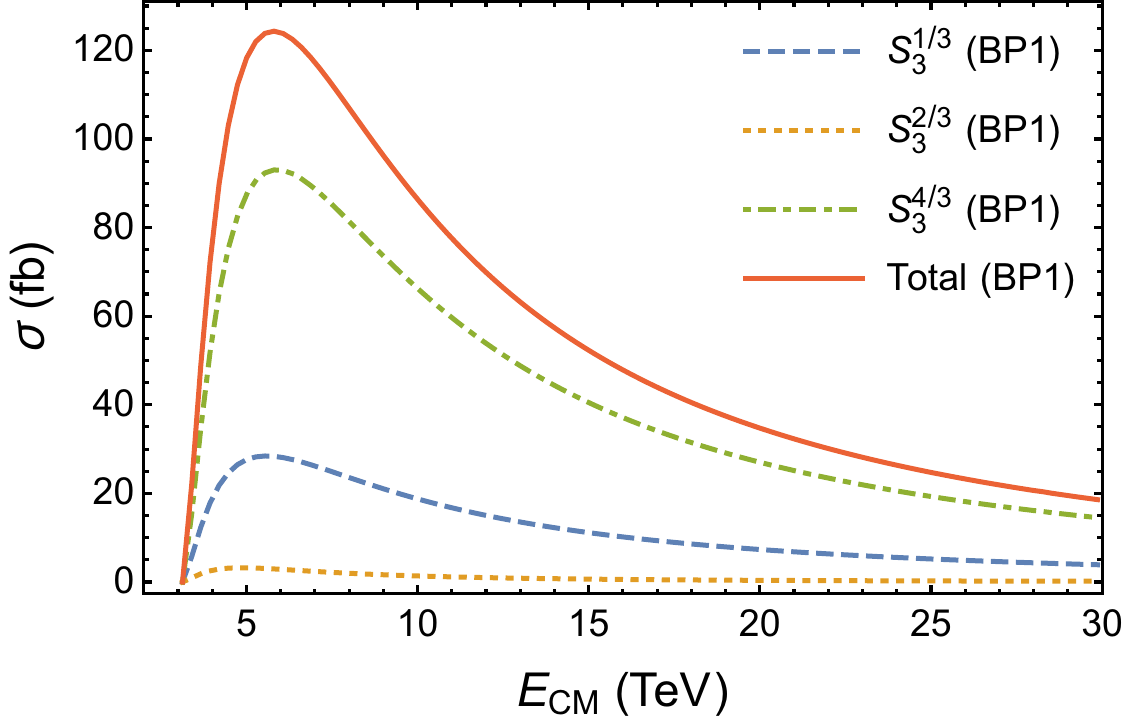}}
	\hfil
	\subfigure[]{\includegraphics[height=0.23\textheight,width=0.46\textwidth]{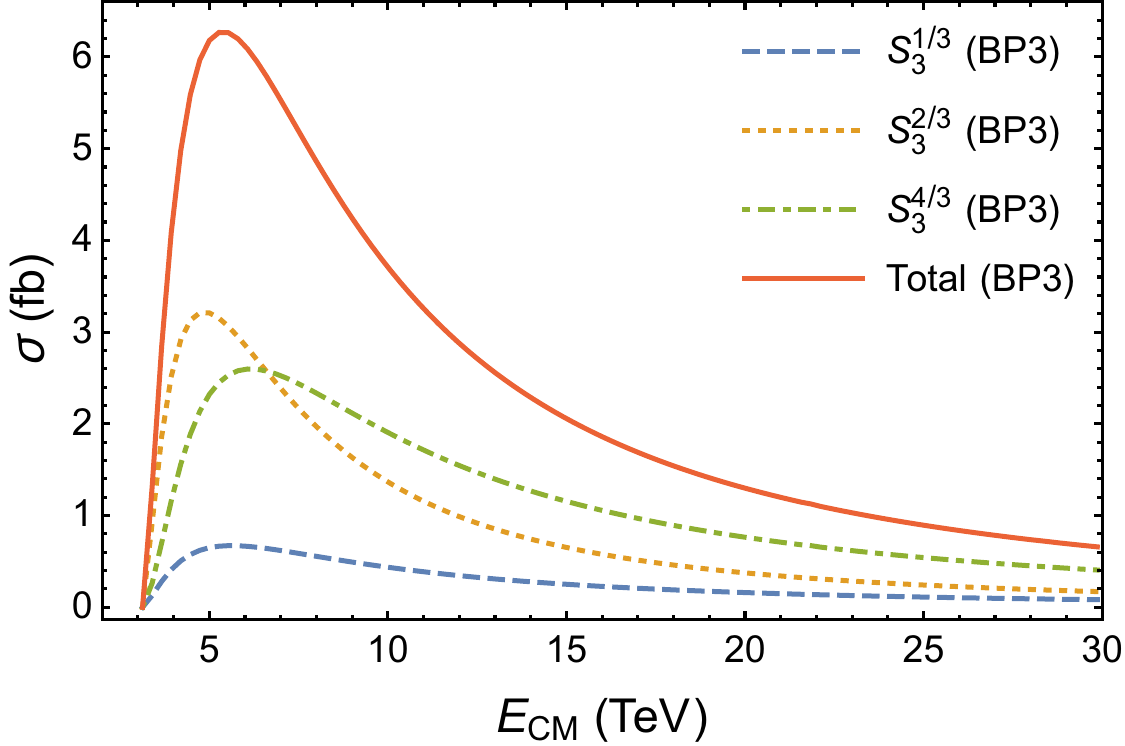}}
	
	\hspace*{3cm}\subfigure[]{\includegraphics[height=0.25\textheight,width=0.70\textwidth]{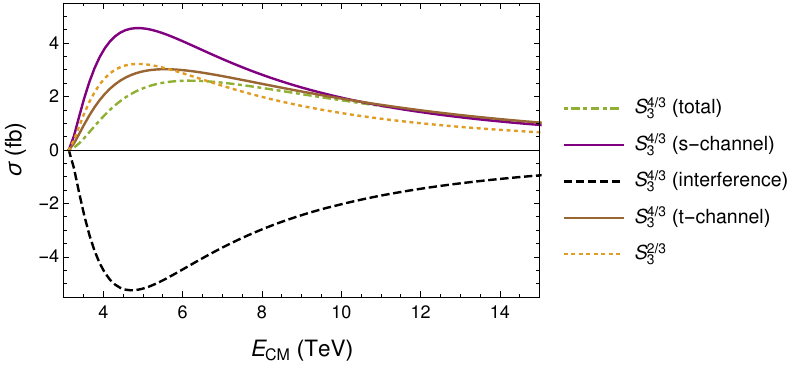}}
	\caption{The variation of cross-sections for the pair production of $S_3$ leptoquark  with the centre-of-mass energy at a multi-TeV muon collider for BP1 (in (a)) and BP3 (in (b)). The blue (dashed), yellow (dotted) and green (dot-dashed)  curves indicate individual contributions arising from $S_3^{1/3}$, $S_3^{2/3}$ and $S_3^{4/3}$ components of $S_3$ respectively, and the red (solid) line signifies the total production cross-section for $S_3$ leptoquark. The panel (c) zooms the BP3 case in the low energy region showing separately the $s$-channel (in purple solid), $t$-channel (in brown solid), interference of $s$- and $t$- channels (in black dashed) for $S_3^{4/3}$, as well as the total contributions of $S_3^{4/3}$ (in green dot-dashed) and  $S_3^{2/3}$ (in yellow dotted) in the production cross-sections.}
	\label{fig:muon_cross}
\end{figure}
%%%%%%%%%%%%%%%%%%%%%%%%%%%%%%%%%%%%%%%%%%%%%%%%%%%%%%%%%%%%%

This section is devoted to explore leptoquarks at a proposed muon collider about which a growing interest is noticed at recent times. The reach of a multi-TeV muon collider is expected to be 90\,ab$^{-1}$ with the centre-of-mass energy of 30 TeV \cite{AlAli:2021let}. Due to the absence of initial state QCD radiation, reduced synchrotron radiation compared to electron collider and known centre-of-mass frame, makes it a superior precision machine. In this section, we study the feasibility of producing leptoquarks in pair at muon collider. It is important to mention here that, in the case of $\mu^+ \mu^-$ collisions, it is not possible to have a single leptoquark produced at the final state. The possibility of resonant production of a single leptoquark from a muon-quark fusion is mentioned in ref. \cite{Azatov:2022itm}, which can arise only when the quark contribution in the muon PDF is considered. These contributions are very tiny, and in the context of this paper, such small estimates are not very relevant for a detailed collider study. Hence, the pair production is the only possibility, where Yukawa-type couplings involving second generation leptons can play the major role via $t$-channel process. The initial setup and the kinematic cuts remain the same as described in \autoref{sec:setup_colliders}. For our choices of benchmark points, given in \autoref{tab:all}, these production processes occur through the Feynman diagrams shown in \autoref{fig:muon_coll}. It is worthwhile to remind that the benchmark points are motivated from the tensions observed in $B$-decays, where the leptoquark $S_1$ couples only to third generation leptons aiming to reduce the $b\to c \tau \bar{\nu}$ discrepancy~\cite{HFLAV:2019otj} and as a result, $S_1$ gets produced only through a photon and a $Z^0$-boson mediated $s$-channel diagram (\autoref{fig:muon_coll}(a)). While by construction of the benchmark points the leptoquark $S_3$ couples to muons contributing to $b\to s \mu\mu$ anomalies~\cite{Aaij:2019wad,Aaij:2021vac,Aaij:2020nrf,Aaij:2017vbb,LHCb:2021zwz} and  will be the prime candidate of our study at muon collider. Apart from the $s$-channel diagrams, $S_3$ can be produced via the quark mediated $t$-channel diagrams (\autoref{fig:muon_coll}(b), \autoref{fig:muon_coll}(c)) as well. The $t$-channel diagram for $S_3^{4/3}$ component of $S_3$ goes through a $s$-quark (BP1, BP2) or a $b$-quark (BP3), whereas for $S_3^{1/3}$ component a  $c$-quark (BP1, BP2) or a $t$-quark (BP3) serves the purpose. It is noteworthy that $S_3^{2/3}$ does not couple to any charged lepton due to the structure of the interaction Lagrangian in \autoref{eq:LS32}, and hence it is produced at muon collider through the $s$-channel diagrams (\autoref{fig:muon_coll}(a)) only.

As BP1 and BP2, quoted in \autoref{tab:all}, differ mainly in the mass of the leptoquark, in this section we choose to present the results only for BP1 for simplicity, and BP3 as well. The variation of production cross-sections for $S_3$ leptoquark with the centre-of-mass energy of the muon collider is presented in \autoref{fig:muon_cross}(a) and \autoref{fig:muon_cross}(b) for BP1 and BP3, respectively. The contributions arising from different components of $S_3$ leptoquark are separately presented with different colour codes as specified in the plot legend. For BP1, $S_3^{4/3}$ shows prepotent effects while $S_3^{1/3}$ remains sub-dominant. In this case, the effects of $t$-channel diagrams are superior to the contributions from $s$-channel processes. However, for BP3, $S_3^{2/3}$ dominates at low centre-of-mass energy and as energy starts increasing, $S_3^{4/3}$ becomes the main contributor to the total cross-section  mostly via $s$-channel contribution.  Due to smaller values of leptoquark Yukawa-type couplings in BP3, $t$-channel processes are suppressed compared to BP1. Note that the interference of $t$- and $s$-channel diagrams in \autoref{fig:muon_cross}(c)  introduces negative contribution, which are large at lower energies and are substantial even at higher energies. This  keeps the cross-sections of $S_3^{4/3}$ and $S_3^{1/3}$ of the same order and results into a crossover of cross-sections for $S_3^{4/3}$ and $S_3^{2/3}$ around 6.5 TeV. It is easy to see from the two figures that the total production cross-section for $S_3$ in BP1 scenario is much higher than the BP3 case as the $Y_{S_3}^{22}$ coupling is significantly smaller in BP3 compared to BP1 (see \autoref{tab:all}).  On the other hand, we have chosen the hardness cut of $1.2$ TeV in our simulation, as discussed in \autoref{321}, in such a way that the effects of $s$-channel processes could be neglected. Thus contributions from the $S_3^{2/3}$  in \autoref{fig:muon_cross}(a) and \autoref{fig:muon_cross}(b) and similarly for $S_1$ leptoquark become negligible.

%%%%%%%%%%%%%%%%%%%Cross-section@MuonC%%%%%%%%%%%%%%%%%%%%%
\begin{table}[h!]
	\renewcommand{\arraystretch}{2}
	\centering
	\begin{tabular}{|c||c|c||c|c||c|c|}
		\hline 
		Bench-		&\multicolumn{2}{|c||}{$\sigma(\mu^+\mu^-\to S_3^{4/3}S_3^{-4/3})$  }&\multicolumn{2}{|c||}{$\sigma(\mu^+\mu^-\to S_3^{1/3}S_3^{-1/3})$ }&\multicolumn{2}{|c|}{$\sigma(\mu^+\mu^-\to S_3^{2/3}S_3^{-2/3})$ }\\
		mark&\multicolumn{2}{|c||}{in fb with $E_{\rm{CM}}$ in TeV}&\multicolumn{2}{|c||}{in fb with $E_{\rm{CM}}$ in TeV}&\multicolumn{2}{|c|}{in fb with $E_{\rm{CM}}$ in TeV}\\
		\cline{2-7}
		\makecell{Points \\ ($M_{S_3}$)}	& 8 TeV &30 TeV & 8 TeV &30 TeV & 8 TeV &30 TeV\\
		\hline\hline		
		\makecell{BP1 \\ (1.5 TeV)}&\hspace*{5mm}80.74\hspace*{5mm}&14.75&\hspace*{5mm}23.36\hspace*{5mm}&3.95&\hspace*{5mm}1.94\hspace*{5mm}&0.17\\
		\hline
		\makecell{BP3 \\ (1.5 TeV)}&2.32&0.41&0.55&0.08&1.94&0.17\\
		\hline
	\end{tabular}
	\caption{The cross-sections for pair production of $S_3$ at a multi-TeV muon collider for two different benchmark points BP1 and BP3 (specified in \autoref{tab:all}) at the centre-of-mass energies of 8\,TeV and 30\,TeV. Here $\sqrt{s}$ is used as the renormalization/factorization scale.}
	\label{tab:muon_cross}
\end{table}
%%%%%%%%%%%%%%%%%%%%%%%%%%%%%%%%%%%%%%%%%%%%%%%%%%%%%

For our analysis, we pick two centre-of-mass energies of  8 TeV and 30 TeV with the integrated luminosities of 1000 \fbi and 10000 \fbi, respectively. The cross-sections for pair production of different components of $S_3$ at these two centre-of-mass energies are tabulated in \autoref{tab:muon_cross}. Interestingly enough, the cross-sections for pair production of $S_3^{1/3}$ are significantly smaller than that of $S_3^{4/3}$. Although apparently it seems that the ratio of these two cross-sections at some particular centre-of-mass energy will be $1:4$ due to the extra $\sqrt2$ factor in the interaction vertex of $S_3^{4/3}$ with quarks and leptons, the presence of $s$-channel diagrams and masses of $t$-channel propagators cause a deviation from this $1:4$ ratio. On the other hand, the cross-sections for $S_3^{4/3}$ and $S_3^{1/3}$ in BP3 case are around 40 times smaller than that in BP1 due to magnitude of $Y_{S_3}^{22}$ as mentioned previously. The production cross-section for $S_3^{2/3}$ at muon collider remains the same in BP1 and BP3 since this process involves $s$-channel gauge interactions only.

\subsection{Kinematic distributions and topologies}\label{sec:kine_S3_muon}

%%%%%%%%%%%%%%%%%%%%%%nj  and nl distributions  %%%%%%%%%%%%%%
\begin{figure}[h!]
	\centering
	\subfigure[]
	{\includegraphics[scale=0.4]{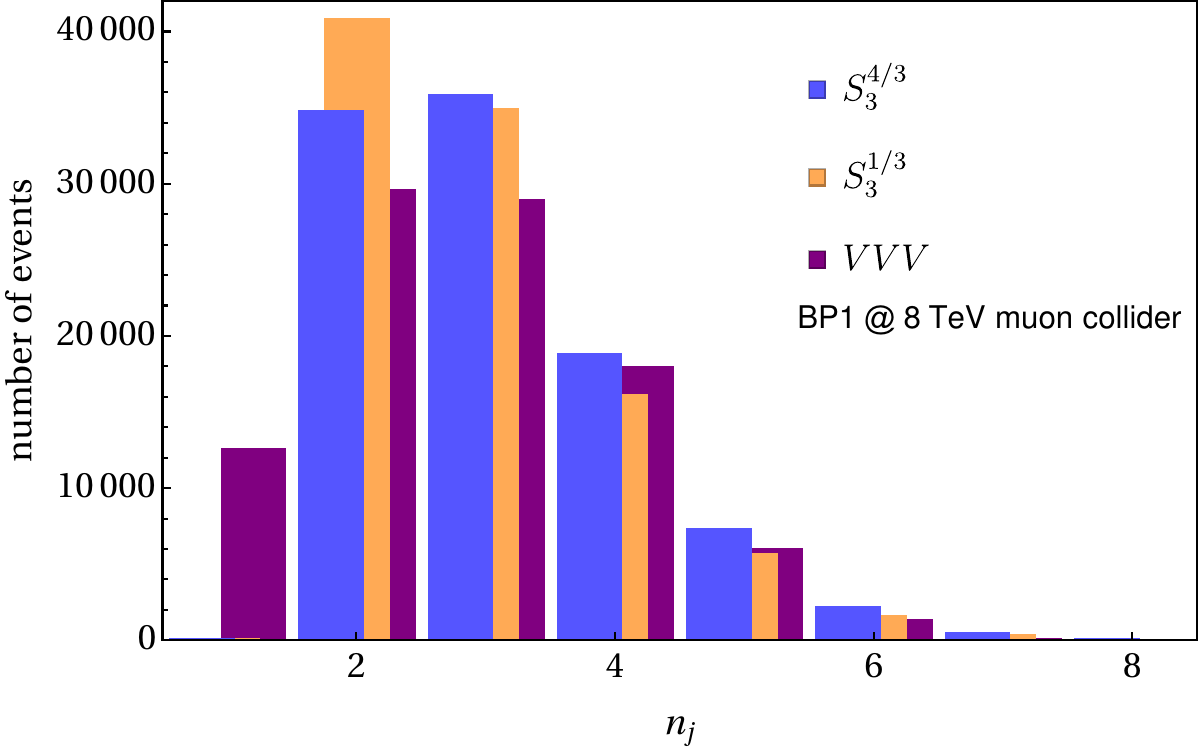}}
	\hfil
	\subfigure[]{\includegraphics[scale=0.4]{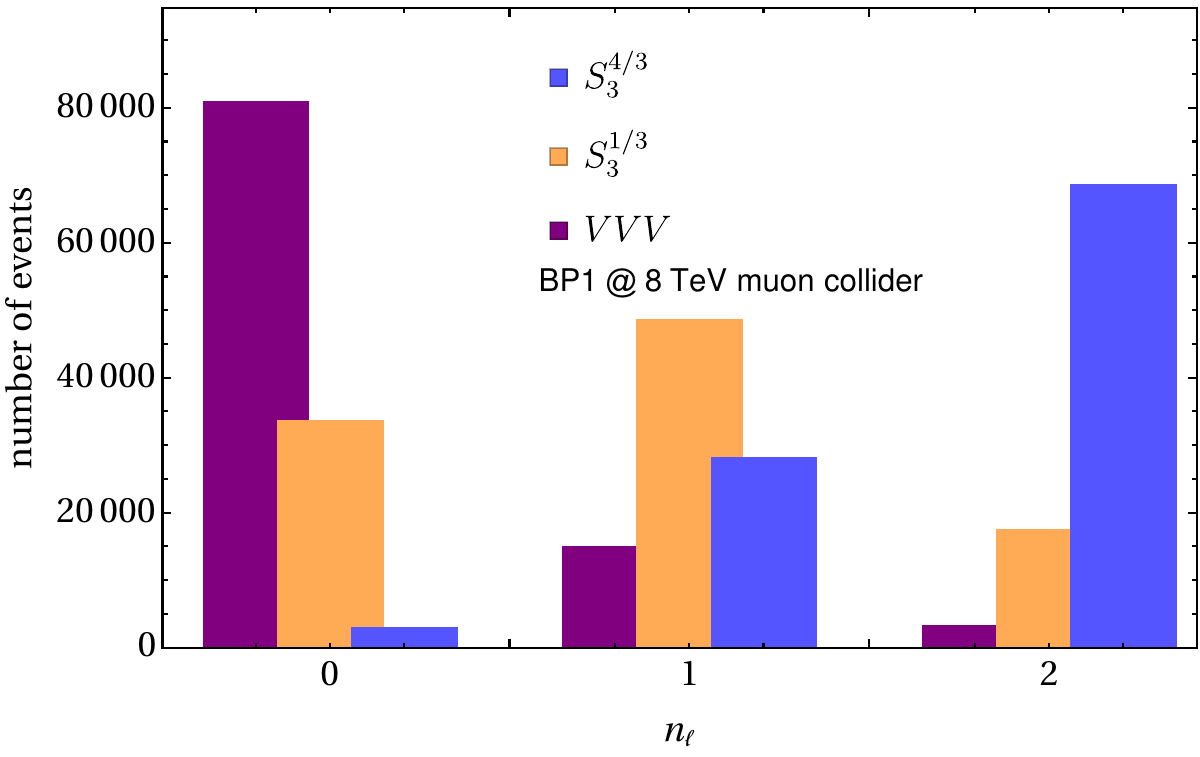}}
	\caption{The jet multiplicity ($n_j$ in (a)) and lepton multiplicity ($n_\ell$ in (b)) distributions for the pair production of $S_3^{4/3}$ and $S_3^{1/3}$ for BP1 along with the SM background from triple gauge boson at a muon collider with 8\,TeV centre-of-mass energy.}
	\label{fig:mu_jet_lep}
\end{figure}
%%%%%%%%%%%%%%%%%%%%%%%%%%%%%%%%%%%%%%%%%%%%%%%%%%%%

%%%%%%%%%%%%%%%%%p^j_T  distribution %%%%%%%%%%%%%%%%%%%%%%%%
\begin{figure}[h!]
	\centering
\includegraphics[scale=0.5]{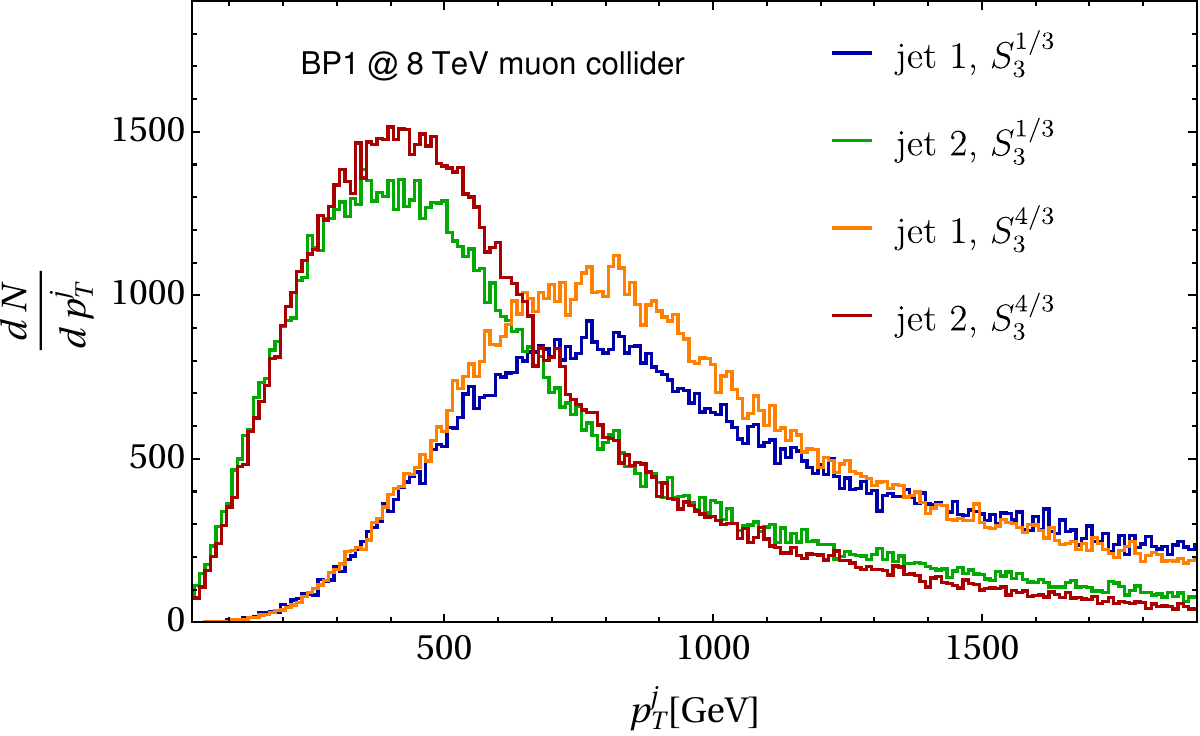}
	\caption{The jet $p_T$ distribution of the pair production of $S_3^{4/3}$ and  $S_3^{1/3}$ in BP1  at muon collider with 8\,TeV centre-of-mass energy.}
	\label{fig:mu_jet_pt}
\end{figure}
%%%%%%%%%%%%%%%%%%%%%%%%%%%%%%%%%%%%%%%%%%%%%%%%%%%
As discussed in the previous \autoref{sec:kine_S3}, we start with the comparison of various kinematic distributions of the $S_3$ leptoquark and the dominant SM backgrounds at muon collider in order to understand the different interplay between hadron and muon collider. To demonstrate, we select BP1 scenario with 8 TeV of centre-of-mass energy. At this point, it is interesting to mention that triple gauge boson modes act as dominant SM background for BP1 and it can be easily observed from the results quoted in \autoref{mu_2j2mu} and \autoref{mu_2c2mu} which will be discussed in the next subsection.

\autoref{fig:mu_jet_lep}(a) describes the jet multiplicity distribution ($n_j$) for pair production of $S_3^{4/3}$ (in blue) and $S_3^{1/3}$ (in orange) along with the dominant SM background of triple gauge boson (in purple) with 8\,TeV centre-of-mass energy. While all three distributions peak at around two or three jets, there are negligible number of mono-jet events for the signal processes, as both the pair produced leptoquarks must give one jet each. In contrast, the $VVV$ background has significant number of monojet events, owing to pure leptonic decay modes of the vector bosons. In parallel, we have shown the lepton multiplicity distributions ($n_\ell$) for $S_3^{4/3}$, $S_3^{1/3}$ and triple gauge boson background in \autoref{fig:mu_jet_lep}(b). As expected, $S_3^{4/3}$ displays peak with two leptons while $S_3^{1/3}$  exhibits substantial contributions mainly to no-lepton and mono-lepton channels. This is due to the reason that $S_3^{4/3}$ component of $S_3$ in BP1 decays to $s\mu$ mode with 100\% branching ratio (see \autoref{brss3}), whereas, $S_3^{1/3}$ component decays to $c\mu$ and $s\nu$ with equal probability. However, the SM background coming from triple gauge bosons diminishes gradually with increase in lepton number at the final state as the weak gauge bosons mostly decay into jets.

The jet transverse momentum ($p^j_T$)  distribution at 8 TeV muon collider has been depicted in \autoref{fig:mu_jet_pt}, where the two leading jets from the pair production of each of the components $S_3^{1/3}$ and $S_3^{4/3}$ are depicted. In both cases, the hardest jets ($j_1$), shown in blue for $S_3^{1/3}$ and orange for $S_3^{4/3}$ peak around half of the leptoquark mass (i.e. 750 GeV), as expected. The second hardest jets ($j_2$) are shown in green for $S_3^{1/3}$ and red for $S_3^{4/3}$, and both of them reach their maxima at about 400 GeV.

%%%%%%%%%%%%%%%%% p^l_T and misspt distributions %%%%%%%%%%%%%%%%%%%%%%%
\begin{figure}[h!]
	\centering
	\subfigure[]{\includegraphics[scale=0.4]{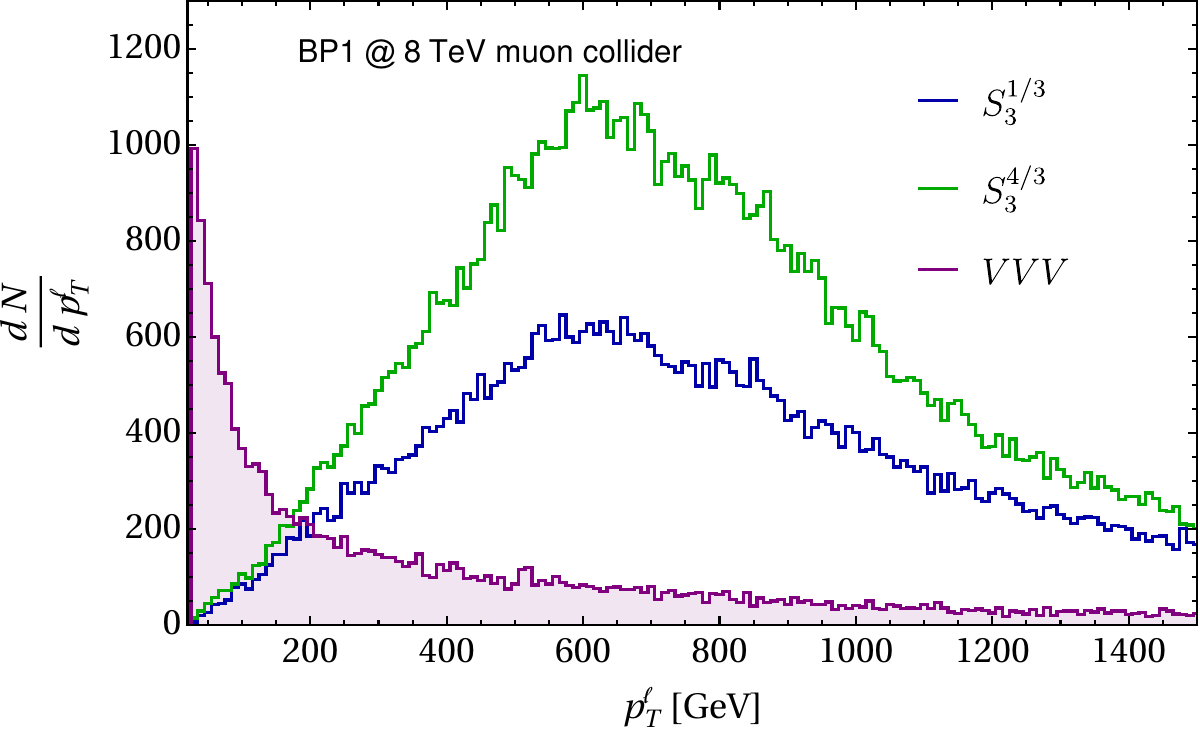}}
	\hfil
	\subfigure[]{\includegraphics[scale=0.4]{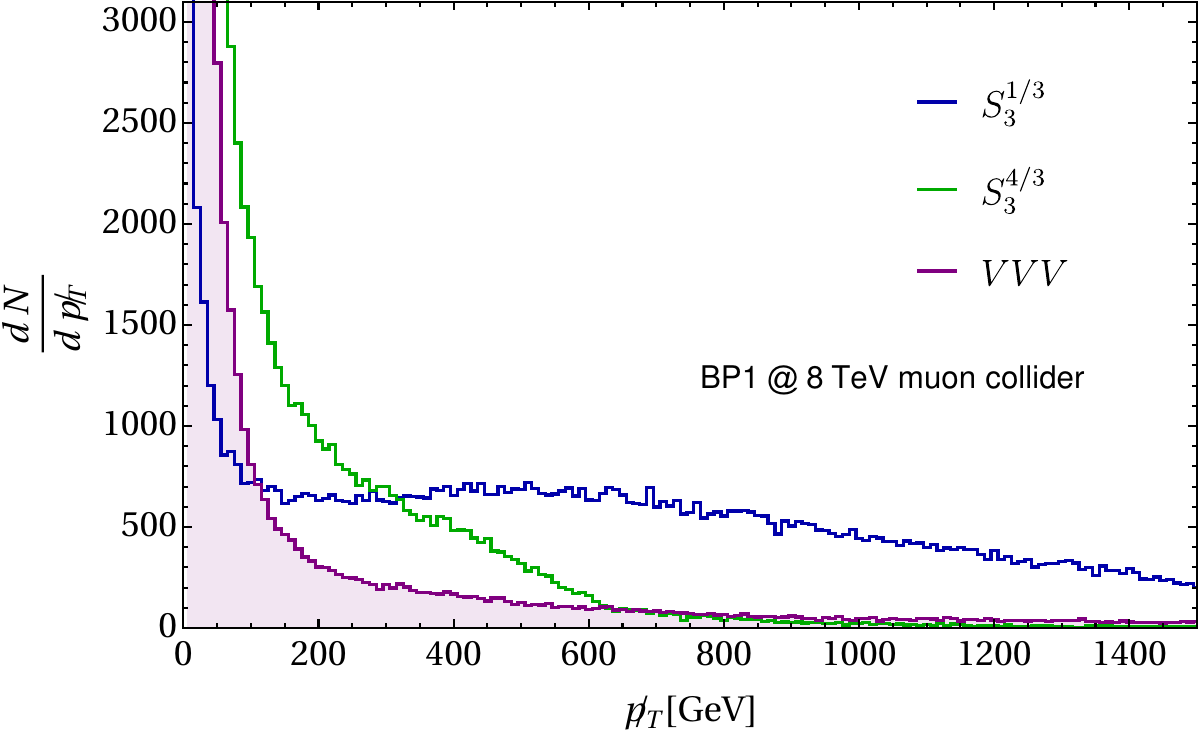}}
	\caption{The lepton $p_T$ ( $p^\ell_T$ in (a)) and missing $p_T$ ( $\ptmiss$ in (b)) distributions for $S_3^{4/3}$, $S_3^{1/3}$ and the SM background from triple gauge boson at a muon collider with centre-of-mass energy of 8\,TeV.}
	\label{fig:mu_lep_mis_pt}
\end{figure}
%%%%%%%%%%%%%%%%%%%%%%%%%%%%%%%%%%%%%%%%%%%%%%%%%%%%
The transverse momentum  ($p^\ell _T$)   distributions for light charged leptons in the pair production channels of $S_3^{4/3}$ and $S_3^{1/3}$ along with the SM background arising from triple gauge boson have been depicted in \autoref{fig:mu_lep_mis_pt}(a). Distributions for both the signals (blue for $S_3^{1/3}$ and green for $S_3^{4/3}$) reach their maxima at 600 GeV, which is slightly lower than half of the leptoquark mass (i.e. 750 GeV). However, the distribution for the dominant SM background (in purple) peak at around 40 GeV, and shows a long tail with very less events. Similarly the missing transverse momentum $\ptmiss$ distributions are displayed in \autoref{fig:mu_lep_mis_pt}(b). The distribution for $VVV$ background again peaks at around 40 GeV, showing a long, thin tail. The $\ptmiss$ distributions for $S_3^{4/3}$ dies out comparatively quicker as it does not involve any neutrino in its decay channel. $S_3^{1/3}$, which decays into $s\nu$ with 50\% branching ratio, shows a relatively large tail.

\begin{eqnarray}
\label{eq:mu_S43_BP1}
	\text{BP1}: \qquad\qquad\mu^+\mu^-&\to S_3^{+4/3} S_3^{-4/3}\to& 2-\rm{jet}+2\mu,\\
\label{eq:mu_S13_BP1}
	\mu^+\mu^-&\to S_3^{+1/3} S_3^{-1/3}\to & 2c-\rm{jet}+2\mu,\\[3mm]
\label{eq:mu_S43_BP3}
	\text{BP3}: \qquad\qquad\mu^+\mu^-&\to S_3^{+4/3} S_3^{-4/3}\to& 2b-\rm{jet}+2\mu,\\
\label{eq:mu_S13_BP3}
	\mu^+\mu^-&\to S_3^{+1/3} S_3^{-1/3}\to & 2b-\rm{jet}+4-jet+2\mu.
\end{eqnarray}

Now we proceed to study the detailed phenomenology of the two benchmark scenarios BP1 and BP3. After the pair production, $S_3^{4/3}$ decays into $s\mu$ ($b\mu$) with 100\% branching fraction whereas, $S_3^{1/3}$ decays into $c\mu$ and $s\nu$ ($t\mu$ and $b\nu$) finalstates each with 50\% branching fractions for BP1 (BP3), as displayed in \autoref{brss3}(\autoref{brss3_bp3}). Thus, for $S_3^{4/3}$, we have di-jet plus di-muon (\autoref{eq:mu_S43_BP1}) and two $b$-jets plus di-muon (\autoref{eq:mu_S43_BP3}) signals at the muon collider for BP1 and BP3, respectively. However, for $S_3^{1/3}$, several finalstates are plausible depending on its decay channels. Here, we only focus on those finalstates with no missing energy. It helps us to reduce the contamination from $S_3^{2/3}$ which despite of having a very low production cross-section, finally decays into finalstates with one neutrino for both the benchmark cases. Therefore for $S_3^{1/3}$, we consider two $c$-jets plus di-muon and two $b$-jets plus tetra-jet with di-muon topologies for BP1 and BP3, respectively. In the following few subsections we describe the simulated results for these four finalstates. We remind that we do not look for signals of $S_3^{2/3}$ in this section, as it gets produced through $s$-channel contributions only. As far as backgrounds are concerned, the $\mu^+\mu^- \to Z\ell^+\ell^-$ process can contribute to the aforementioned finalstates, along with the usual backgrounds of $t\bar{t},\, VV,\, VVV$, and $t\bar{t}V$. Similar to our analysis at the LHC, a cut on the total hardness variable $p_T^H \geq 1.2$ TeV is applied to both the signal and the background, which reduces the background contribution to the finalstates.

\subsection{$2-\rm{jet}+2\mu$}
This finalstate arises for $S_3^{4/3}$ in BP1 scenario (see \autoref{eq:mu_S43_BP1}). The complete finalstate with other cuts is given as:

\begin{center}
	$n_{j} = 2$, $n_\mu = 2$ \& $p_T^H \geq$ 1200 GeV.
\end{center}

%%%%%%%%%%%%%%%%%%%%% 2j+2mu %%%%%%%%%%%%%%%%%%%%%%%%%%%%%

\begin{table}[h!]
	\renewcommand{\arraystretch}{1.1}
	\centering
	\begin{tabular}{|c|c||c|c||c|c|c|c|c|}
		\hline
		\multirow{2}{*}{ $\sqrt s$ in}&&\multicolumn{7}{c|}{$2-\rm{jet}+2\mu$}\\
		\cline{3-9}
		\multirow{2}{*}{TeV}	& Mode &\multicolumn{2}{c||}{Signal}&\multicolumn{5}{c|}{Backgrounds}\\
		\cline{3-9}
		&&BP1& BP3 &$t\bar t$&$VV$&$VVV$&$t\bar t V$&$Z\ell^+\ell^-$\\
		\hline
		\hline	
		\multirow{2}{*}{8}&$S_3^{4/3}$&23304.06&779.94&\multirow{2}{*}{0.11}&\multirow{2}{*}{0.00}&\multirow{2}{*}{85.72}&\multirow{2}{*}{2.00}&\multirow{2}{*}{5.91}\\
		\cline{2-4}
		&$S_3^{1/3}$&1784.73&30.55&&&&&\\
		\cline{2-4}
		%&$S_3^{2/3}$&0.00&0.74&&&&\\
		\hline
		\multicolumn{2}{|c||}{Total}&25088.79&810.49&\multicolumn{5}{c|}{93.74}\\
		\hline
		\multicolumn{2}{|c||}{Significance($\sigma$)}&158.09&26.95&\multicolumn{5}{c|}{ }\\
		\cline{1-4}
		\multicolumn{2}{|c||}{$\mathcal{L}_{5\,\sigma}$ (fb$^{-1}$)}&1.00&34.19&\multicolumn{5}{c|}{ }\\
		\hline
		\hline
		
		\multirow{2}{*}{30}&$S_3^{4/3}$&23988.13&506.18&\multirow{2}{*}{0.00}&\multirow{2}{*}{0.00}&\multirow{2}{*}{139.01}&\multirow{2}{*}{2.16}&\multirow{2}{*}{25.73}\\
		\cline{2-4}
		&$S_3^{1/3}$&1628.46&34.66&&&&&\\
		\cline{2-4}
		%		&$S_3^{2/3}$&0.00&0.05&&&&\\
		\hline
		\multicolumn{2}{|c||}{Total}&25616.59&540.84&\multicolumn{5}{c|}{166.90}\\
		\hline
		\multicolumn{2}{|c||}{Significance($\sigma$)}&159.53&20.32&\multicolumn{5}{c|}{ }\\
		\cline{1-4}
		\multicolumn{2}{|c||}{$\mathcal{L}_{5\,\sigma}$ (fb$^{-1}$)}&9.82&604.89&\multicolumn{5}{c|}{ }\\
		\hline
		
	\end{tabular}
	\caption{The number of events for $2-\rm{jet}+2\mu$ finalstate (\autoref{eq:mu_S43_BP1}) for the benchmark points and dominant SM backgrounds at a multi-TeV muon collider with the centre-of-mass energy of 8\,TeV and 30\,TeV at an integrated luminosity of 1000\,\fbi and 10000\,\fbi\!\!, respectively. The required luminosities to achieve a $5\,\sigma$ signal ($\mathcal{L}_{5\,\sigma}$) are also shown for both the cases.
	}\label{mu_2j2mu}
\end{table}

 Here, similar to many of the finalstates in the LHC/FCC analysis, we have put the hardness cut $p_T^H \geq 1.2$ TeV to reduce the background contamination. The signal and background analyses for this finalstate at  8 TeV and 30 TeV centre-of-mass energies with 1000 \fbi and 10000 \fbi of integrated luminosities are tabulated in \autoref{mu_2j2mu}. Triple gauge boson is the dominant background here, although tiny. The signal gets some contribution from $S_3^{1/3}$ mode, where the $c-$jets are misidentified with light$-$jets. The results are very inspiring here since we can achieve $\sim 158 \sigma$ of signal significance for BP1 at both of the centre-of-mass energies with the specified luminosities. Therefore, significance of $5\sigma$ can be achieved at very early stage for both the centre-of-mass energies. It is also interesting to notice that with the specified luminosities at both the centre-of-mass energies one can attain more than $20\sigma$ significance for BP3 as well, in which the $b$-jet remains untagged. It is worth mentioning here that the reduction in production cross-sections at higher energy is compensated by our choice of enhanced luminosity (10000\,\fbi) at 30\,TeV simulation. Thus the signal significance turns out to be very similar between 8\,TeV and 30\,TeV collisions for both the benchmark points.

\subsection{$2c-\rm{jet}+2\mu$}

%%%%%%%%%%%%%%%%%%%%%%%%%%%%%%%% 2c+2mu %%%%%%%%%%%%%%%%%%%%%%%%%%%%%%%%

\begin{table}[h!]
	\renewcommand{\arraystretch}{1.1}
	\centering
	\begin{tabular}{|c|c||c|c||c|c|c|c|c|}
		\hline
		\multirow{2}{*}{ $\sqrt s$ in}&&\multicolumn{7}{c|}{$2c-\rm{jet}+2\mu$  }\\
		\cline{3-9}
		\multirow{2}{*}{TeV}	& Mode &\multicolumn{2}{c||}{Signal}&\multicolumn{5}{c|}{Backgrounds}\\
		\cline{3-9}
		&&BP1& BP3 &$t\bar t$&$VV$&$VVV$&$t\bar t V$&$Z\ell^+\ell^-$\\
		\hline
		\hline	
		\multirow{2}{*}{8}&$S_3^{4/3}$&0.81&4.29&\multirow{2}{*}{0.00}&\multirow{2}{*}{0.00}&\multirow{2}{*}{0.97}&\multirow{2}{*}{0.00}&\multirow{2}{*}{0.00}\\
		\cline{2-4}
		&$S_3^{1/3}$&747.86&0.07&&&&&\\
		\cline{2-4}
		%		&$S_3^{2/3}$&0.00&0.16&&&&\\
		\hline
		\multicolumn{2}{|c||}{Total}&748.67&4.36&\multicolumn{5}{c|}{0.97}\\
		\hline
		\multicolumn{2}{|c||}{Significance($\sigma$)}&27.34&1.88&\multicolumn{5}{c|}{ }\\
		\cline{1-4}
		\multicolumn{2}{|c||}{$\mathcal{L}_{5\,\sigma}$ (fb$^{-1}$)}&33.43&7009.62&\multicolumn{5}{c|}{ }\\
		\hline
		\hline
		
		\multirow{2}{*}{30}&$S_3^{4/3}$&2.95&28.85&\multirow{2}{*}{0.00}&\multirow{2}{*}{0.00}&\multirow{2}{*}{7.63}&\multirow{2}{*}{0.00}&\multirow{2}{*}{0.00}\\
		\cline{2-4}
		&$S_3^{1/3}$&831.41&0.17&&&&&\\
		\cline{2-4}
		%		&$S_3^{2/3}$&0.00&0.00&&&&\\
		\hline
		\multicolumn{2}{|c||}{Total}&834.36&29.02&\multicolumn{5}{c|}{7.63}\\
		\hline
		\multicolumn{2}{|c||}{Significance($\sigma$)}&28.75&4.79&\multicolumn{5}{c|}{ }\\
		\cline{1-4}
		\multicolumn{2}{|c||}{$\mathcal{L}_{5\,\sigma}$ (fb$^{-1}$)}&302.37&10897.8&\multicolumn{5}{c|}{ }\\
		\hline
		
	\end{tabular}
	\caption{The number of events for $2c-\rm{jet}+2\mu$ finalstate (\autoref{eq:mu_S13_BP1}) for the benchmark points and dominant SM backgrounds at a multi-TeV muon collider with the centre-of-mass energy of 8\,TeV and 30\,TeV at an integrated luminosity of 1000\,\fbi and 10000\,\fbi\!\!, respectively. The required luminosities to achieve a $5\,\sigma$ signal ($\mathcal{L}_{5\,\sigma}$) are also shown for both the cases.}\label{mu_2c2mu}
\end{table} 

The finalstate $2c-\rm{jet}+2\mu$ emerges for BP1 scenario when the $S_3^{1/3}$ component of $S_3$ is produced in pair and each of them decays into $c\mu$ states (in \autoref{eq:mu_S13_BP1}). As mentioned earlier, this is not the only finalstate accessible at muon collider for $S_3^{1/3}$ with BP1, rather we choose this finalstate since it does not involve any missing energy. BP3 can  contribute only when the  $b$-jets are miss-tagged as $c$-jets, thus  is subdominant. The complete finalstate is described as follows:
\begin{center}
	$n_{c-\rm{jet}} = 2,\, n_{b-\rm{jet}} = 0, \, n_\mu =2$ \& $p_T^H \geq$ 1200 GeV.
\end{center}

 In addition to the hardness cut, $b$-jet veto potentially  reduces BP3 contribution along  with  the dominant  $t\bar{t}$ background. The results for this finalstate at the centre-of-mass energies of 8 TeV and 30 TeV with the respective integrated luminosities of 1000 \fbi and 10000 \fbi are illustrated in \autoref{mu_2c2mu}. As the production cross-section of $S_3^{1/3}$ is considerably smaller than that of $S_3^{4/3}$, and furthermore the branching fraction of  $S_3^{1/3}$ to $c\mu$ is only 50\%, the signal numbers for this finalstate remain substantially low compared to the $2-\rm{jet}+2\mu$ finalstate. Although, these number of events are large enough compared to the SM backgrounds which are negligible after imposition of suitable cuts,  and thus rendering $\sim 28\,\sigma$ signal significance at both the centre-of-mass energies. Interestingly, it requires only 34 \fbi and 302 \fbi of integrated luminosities to obtain a $5\,\sigma$ signal significance at the two energies respectively. It is worth mentioning that BP3 scenario can also provide $5\,\sigma$ significance for this finalstate with luminosity less than 10000 \fbi at both the centre-of-mass energies.

\subsection{$2b-\rm{jet}+2\mu$}
%%%%%%%%%%%%%%%%%%%%%%%%%%%%%%%%%%%% 2b+2mu %%%%%%%%%%%%%%%%%%%%%%%%%%

\begin{table}[h!]
	\renewcommand{\arraystretch}{1.1}
	\centering
	\begin{tabular}{|c|c||c|c||c|c|c|c|c|}
		\hline
		\multirow{2}{*}{ $\sqrt s$ in}&&\multicolumn{7}{c|}{$2b-\rm{jet}+2\mu$  }\\
		\cline{3-9}
		\multirow{2}{*}{TeV}	& Mode &\multicolumn{2}{c||}{Signal}&\multicolumn{5}{c|}{Backgrounds}\\
		\cline{3-9}
		&&BP1& BP3 &$t\bar t$&$VV$&$VVV$&$t\bar t V$&$Z\ell^+\ell^-$\\
		\hline
		\hline	
		\multirow{2}{*}{8}&$S_3^{4/3}$&0.00&680.58&\multirow{2}{*}{0.11}&\multirow{2}{*}{0.00}&\multirow{2}{*}{0.78}&\multirow{2}{*}{1.50}&\multirow{2}{*}{0.00}\\
		\cline{2-4}
		&$S_3^{1/3}$&0.00&20.98&&&&&\\
		\cline{2-4}
		%		&$S_3^{2/3}$&0.00&0.50&&&&\\
		\hline
		\multicolumn{2}{|c||}{Total}&0.00&701.56&\multicolumn{5}{c|}{2.39}\\
		\hline
		\multicolumn{2}{|c||}{Significance($\sigma$)}&0.00&26.44&\multicolumn{5}{c|}{ }\\
		\cline{1-4}
		\multicolumn{2}{|c||}{$\mathcal{L}_{5\,\sigma}$ (fb$^{-1}$)}&---&35.76&\multicolumn{5}{c|}{ }\\
		\hline
		\hline
		
		\multirow{2}{*}{30}&$S_3^{4/3}$&2.95&368.00&\multirow{2}{*}{0.00}&\multirow{2}{*}{0.00}&\multirow{2}{*}{0.69}&\multirow{2}{*}{1.81}&\multirow{2}{*}{0.00}\\
		\cline{2-4}
		&$S_3^{1/3}$&0.00&25.17&&&&&\\
		\cline{2-4}
		%		&$S_3^{2/3}$&0.00&0.05&&&&\\
		\hline
		\multicolumn{2}{|c||}{Total}&2.95&393.17&\multicolumn{5}{c|}{2.5}\\
		\hline
		\multicolumn{2}{|c||}{Significance($\sigma$)}&1.26&19.76&\multicolumn{5}{c|}{ }\\
		\cline{1-4}
		\multicolumn{2}{|c||}{$\mathcal{L}_{5\,\sigma}$ (fb$^{-1}$)}&$\gg$10000&639.90&\multicolumn{5}{c|}{ }\\
		\hline
		
	\end{tabular}
	\caption{The number of events for $2b-\rm{jet}+2\mu$ finalstate (\autoref{eq:mu_S43_BP3}) for the benchmark points and dominant SM backgrounds at a multi-TeV muon collider with the centre-of-mass energy of 8\,TeV and 30\,TeV at an integrated luminosity of 1000\,\fbi and 10000\,\fbi\!\!, respectively. The required luminosities to achieve a $5\,\sigma$ signal ($\mathcal{L}_{5\,\sigma}$) are also shown for both the cases.}\label{mu_2b2mu}
\end{table}

The finalstate of two $b-$jets with two muons emerges at muon collider when the $S_3^{4/3}$ component of $S_3$ leptoquark are produced in pair in BP3 scenario (\autoref{eq:mu_S43_BP3}). BP1 fails to contribute much as it renders $s$-jets in the  finalstate as well due to demand  of  only two jets, which are $b$-jets. Thus  finalstate looks like:
	\begin{center}
		$n_{b-\rm{jet}} = 2, \, n_\mu = 2$ \& $p_T^H \geq$ 1200 GeV.
	\end{center}

The signal and background analyses for this finalstate at the similar previously specified setups for the centre-of-mass energy and integrated luminosity are presented in \autoref{mu_2b2mu}. We see from \autoref{tab:muon_cross} that the production cross-sections for both $S_3^{4/3}$ and $S_3^{1/3}$ in BP3 are significantly low compared to BP1 case, and hence the signal significance would also be reduced. However, as the SM backgrounds in this case are also negligible and thus this finalstate results are inspiring too. In fact, one can attain $\sim26.5\,\sigma$ ($20\,\sigma$) significance at 8\,TeV (30\,TeV) energy with the specified integrated luminosity. It implies that less than 50 \fbi (650 \fbi) of luminosity is required to achieve the $5\,\sigma$ significance for this finalstate. Note that there is no significant signal events for this finalstate in BP1 scenario as apart from the demand of two $b-$jets,  a limit on total number of light jets $n_j = 2$ is applied here.

\subsection{$2b-\rm{jet}+2-\rm{jet}+2\mu$}

%%%%%%%%%%%%%%%%%%%%%% 2b+4j+2mu %%%%%%%%%%%%%%%%%%%%%%%%%%%%%%%%

\begin{table}[h!]	\renewcommand{\arraystretch}{1.1}
	\centering
	\begin{tabular}{|c|c||c|c||c|c|c|c|c|}
		\hline
		\multirow{2}{*}{ $\sqrt s$ in}&&\multicolumn{7}{c|}{$2b-\rm{jet}+4-\rm{jet}+2\mu$  }\\
		\cline{3-9}
		\multirow{2}{*}{TeV}	& Mode &\multicolumn{2}{c||}{Signal}&\multicolumn{5}{c|}{Backgrounds}\\
		\cline{3-9}
		&&BP1& BP3 &$t\bar t$&$VV$&$VVV$&$t\bar t V$&$Z\ell^+\ell^-$\\
		\hline
		\hline	
		\multirow{2}{*}{8}&$S_3^{4/3}$&112.22&204.48&\multirow{2}{*}{0.43}&\multirow{2}{*}{0.00}&\multirow{2}{*}{0.00}&\multirow{2}{*}{2.27}&\multirow{2}{*}{0.00}\\
		\cline{2-4}
		&$S_3^{1/3}$&7.24&21.51&&&&&\\
		\cline{2-4}
		%		&$S_3^{2/3}$&0.00&0.08&&&&\\
		\hline
		\multicolumn{2}{|c||}{Total}&119.46&225.99&\multicolumn{5}{c|}{2.70}\\
		\hline
		\multicolumn{2}{|c||}{Significance($\sigma$)}&10.80&14.94&\multicolumn{5}{c|}{ }\\
		\cline{1-4}
		\multicolumn{2}{|c||}{$\mathcal{L}_{5\,\sigma}$ (fb$^{-1}$)}&214.00&111.95&\multicolumn{5}{c|}{ }\\
		\hline
		\hline
		
		\multirow{2}{*}{30}&$S_3^{4/3}$&20.64&19.88&\multirow{2}{*}{0.02}&\multirow{2}{*}{0.00}&\multirow{2}{*}{0.00}&\multirow{2}{*}{2.78}&\multirow{2}{*}{0.00}\\
		\cline{2-4}
		&$S_3^{1/3}$&0.39&3.42&&&&&\\
		\cline{2-4}
		%		&$S_3^{2/3}$&0.00&0.00&&&&\\
		\hline
		\multicolumn{2}{|c||}{Total}&21.03&23.30&\multicolumn{5}{c|}{2.80}\\
		\hline
		\multicolumn{2}{|c||}{Significance($\sigma$)}&4.30&4.56&\multicolumn{5}{c|}{ }\\
		\cline{1-4}
		\multicolumn{2}{|c||}{$\mathcal{L}_{5\,\sigma}$ (fb$^{-1}$)}&$\gg$10000&$\gg$10000&\multicolumn{5}{c|}{ }\\
		\hline
		
	\end{tabular}
	\caption{The number of events for $2b-\rm{jet}+ 4-jet+2\mu$ finalstate (\autoref{eq:mu_S13_BP3}) for the benchmark points and dominant SM backgrounds at a multi-TeV muon collider with the centre-of-mass energy of 8\,TeV and 30\,TeV at an integrated luminosity of 1000\,\fbi and 10000\,\fbi\!\!, respectively. The required luminosities to achieve a $5\sigma$ signal ($\mathcal{L}_{5\sigma}$) are also shown for both the cases.}\label{mu_2b4j2mu}
\end{table}

This particular finalstate appears if $S_3^{1/3}$ is produced at muon collider in pair in BP3 scenario and then both of them decay through $t\mu$ channel (\autoref{eq:mu_S13_BP3}). The top quark would disintegrate into a $b$-quark and a $W$-boson, and eventually the $W$-boson will produce two light jets. Thus, from the pair production of $S_3^{1/3}$, for BP3, we get $2b-\rm{jet}+4-\rm{jet}+2\mu$ finalstate. However, the light jets  coming  from the  $W^\pm$  can be boosted and often form a Fatjet \cite{Bandyopadhyay:2022mej,Sen:2021fha}, which renders us to choose  $2b-\rm{jet}+2-\rm{jet}+2\mu$ finalstate. Interestingly,  for BP1, the partonic finalstates is $2c+ 2\mu$(\autoref{eq:mu_S13_BP1}) owing to  dominant branching  of  $S_3^{1/3}$ into  $c\, \mu$ and though subdominant but can  contribute to the desired finalstate when  the  $c$-jet is miss-tagged as $b$-jet with additional jets  coming from FSR.
A serious contribution from $S_3^{4/3}$ cannot be avoided due to large cross-section  of $S_3^{4/3}$ pair and  100\% branching to  $b\,  \mu$.  Thus the finalstate looks like as  
\begin{center}
	$n_j\geq 4  (n_{b-\rm{jet}} = 2),\, n_\mu = 2$ \& $p_T^H \geq$ 1200 GeV.
\end{center}

 The results for this finalstate at 8 TeV and 30 TeV centre-of-mass energies with the respective 1000 \fbi and 10000 \fbi of integrated luminosities are quoted in \autoref{mu_2b4j2mu}. At 8 TeV centre-of-mass energy with 1000\fbi of data one can reach $\sim15\,\sigma$ of signal significance for BP3 indicating a need of $\sim 110$ \fbi of integrated luminosity to achieve a $5\,\sigma$ signal significance. Surprisingly, one can reach $10.8\,\sigma$ signal significance at the same energy with 1000 \fbi of data for BP1 case as well contributing through the $2-{\rm jet}+2\mu$ channel. However, the results for 30 TeV is not heartening at all since we need integrated luminosity of more than 10000 \fbi to achieve $5\,\sigma$ significance.

\section{Comparison of results and  reach at colliders}
\label{sec:reachplots}
In order to identify the best outcomes of the previous sections and their implications in future searches at the colliders, in this section we explore the particular regions in the NP parameter space where more than $5\sigma$ signal significance can be reached with the specific choices of centre-of-mass energy and integrated luminosity. For this purpose we select those finalstates which have very small model background (i.e. the contamination from other production channels). Then we observe the variation of significance with the parameters of the NP model, namely, the mass of the leptoquark and its couplings with quarks and leptons keeping the centre-of-mass energy and integrated luminosity fixed at the specific choices. It should be noted that though the SM backgrounds remain unaltered for any specific centre-of-mass energy and luminosity, the model background (along with signal) varies with the change in the parameters of the NP model. At this point it is worth mentioning that the significance presented in this section are slightly smaller than those quoted in the corresponding tables in previous sections, as we
separate out the contributions arising from different production modes and then except the  desired signal channel we treat the rest of the signal numbers as background events.

\subsection{Discussion on $S_1$}
\label{sec:reach_S1}

Compiling the results for various different finalstates of $S_1$ leptoquark at the LHC, discussed in \autoref{sec:S1}, as a first step, we note down the variations of different production cross-sections and branching fractions as functions of three parameters, namely, $M_{S_1}$, $Y_{\tiny S_1}^{33}$ and $Z_{\tiny S_1}^{23}$. Then we weigh the signals and model backgrounds presented in any table accordingly to calculate the signal significance for different values of these three NP parameters. In this case, we find from the results quoted in \autoref{sec:S1_btau}, which aim at the finalstate composed of a $b-$jet and $\tau-$jet, the signal numbers in \autoref{btau1} and \autoref{btau3} are dominated by one particular production channel $c/t-g \to S_1\tau$ and $b-g \to S_1 \nu$, respectively. When it comes to the finalstates with a $c-$jet, similar pattern is seen in \autoref{ctau2} as discussed in \autoref{sec:S1_ctau}. In the other two finalstates described in \autoref{btau2} and \autoref{ctau1}, all the production channels contribute comparably, and hence it is not possible to single out any particular contribution with reasonable signal significance. Hence, we examine the cases described in \autoref{btau1}, \autoref{btau3}, and \autoref{ctau2} in the subsequent paragraphs.

\begin{figure}[t!]
	\centering
\includegraphics[scale=0.44]{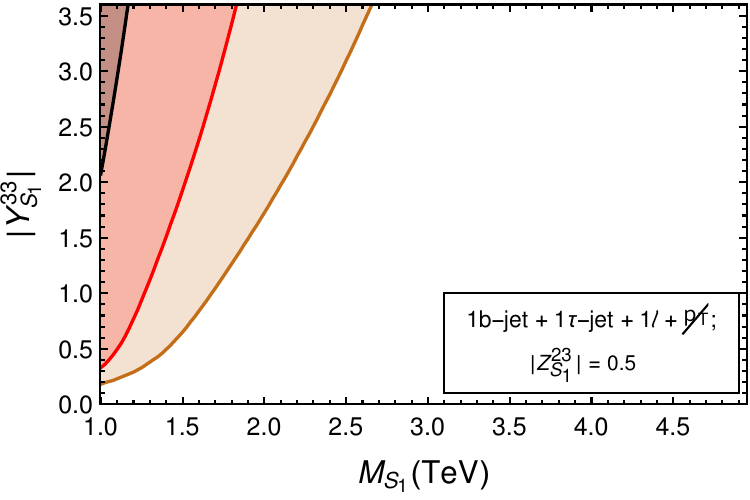}
\hfil
\includegraphics[scale=0.44]{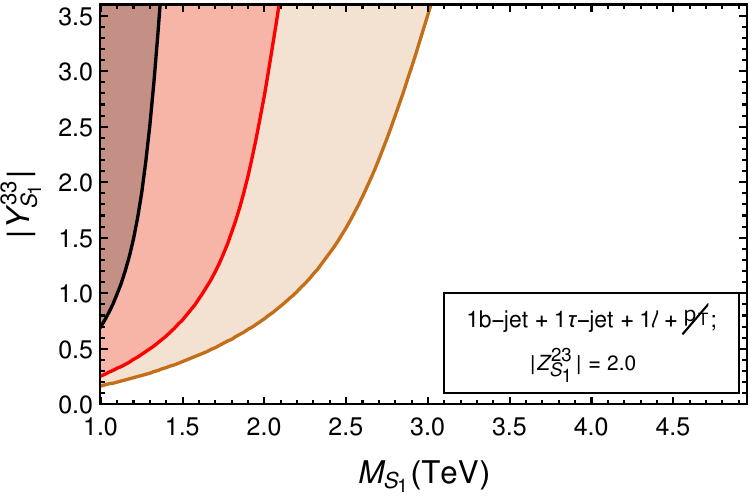}
\hfil
\includegraphics[scale=0.44]{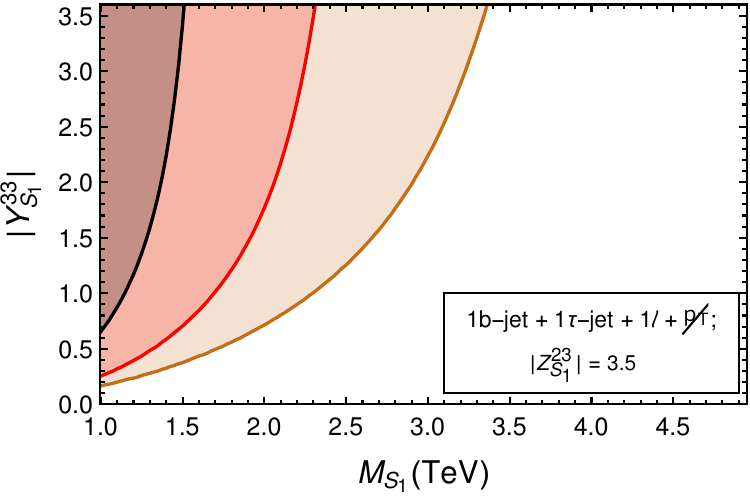}
\includegraphics[scale=0.5]{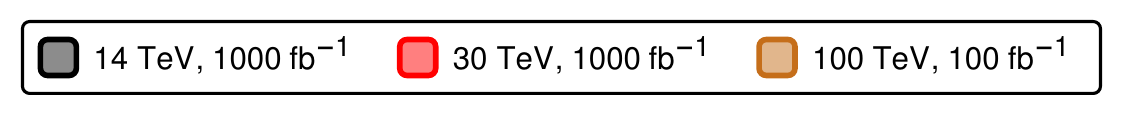}
	\caption{The regions with more than $5\sigma$ signal significance in $M_{S_1}-|Y^{33}_{\tiny S_1}|$ plane for the finalstate $1b-\rm{jet} + 1  \tau-\rm{jet}+1\ell + \ptmiss$ (see \autoref{btau1}) at different centre-of-mass energies at the LHC/FCC. The three different plots (from left) correspond to $|Z^{23}_{S_1}|$ values equal to 0.5, 2.0 and 3.5, respectively. The yellow curve represents the reach for $5\sigma$ signal significance at 100 TeV centre-of-mass energy with 100 \fbi of integrated luminosity. The red and black curves highlight the same signal significance at 30 TeV and 14 TeV centre-of-mass energies, respectively, with 1000 \fbi of integrated luminosity.}
	\label{reach:S1_1}
\end{figure}

The finalstate mentioned in \autoref{btau1} is $1b-\rm{jet} + 1  \tau-\rm{jet}+1\ell + \ptmiss$, for which the $c/t-g\to S_1\tau$ acts as signal and $b-g\to S_1\nu$ serves as model background. While, the $S_1$ production through $c-g$ fusion depends on $Z_{S_1}^{23}$, the other two production modes involve $Y_{S_1}^{33}$ only. Now as the decay vertex of $S_1$ for this finalstate (i.e. $S_1\to t\tau$) contains $Y_{S_1}^{33}$ alone, the total rate depends on both $Y_{S_1}^{33}$ and $Z_{S_1}^{23}$ couplings. The combined effects of all these facts are displayed in \autoref{reach:S1_1} in the $M_{S_1}-|Y_{\tiny S_1}^{33}|$ plane, where the three sub-figures represent three different values of the coupling $|Z_{S_1}^{23}|$ i.e., 0.5, 2.0 and 3.5, respectively. In each plot the yellow region indicates more than $5\sigma$ signal significance with 100 TeV centre-of-mass energies and 100 \fbi integrated luminosity, whereas the red and grey regions depict the same significance at 30 TeV and 14 TeV collisions, respectively, with an integrated luminosity of 1000 \fbi\!\!. Now, it is easy to understand that increasing the mass of leptoquark will decrease the signal events requiring larger values for $|Y_{\tiny S_1}^{33}|$ to reach the same significance. An interesting point to note here that each of the black, red and yellow curves gradually move toward the right side with enhancement in $|Z_{\tiny S_1}^{23}|$ value indicating that with higher value of $|Z_{\tiny S_1}^{23}|$, one needs smaller $|Y_{\tiny S_1}^{33}|$ coupling to reach the same significance for any particular mass of  the leptoquark. This is due to the fact that higher $|Z_{\tiny S_1}^{23}|$ value increases the production cross-section for the signal via $c-g$ fusion while the model background, arising from the other production channels, being independent of $Z_{\tiny S_1}^{23}$ remains unaltered. We find that the 14\,TeV results can only probe $Y_{\tiny S_1}^{33}\sim 2.5$ and above for low leptoquark mass that is close to 1\,TeV$-$1.2\,TeV, for the smallest $Z_{\tiny S_1}^{23}$ value of 0.5. Increase in $Z_{\tiny S_1}^{23}$ leads to the feasibility of probing $Y_{\tiny S_1}^{33}\sim 1$ in the same low mass range of the leptoquarks. On the other hand, for this finalstate, considering the highest value of $Z_{\tiny S_1}^{23}= 3.5$, the 30 TeV and 100 TeV searches can reach up to leptoquark masses of 1.8 TeV and 2.4 TeV, respectively, probing $Y_{\tiny S_1}^{33}\sim 1$. It is also inferred from this discussion that, a minimal change in the chosen benchmark values of the Yukawa-type couplings can alter the signal significance substantially. For example, in reference to the finalstate studied in \autoref{btau1} and discussed in \autoref{reach:S1_1} for the $S_1$ leptoquark, we find that, if we fix $m_{S_1} = 1.5$ TeV, a change of $\pm 0.1$ in the value of $Y_{S_1}^{33} = 0.91$ (in BP1) can change the signal significance by $\pm (12\%-13\%)$ at the 30 TeV LHC. On the other hand, a similar change of $\pm 0.1$ in the value of $|Z_{S_1}^{33}| = 0.5$ (in BP1) alters the signal significance at the 30 TeV LHC by $\pm (15\% - 16\%)$.

\begin{figure}[h!]
	\centering
	\includegraphics[scale=0.44]{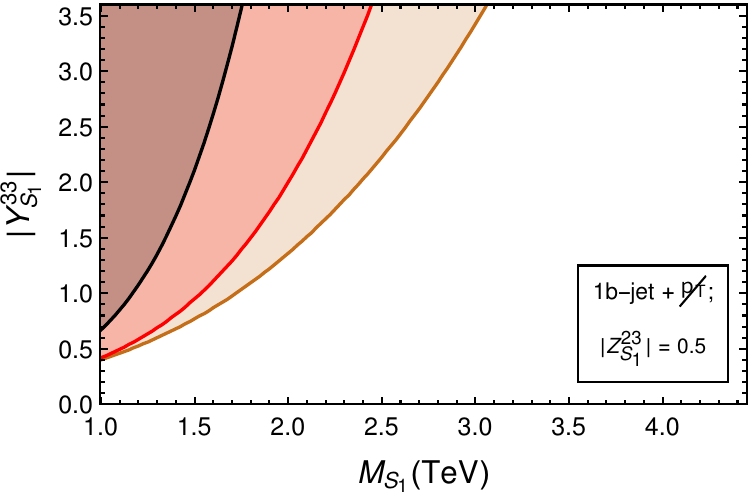}
	\hfil
	\includegraphics[scale=0.44]{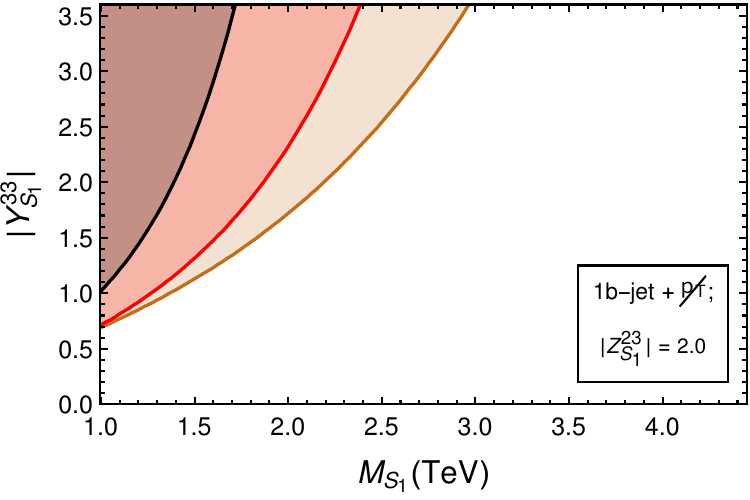}
	\hfil
	\includegraphics[scale=0.44]{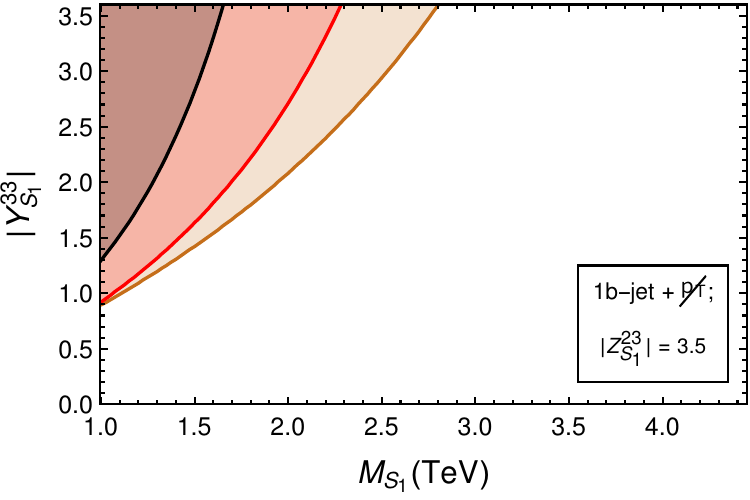}
	\includegraphics[scale=0.5]{plots/label1}
	\caption{The regions with more than $5\sigma$ signal significance in $M_{S_1}-|Y^{33}_{\tiny S_1}|$ plane for the finalstate $1b-\rm{jet} + \ptmiss$ (see \autoref{btau3}) at different centre-of-mass energies at the LHC/FCC. The three different plots (from left) correspond to $|Z^{23}_{S_1}|$ values equal to 0.5, 2.0 and 3.5, respectively. The yellow curve represents the reach for $5\sigma$ signal significance at 100 TeV centre-of-mass energy with 100 \fbi of integrated luminosity. The red and black curves highlight the same signal significance at 30 TeV and 14 TeV centre-of-mass energies, respectively, with 1000 \fbi of integrated luminosity.}
	\label{reach:S1_2}
\end{figure}

The next finalstate we consider to explore the reaches of $S_1$ at the LHC/FCC is $1b-\rm{jet} + \ptmiss$, corresponding to the results shown in \autoref{btau3}. In this case $b-g\to S_1\nu$ process provides the signal, whereas, events from $c/t-g\to S_1\tau$ act as model background. Therefore, the production vertex for signal as well as the decay vertex of $S_1$ depend only on one coupling $Y_{S_1}^{33}$, while the model background channels involve both $Y_{S_1}^{33}$ and $Z_{S_1}^{23}$. The $5\sigma$ reach of signal significance for this finalstate for three different $|Z_{S_1}^{23}|$ values equal to 0.5, 2.0 and 3.5 are presented in \autoref{reach:S1_2} in three different panels, respectively. The colour codes are the same as of \autoref{reach:S1_1}. In this case, unlike the previous scenario, we notice that the black, red and yellow curves shift upwards as we look at the three plots from left to right indicating necessity of higher $|Y_{S_1}^{33}|$ values with the increase in $|Z_{S_1}^{23}|$ coupling to maintain the same significance for any particular mass of the leptoquark. The reason behind this is that the cross-section for model background from $c-g$ fusion is enhanced with the increase in $|Z_{S_1}^{23}|$ value while the signal events remain unaffected. For $|Z_{S_1}^{23}|=0.5 $, we find that this finalstate can probe $Y_{S_1}^{33}\sim 1$ when the leptoquark mass is around 1.2 \,TeV scale at the 14\,TeV LHC, and can go up to 1.6 TeV,  2\,TeV masses with higher centre-of-mass energies of 30\,TeV and 100\,TeV,  respectively.

\begin{figure}[h!]
	\centering
	\includegraphics[scale=0.44]{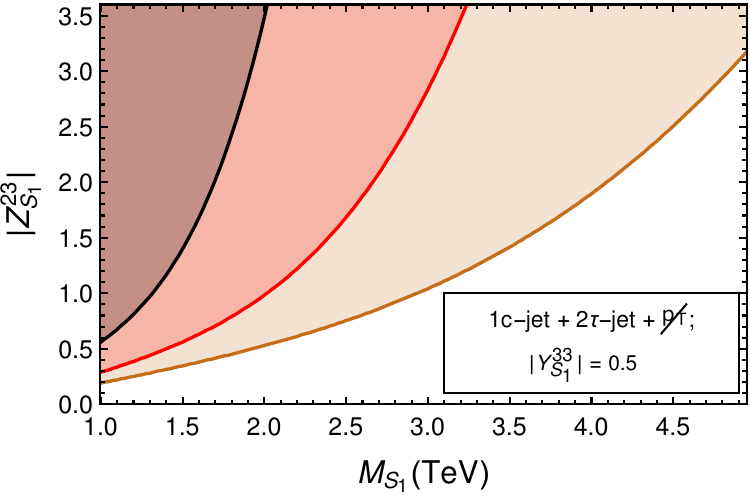}
	\hfil
	\includegraphics[scale=0.44]{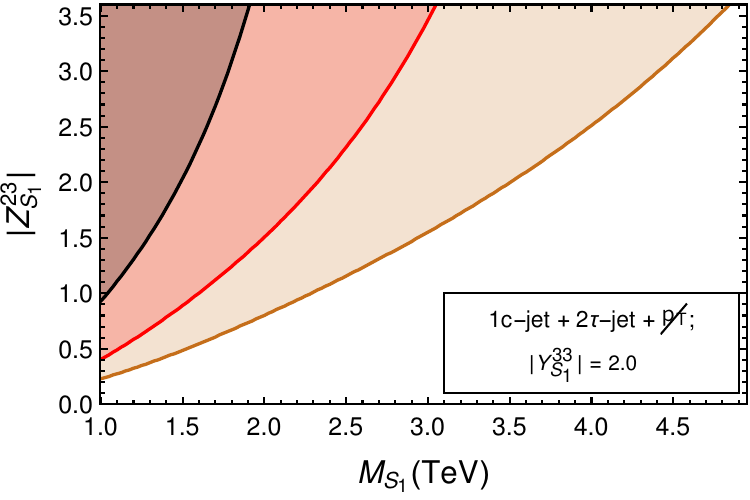}
	\hfil
	\includegraphics[scale=0.44]{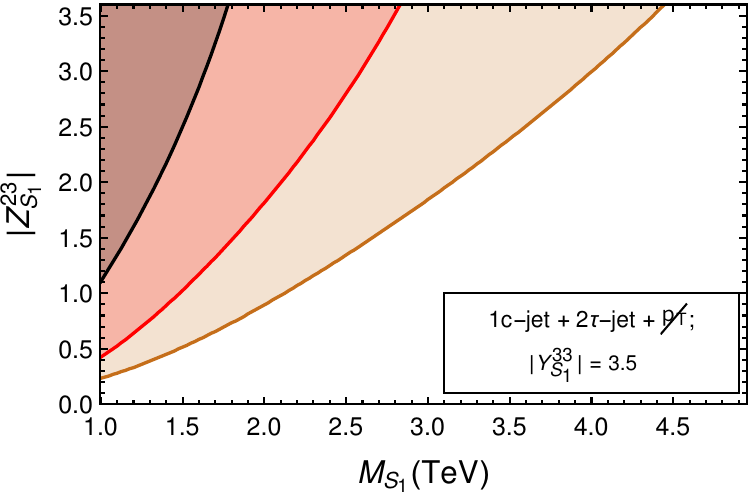}
	\includegraphics[scale=0.5]{plots/label1}
	\caption{The regions with more than $5\sigma$ signal significance in $M_{S_1}-|Z^{23}_{\tiny S_1}|$ plane for the finalstate $1c-\rm{jet} + 2\tau-\rm{jet}+ \ptmiss$ (see \autoref{ctau2}) at different centre-of-mass energies at the LHC/FCC. The three different plots (from left) correspond to $|Y^{33}_{S_1}|$ values equal to 0.5, 2.0 and 3.5, respectively. The yellow curve represents the reach for $5\sigma$ signal significance at 100 TeV centre-of-mass energy with 100 \fbi of integrated luminosity. The red and black curves highlight the same signal significance at 30 TeV and 14 TeV centre-of-mass energies, respectively, with 1000 \fbi of integrated luminosity.}
	\label{reach:S1_3}
\end{figure}

Next we move to the finalstate comprising of $1 c-\rm{jet} + 2\tau-\rm{jet} + \ptmiss$, whose signal and background event numbers are described in \autoref{ctau2}. This finalstate essentially shows a complementary behaviour to the previous two for which we studied the reach for the $S_1$ leptoquark. Similar to \autoref{btau1}, the $c/t-g \to S_1 \tau$ mode acts as the signal while $b-g\to S_1 \nu$ provides the model background. In the $c-g \to S_1 \tau$ case, the production cross-section depends on $Z_{S_1}^{23}$, while the production of $t-g \to S_1 \tau$ has a $Y_{S_1}^{33}$ dependence. The model background i.e. $b-g\to S_1 \nu$ production also varies with $Y_{S_1}^{33}$. However, in all three cases, the decay vertex $S_1 \to c\tau$ is purely dependent on $Z_{S_1}^{23}$. Thus, the cumulative effects of the $Z_{S_1}^{23}$ and $Y_{S_1}^{33}$ couplings are presented for this case in the $M_{S_1}-|Z_{\tiny S_1}^{23}|$ plane, depicted in \autoref{reach:S1_3}. The three panels of \autoref{reach:S1_3} correspond to the $5\sigma$ reach in this finalstate for three different $Y_{S_1}^{33}$ values equalling 0.5, 2.0, and 3.5, respectively. While this finalstate shows a similar behaviour of the lines moving upwards with the increase in $Y_{S_1}^{33}$, which we observed in case of \autoref{reach:S1_2}. This is accounted for by the enhancement of model background from the $b-g\to S_1 \nu$ due to the increment in $Y_{S_1}^{33}$. However, compared to \autoref{reach:S1_2}, we witness the possibility of a $5\sigma$ reach for a larger parameter space. For the lowest $Y_{S_1}^{33}$ value of 0.5, the 14 TeV LHC can probe $Z_{S_1}^{23} \sim 1$ up to a leptoquark mass value of $\sim 1.3$ TeV. For higher centre-of-mass energies of 30 TeV and 100 TeV, this reach increases to the leptoquark masses of $\sim 2$ TeV and $\sim 3$ TeV, respectively. From the combined analysis of these three aforementioned finalstates, we see that, compared to $Y_{S_1}^{33}$, the $Z_{S_1}^{23}$ coupling can be probed at similar orders, with a $5\sigma$ significance for a wider range of the leptoquark mass.

\subsection{Discussion on $S_3$}
\label{sec:reach_S3}

We learn from the phenomenological study performed in \autoref{sec:S3}, \autoref{sec:lfv} and \autoref{sec:muon} that the leptoquark $S_3$ is quite interesting as various different components of it give rise to quite unique signatures at colliders. The circumstance to discriminate these components becomes easier when we look for the analysis performed with BP1 and BP2 at the LHC/FCC. As the production cross-section is low in the case of BP3, we have obtained lower signal significance for it compared to the other two scenarios (BP1 and BP2), and thus is not a very favorable case to study the reach at the colliders. In subsequent subsections, we discuss the outcomes both at hadron and muon colliders separately.

\subsubsection{For the LHC/FCC}
\label{sec:reach_S3_LHC}

It can be noted from \autoref{tab:all} that in the case of BP1 and BP2, as $Y_{\tiny S_3}^{32}$ is very tiny, the phenomenology is mainly determined by the coupling $Y_{\tiny  S_3}^{22}$. This simplifies the situation due to the fact that as long as $|Y_{\tiny  S_3}^{22}|$ is greater that 0.03 (10 times larger than $Y_{\tiny S_3}^{32}$), the effect of $Y_{\tiny S_3}^{32}$ is insignificant. That means keeping all the other parameters unchanged, the branching fractions for different components of $S_3$ remain almost unaltered. Therefore, ignoring the effects of tiny $Y_{\tiny S_3}^{32}$, we adopt $|Y_{\tiny S_3}^{22}|\geq0.03$, and hence, we are left with only two parameters in this case, which are $M_{S_3}$ and $Y_{\tiny S_3}^{22}$.

\begin{figure}[h!]
	\centering
    \includegraphics[scale=0.65]{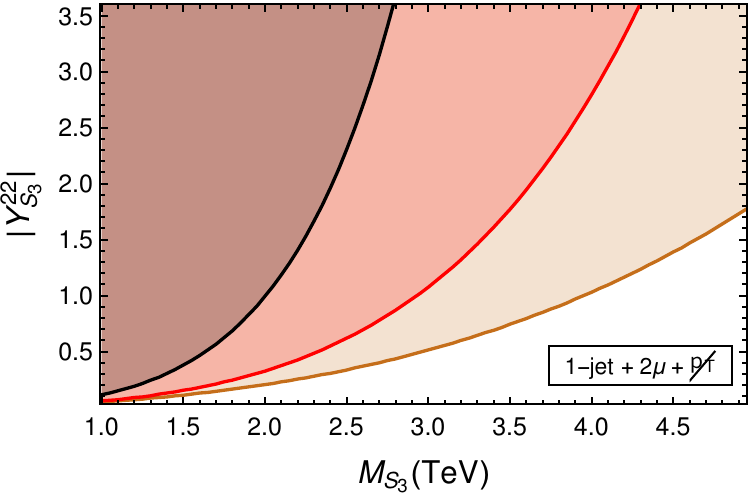}
	\hfil
	\includegraphics[scale=0.65]{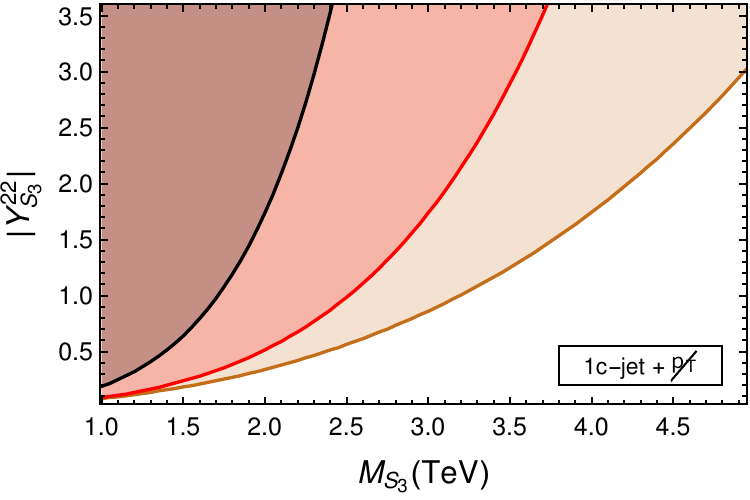}
	
	\includegraphics[scale=0.5]{plots/label1}
	\caption{The regions with more than $5\sigma$ signal significance in $M_{S_3}-|Y^{22}_{\tiny S_3}|$ plane for the finalstates $1-\rm{jet} +2\mu+ \ptmiss$ (in left panel) and $1c-\rm{jet}+\ptmiss$ (in right panel) at different centre-of-mass energies at the LHC/FCC. The yellow curve represents the reach for $5\sigma$ signal significance at 100 TeV centre-of-mass energy with 100 \fbi of integrated luminosity. The red and black curves highlight the same signal significance at 30 TeV and 14 TeV centre-of-mass energies, respectively, with 1000 \fbi of integrated luminosity. The signal and SM background numbers for these two final sates are highlighted in \autoref{tab:s343} and \autoref{tab:s323}, respectively.}
	\label{reach:S3_2343}
\end{figure}

Now we first consider the two finalstates  $1-\rm{jet} +2\mu+ \ptmiss$ and $1c-\rm{jet} + \ptmiss$, tabulated in \autoref{tab:s343} and \autoref{tab:s323}, respectively. For the first one, the signal events emerge from $s-g\to S_3^{4/3}\mu$ mode and the model background comes from $c-g\to S_3^{1/3}\mu$ channel making this finalstate an unique signature for the $S_3^{4/3}$ component. The second case corresponds to the signature for $S_3^{2/3}$ where the signal events arises from $c-g\to S_3^{2/3}\nu$ mode, while the model background appears from $s-g\to S_3^{1/3}\nu$ channel. The $5\sigma$ reach for these two finalstates with varying $M_{S_3}$ and $Y_{S_3}^{22}$ are presented in the left and right panels of \autoref{reach:S3_2343}, respectively. The yellow region signifies signal significance of more than $5\sigma$ at 100\,TeV centre-of-mass energy with 100\,\fbi integrated luminosity, and the respective red and the grey region indicate the same significance at the 30 TeV and 14 TeV centre-of-mass energies with 1000 \fbi of luminosity. It can be seen that the finalstate $1-\rm{jet} +2\mu+ \ptmiss$ probes larger parameter space than the finalstate $1c-\rm{jet} + \ptmiss$ as higher significance can be attained with the former one for same values of $M_{S_3}$ and $Y_{S_3}^{22}$. We find that the 14\,TeV results for $1-\rm{jet} +2\mu+ \ptmiss$ is quite promising as it can probe $Y_{\tiny S_3}^{22}\sim1$ until 1.8\,TeV mass of the leptoquark $S_3$ and with higher centre-of-mass energies like 30\,TeV and 100\,TeV, the same coupling value can be probed until $\sim$3\,TeV and $\sim$4\,TeV mass of $S_3$, respectively. In the case of $1c-\rm{jet} + \ptmiss$ finalstate, with $Y_{\tiny S_3}^{22}\sim1$, the mass reach for $S_3$ for the three centre-of-mass energies 14\,TeV, 30\,TeV and 100\,TeV are $\sim$1.7\,TeV, $\sim$2.7\,TeV and $\sim$3.5\,TeV, respectively. It is worthwhile to point out that both these two channels have much higher reach in the $S_3$ mass axis compared to the cases discussed in the previous subsection (\autoref{sec:reach_S1}) for $S_1$ leptoquark for an $\mathcal{O}(1)$ value of the corresponding Yukawa type coupling(s). The effect of deviation from the chosen benchmark values of the Yukawa-type couplings on the signal significance is very pronounced in case of $S_3$ as well. Taking the example of the finalstate analyzed in \autoref{tab:s343}, as well as discussed in \autoref{reach:S3_2343}(a), a  change of $\pm 0.1$ in the value of $Y_{S_3}^{22} = 0.5$ (in BP1) can affect the obtained signal significance at the 30 TeV LHC by $\pm28\%$,  for the fixed choice of $m_{S_3} = 1.5$ TeV.  

\begin{figure}[h!]
	\centering
\includegraphics[scale=0.65]{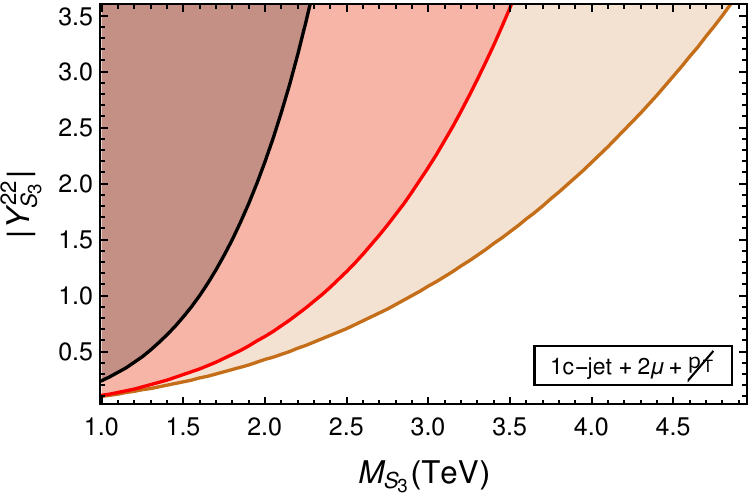}
\hfil
\includegraphics[scale=0.65]{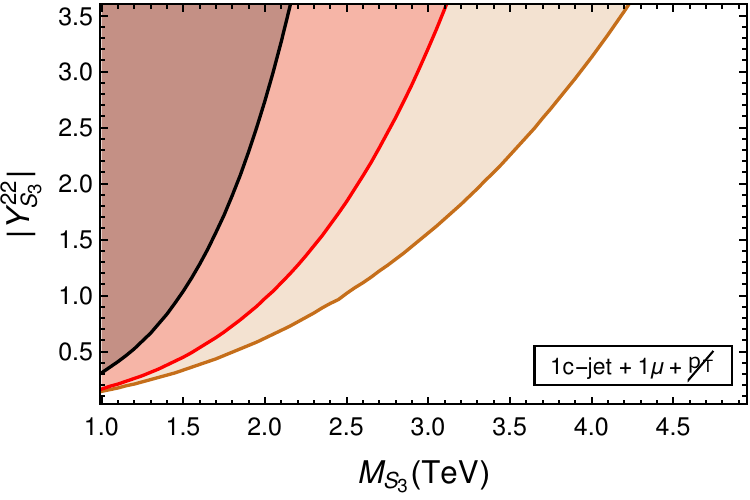}
	
\includegraphics[scale=0.5]{plots/label1}
	\caption{The regions with more than $5\sigma$ signal significance in $M_{S_3}-|Y^{22}_{\tiny S_3}|$ plane for the finalstates $1c-\rm{jet} +2\mu+ \ptmiss$ (in left panel) and $1c-\rm{jet}+ 1 \mu +\ptmiss$ (in right panel) at different centre-of-mass energies at the LHC/FCC. The yellow curve represents the reach for $5\sigma$ signal significance at 100 TeV centre-of-mass energy with 100 \fbi of integrated luminosity. The red and black curves highlight the same signal significance at 30 TeV and 14 TeV centre-of-mass energies, respectively, with 1000 \fbi of integrated luminosity. The signal and SM background numbers for these two final sates are highlighted in \autoref{s3132mu1c} and \autoref{s3131mu1c}, respectively. }
	\label{reach:S3_13}
\end{figure}

Having discussed the status of the two components of $S_3$, namely, $S_3^{4/3}$ and $S_3^{2/3}$, we now focus on the finalstates corresponding to $S_3^{1/3}$ component. For this purpose we select the two following decay topologies: $1c-\rm{jet} +2\mu+ \ptmiss$ (see \autoref{s3132mu1c}) and $1c-\rm{jet} +1\mu+ \ptmiss$ (see \autoref{s3131mu1c}). The finalstate $1c-\rm{jet} +2\mu+ \ptmiss$ mainly arises from the channel $c-g\to S_3^{1/3}\mu$ where the mode $s-g\to S_3^{4/3}\mu$ acts as model background. Likewise, the finalstate $1c-\rm{jet} +1\mu+ \ptmiss$ is generated from the production channel $s-g\to S_3^{1/3}\nu$ whereas the modes $s-g\to S_3^{4/3}\mu$ and $c-g\to S_3^{1/3}\mu$ function as model backgrounds. The left and right panels of \autoref{reach:S3_13} illustrate the $5\sigma$ reach for these two finalstates, respectively, at three different centre-of-mass energies and the similar luminosity choices as described in the last paragraphs. We can see from the left panel of \autoref{reach:S3_13} that the presence of $c-$jet in the finalstate reduces the signal significance compared to the left panel of \autoref{reach:S3_2343} that has a similar finalstate except for a replacement of the $c-$jet with a light-jet. This is due to the fact that we have an enhancement factor for the $S_3^{4/3}$ channel (i.e., $1-\rm{jet} +2\mu+ \ptmiss$) arising from the interaction vertex and also a suppression factor in $1c-\rm{jet} +2\mu+ \ptmiss$ originating from the branching fraction of $S_3^{1/3}$. As depicted in \autoref{reach:S3_13}, the finalstate with two muons (left panel) yield a better reach than that with one muon (right panel), due to it having less SM background events. In the di-muon finalstate, for $Y_{\tiny S_3}^{22}\sim 1$, we can probe the leptoquark mass up to $\sim$1.6 TeV, $\sim$2.4 TeV, and $\sim$3.0 TeV, respectively for the centre-of-mass energies of 14\,TeV, 30\,TeV and 100\,TeV. For the single muon finalstate, these mass reaches reduce to $\sim$1.5 TeV, $\sim$2.0 TeV, and $\sim$2.5 TeV.

\subsubsection{For a muon collider}\label{sec:reach_S3_muon}
\begin{figure}[h!]
	\centering
	\includegraphics[scale=0.75]{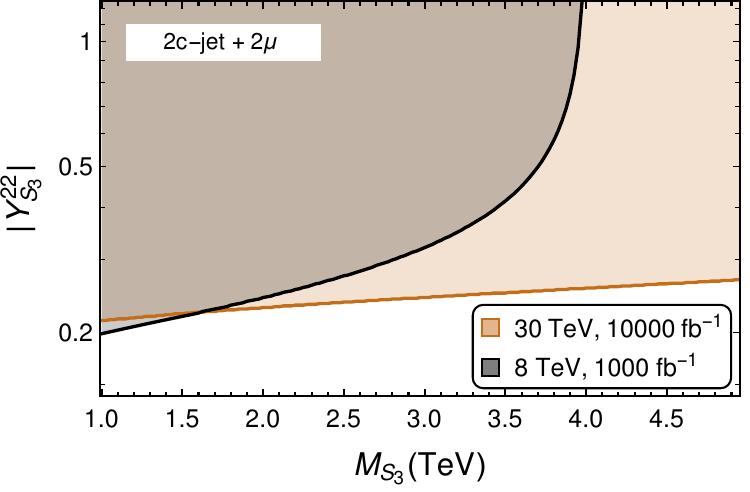}
	\caption{The regions with more than $5\sigma$ significance in $M_{S_3}-|Y^{22}_{\tiny S_3}|$ plane for the finalstate $2c-\rm{jet} +2\mu$ (see \autoref{mu_2c2mu}) at two different centre-of-mass energies at a multi-TeV muon collider. The yellow (black) curve represents the reach for $5\sigma$ signal significance at 30 TeV (8\,TeV) centre-of-mass energy with 10000\,\fbi (1000\,\fbi\!) integrated luminosity.}
	\label{reach:mu_S3}
\end{figure}

We continue to explore the similar outcomes at a multi-TeV muon collider. Here, the most encouraging finalstate is $2-\rm{jet} +2\mu$ (see \autoref{mu_2j2mu}) where we find enormously healthy signal numbers that arise from the $S_3^{4/3}$ component of $S_3$, rendering a huge significance for such a signal. Thus one can achieve the $5\sigma$ signal significance with very small value of $|Y^{22}_{S_3}|$ coupling and for large  mass of the $S_3$ leptoquark. A similar scenario occurs for $2b-\rm{jet} +2\mu$ finalstate in BP3 too. Therefore, we focus on $2c-\rm{jet} +2\mu$ finalstate in BP1 scenario. As already shown in \autoref{mu_2c2mu}, $S_3^{1/3}$ provides signal events for this finalstate while $S_3^{4/3}$ behaves as a model background. The $5\sigma$ reach plot, in the $M_{S_3}-|Y^{22}_{\tiny S_3}|$ plane, for this finalstate is depicted in \autoref{reach:mu_S3}. The yellow region signifies the parameter space with signal significance of more than $5\sigma$ level with the centre-of-mass energy being 30 TeV and an integrated luminosity of 10000 \fbi, whereas the same signal significance with 8 TeV centre-of-mass energy and 1000 \fbi integrated luminosity is shown in grey. It should be kept in mind that at 8 TeV centre-of-mass energy, leptoquark of mass greater than 4 TeV can not be produced in pairs. Therefore, we find a sharp rise of the black curve while approaching the mass of 4 TeV indicating no sensitivity after that mass scale. On the other hand, the $5\sigma$ reach for 30 TeV energy with an integrated luminosity of 10000 \fbi (shown by the yellow curve) remains almost flat for the small value of $|Y^{22}_{S_3}|$ until very large mass of the leptoquark. It is apparent from the discussions that the muon collider has much more sensitivity to probe the small coupling values up to the very large mass of the leptoquark compared to the hadron collider.

\subsection{Discussion on uncertainties}\label{uncer}
In this subsection we discuss the systematic uncertainties in context of hadron colliders that might affect the  signal significance of the  finalstates which are discussed in this article. These  include systematic uncertainties \cite{CMS:2007sch,Hubisz:2008gg}  due to $b$-jet tagging  of 15\% \cite{ATLAS:2021xox}, $c$-jet tagging  of 7.5 \% \cite{ATLAS:2021zwx}, $\tau$-jet tagging of 8\% \cite{Theveneaux-Pelzer:2014aia}, jet scale uncertainty of 3\% \cite{ATLAS:2020cli}, luminosity uncertainty 2\% and the parton distribution function uncertainty of  10\% \cite{Bandyopadhyay:2007cp}. We add them in quadrature to estimate the systematic uncertainties for $b-\rm{jet}\,+\,\tau-\rm{jet}$, $c-\rm{jet}\,+\,\tau-\rm{jet}$, $b-\rm{jet}\,+\, c-\rm{jet}$, $b-\rm{jet}$ and $c-\rm{jet}$ finalstates as 20\%, 15\%, 20\%, 18\% and 13\%, respectively. This can affect the signal significance roughly $+25\%$ to $-25\%$ depending on the finalstates. 

Finally we proceed to estimate the contamination arising from the leptoquark pair production mainly mediated by the strong interaction processes. For our chosen finalstates such  contamination can happen  when one or more  $b-, \, c-, \tau-$jets or charged leptons are missed from the pair production and in principle fake as a signal originating from the single leptoquark production.  Given the fact, we can measure the leptoquark mass via the invariant mass reconstruction of $c\mu, \, s\mu$, or mass edge of $c\tau$ (as shown in  \autoref{sec:edge}) or via the invariant  mass edge of   $c$-jet and missing energy, we can estimate such model backgrounds for a given benchmark point.  Determination of jet charges along with the finalstates can also identify  the  different  excitations of  leptoquarks \cite{Bandyopadhyay:2020jez, Bandyopadhyay:2020wfv}, which in turn can isolate singlet and triplet leptoquarks. The appraisal of model contamination can thus be more precise. We find such  contamination  can reduce the signal significance from a few percent to at most 25\%.  However, note that the leptoquark pair production although mainly generated from strong  interaction processes, the subsequent decays of leptoquarks are governed by the leptoquark Yukawa-type couplings. In that regard, one may also include such effect into signal contributions, which will further enhance the signal  significance. Therefore, we think for any early hint of a leptoquark signature these effects might as well be considered as a signal. Then later in case we are certain about the existence of the leptoquark, for the precision measurement of the  leptoquark Yukawa-type coupling, pair production can be regarded as model contamination. 

\section{Conclusion}
\label{sec:summary}

In this article we study the phenomenology of two scalar leptoquarks via single production channels mediated by quark gluon fusions. The leptoquarks carry color as well as electromagnetic charge, while the leptoquark $S_1$ is singlet and $S_3$ is triplet under the weak gauge group. The decays of these leptoquarks are dictated by specific non-vanishing couplings to fermions where the choice is governed by the series of  discrepancies observed in $B$-decays. Rather constraining
the parameter space explaining such tensions, we have demonstrated that our analysis is general enough and can easily be adopted to any scenario from the collider search perspective.

The pair productions of the leptoquarks at hadron collider are mostly dominated via QCD processes like gluon fusions, however, the single leptoquark productions which can probe the Yukawa-type couplings of the leptoquark to a quark and a lepton become efficient at high energies. The current and upcoming searches at the LHC/FCC play the key role here.  Whereas, interestingly a multi-TeV muon collider can be effective in probing these same Yukawa-type couplings through pair productions of the leptoquarks. We first consider different finalstates bearing distinguishable signatures arising from the $S_1$ leptoquark and three different components of the $S_3$ leptoquark.  In case of a TeV mass range $S_1$, we find among several decay topologies, $1b-\rm{jet} + 1  \tau-\rm{jet}+1\ell + \ptmiss$ and $1b +\ptmiss$ are the most promising ones that include a $b-$jet, which can probe the Yukawa-type coupling $Y^{33}_{\tiny S_1}$ as low as $0.2$ and $0.4$, respectively, for $Z^{23}_{\tiny S_1} = 0.5$ at the LHC/FCC at 30 TeV and 100 TeV energies with upgraded luminosity. Whereas, the finalstate of $1c-\rm{jet} + 2\tau-\rm{jet} + \ptmiss$ can probe minimum values of $Z^{23}_{\tiny S_1} = 0.3$ and 0.2, at 30 TeV and 100 TeV centre-of-mass energies, respectively for $Y^{33}_{\tiny S_1} = 0.5$. In this finalstate, the 100 TeV FCC is shown to have the possibility of probing $S_1$ mass exceeding 5 TeV, for large enough $Z^{23}_{\tiny S_1}$ values $\sim 3.0$.  We have also illustrated that when $S_1$ is produced in association with a visible particle (say a charged lepton), the further decay $S_1 \to \bar{c}\, \tau^+ \to \bar{c} \pi^+ \bar{\nu}$  leads to invariant mass edge at the $S_1$ mass, which can be instrumental in determination of the leptoquark mass scale at the LHC. In all the finalstates pertaining to $S_1$, the number of signal and SM background events are presented at centre-of-mass energies of 30 TeV and 100 TeV, owing to the low signal significance at the 14 TeV LHC.

The phenomenology is richer in the case of $S_3$ leptoquark where three different components, namely $S^{4/3}_3,\, S^{2/3}_3$ and $S^{1/3}_3$ are produced with the same tree-level mass. For our choices of the benchmark points these components often decay into finalstates consisting of muons compared to tau leptons as observed in the case of $S_1$. We notice that $S^{4/3}_3$ and $S^{2/3}_3$ components have distinct signatures; $1-\rm{jet}+2\mu +\ptmiss \leq 30 \,\rm{GeV}$ and  $1c-\rm{jet} +\ptmiss \geq 200\,  \rm{GeV}$, respectively, which can probe very low values ($\lesssim \mathcal{O}(10^{-1})$) of the Yukawa-type coupling $Y^{22}_{\tiny S_3}$ for a TeV mass scale $S_3$ at the upcoming upgrades of the  LHC. On the other hand $S^{1/3}_{3}$ has four modes to search for, and focusing on the most encouraging ones $2\mu + 1c-\rm{jet} + \ptmiss \leq 30\,  \rm{GeV}$ and  $1\mu + 1c-\rm{jet} + \ptmiss \geq 500\,  \rm{GeV}$, we find similar small values of $Y^{22}_{\tiny S_3}$ can be explored at the LHC/FCC. Additionally, we also briefed about the lepton flavour violating signatures in the decay caused due to the off-diagonal Yukawa-type coupling $Y^{32}_{S_3}$. In majority of the finalstates from single production of $S_3$ leptoquarks, the signal and background event numbers are presented at three different centre-of-mass energies of 14 TeV, 30 TeV, and 100 TeV at the LHC/FCC. However, in case of the lepton flavour violating finalstates, the 14 TeV event numbers are not listed, citing low signal significance. The results exhibit a maximum reach of more than 5 TeV mass of the $S_3$ leptoquark at the 100 TeV FCC, if the Yukawa-type coupling of leptoquarks are large,  namely, close to  the perturbativity limit.

We also explore the possibilities for direct searches of  the leptoquarks at a multi-TeV muon collider considering two different centre-of-mass energies;  8\,TeV and 30\,TeV. Here in most cases, we rely on the pair productions via $t$-channel processes (through quarks) to probe the relevant Yukawa-type couplings, except for the $S^{2/3}_{3}$ component of $S_3$ which can only be produced via $s$-channel exchange of photon and $Z$-boson. The situation for $S_1$ leptoquark is very similar to that of $S^{2/3}_{3}$ component, as $S_1$ does not couple to muon for the chosen benchmark scenarios and thus can only be produced through the mentioned $s$-channel processes. Therefore, with the main intention to probe the Yukawa-type couplings of the leptoquarks, we analyze the pair productions of  $S^{4/3}_{3}$ and $S^{1/3}_{3}$ components via $t$-channel contributions. The distinctive feature of these two components are found to be prominent here as well. For $S^{4/3}_{3}$,  the finalstate consisting of $2-{\rm jet}+2\mu$ (for BP1) and $2b-{\rm jet}+2\mu$ (for BP3) can probe the $Y^{22}_{S_3}$ coupling up to its perturbativity limit for $\mathcal{O}(10\,{\rm TeV})$ mass leptoquark with a very early data at muon collider. The reach calculated for the topology $2c+2\mu $ shows a lower sensitivity of $Y^{22}_{S_3} \sim 0.2$. To conclude, we find that the prospect of  the scalar leptoquarks and their different $SU(2)_L$ components can be distinguished and segregated with the complementarity of hadron and muon colliders.

\subsection*{Acknowledgments}
The authors thank Rahul Sinha for useful suggestions. P.B. and A.K. acknowledge SERB CORE Grant CRG/2018/004971 and MATRICS Grant MTR/2020/000668 for the financial support. S.P. acknowledges the Council of Scientific and Industrial Research (CSIR), India for funding his research (File no: 09/1001(0082)/2020-EMR-I). A.K also acknowledges the partial support by MCIN/AEI/10.13039/501100011033 Grant No. PID2020-114473GB-I00,
and Grant PROMETEO/2021/071 (Generalitat Valenciana).

	\begin{appendices}
\section{NLO QCD $K$-factors of SM backgrounds at the LHC/FCC}
	\label{sec:nlobg}

In \autoref{tab:nlobg}, we present the NLO QCD $K$-factors for the five dominant SM backgrounds considered at the analysis for the LHC/FCC. The calculation is performed in \texttt{MadGraph5\_AMC@NLO}\cite{Alwall:2011uj}, following prescriptions from ref. \cite{Alwall:2014hca}. The renormalization and factorization scales are set as the dynamic variable of $\sqrt{\hat{s}}$, and the PDF considered is  \texttt{NNPDF$\_$lo$\_$as$\_$0130$\_$qed}\cite{nnpdf}. The outcomes are compared with the various results from refs .\cite{Campbell:2011bn, Binoth:2008kt, Azzi:2019yne}.
	
	\begin{table}[h]
		\centering
		\renewcommand{\arraystretch}{1.3}
		\begin{tabular}{|c|c|c|c|}
			\hline
			\multirow{2}{*}{Background} & \multicolumn{3}{c|}{$K$-factors at three $E_{CM}$ values}\\
			\cline{2-4}
			& 14 TeV & 30 TeV & 100 TeV \\
			\hline
			$t\bar{t}$ & 1.52 & 1.51 & 1.52 \\
			\hline
			$VV$ &1.49 & 1.58 & 1.81 \\
			\hline
			$VVV$ & 1.77 & 2.05 & 2.74 \\
			\hline
			$t\bar{t}V$ & 1.58 & 1.59 & 1.60 \\
			\hline
			$tVV$ & 1.67 & 1.77 & 1.99 \\
			\hline
		\end{tabular}
		\caption{NLO QCD $K$-factors of the SM backgrounds at three different centre-of-mass energies at the LHC/FCC. \texttt{NNPDF$\_$lo$\_$as$\_$0130$\_$qed}\cite{nnpdf} has been taken as the PDF, with a dynamic scale choice of $\sqrt{\hat{s}}$ using \texttt{MadGraph5\_AMC@NLO}\cite{Alwall:2011uj}.}
		\label{tab:nlobg}
	\end{table}

\end{appendices}
\bibliography{References}
\bibliographystyle{Ref}

\end{document}